%% file: main.tex
\documentclass{article}

\usepackage{arxiv}
\input{macros}
\usepackage[utf8]{inputenc} 
\usepackage[T1]{fontenc}    
\usepackage{hyperref}       
\usepackage{url}            
\usepackage{booktabs}       
\usepackage{amsfonts}       
\usepackage{nicefrac}       
\usepackage{microtype}      
\usepackage{lipsum}

\title{Capuchin: Causal Database Repair for Algorithmic Fairness \thanks{This is an extended version of a paper that appeared at the Proceedings of the 2019 International Conference on Management of Data \cite{salimi2019interventional}.}}
\author{
  Babak Salimi \\
Computer Science and Engineering \\ University of Washington \\ Seattle WA\\
\texttt{bsalimi@cs.washington.edu}\\  \\
   \And
Luke Rodriguez \\
 Information School \\  University of Washington, \\ Seattle WA \\
 \texttt{rodriglr@uw.edu} \\
     \And
    Bill Howe \\
 Information School \\  University of Washington, \\ Seattle WA \\
\texttt{billhowe@uw.edu} \\
     \And
    Dan Suciu \\
Computer Science and Engineering \\ University of Washington \\ Seattle WA\\
\texttt{suciu@cs.washington.edu}} 
\begin{document}
\maketitle
\begin{abstract}
Fairness is increasingly recognized as a critical component of machine learning systems.  
However, it is the underlying data on which these systems are trained that often reflect discrimination, suggesting a database repair problem.  
Existing treatments of fairness rely on statistical correlations that can be fooled by statistical anomalies, such as Simpson's paradox.
Proposals for causality-based definitions of fairness can correctly model some of these situations, but they
require specification of the underlying causal models.  
In this paper, we formalize the situation as a database repair problem, proving sufficient conditions for fair classifiers in terms of
admissible variables as opposed to a complete causal model. We show that these conditions correctly capture subtle fairness violations.  
We then use these conditions as the basis for database repair algorithms that provide provable 
fairness guarantees about classifiers trained on their training labels.
We evaluate our algorithms on real data, demonstrating improvement over the state of the art
on multiple fairness metrics proposed in the literature while retaining high utility.
\end{abstract}

\input{intro}

\input{preli}

\input{transpar}

\input{repair}

\input{discussion}

\input{experiment}
\input{conc}

\clearpage 
\bibliographystyle{plain}
\bibliography{ref}
\input{appx}

\end{document}

%% file: macros.tex
\usepackage{amsfonts}
\usepackage{booktabs}
\usepackage{dsfont}
\usepackage{siunitx}
\usepackage{multirow}
\usepackage{graphicx}
\usepackage{listings}
\usepackage{commath}
\usepackage{epstopdf}
\usepackage{hyperref}
\usepackage{color}
\usepackage{url}
\usepackage{mathtools}
\usepackage{caption}
\usepackage{alltt}
\usepackage{mathrsfs}
\usepackage{float}
\usepackage{subcaption}
\usepackage{graphicx,fancyvrb}
\usepackage[colorinlistoftodos]{todonotes}
\usepackage[linesnumbered,ruled,vlined]{algorithm2e}
\usepackage{algpseudocode}
\usepackage{amsmath,amssymb}
\usepackage{booktabs}
\usepackage{caption}
\usepackage{float}
\usepackage{capt-of}
\usepackage{booktabs}
\usepackage{colortbl}
\usepackage{xcolor}
\usepackage{xfrac}
\usepackage{footnote}
\usepackage{array}
\usepackage{arydshln}
\usepackage{amsmath,amsfonts,amssymb}
\usepackage{todonotes}

\DeclarePairedDelimiterX{\infdivx}[2]{(}{)}{%
	#1\;\delimsize|\delimsize|\;#2%
}
\newcommand{\kld}[2]{\ensuremath{D_{KL}\infdivx{#1}{#2}}\xspace}

\newcommand{\reva}[1]{{ { { #1}}}}
\newcommand{\revb}[1]{{{ { #1}}}}

\newcommand{\revd}[1]{{{{ #1}}}}
\newcommand{\reve}[1]{{{ { #1}}}}

\definecolor{mygreen}{rgb}{0,0.6,0}
\definecolor{mygray}{rgb}{0.5,0.5,0.5}
\definecolor{mymauve}{rgb}{0.58,0,0.82}

\usepackage{listings}
\lstnewenvironment{VerbatimText}[1][]{
    
    \lstset{fancyvrb=true,basicstyle=\footnotesize,captionpos=b,xleftmargin=2em,#1}
}{}

\newcommand{\qii}{\delta}

\usepackage{booktabs}
\usepackage{pgfplots}
\usepackage{pgfplotstable}

\PassOptionsToPackage{hyphens}{url}\usepackage{hyperref}

\newcommand{\cm}{{{\mb M }}}

\newcommand{\cg}{ G}

\newcommand{\nipsadult}{{{Binned Adult}}}
\newcommand{\nipscompas}{{{Binned COMPAS}}}

\usepackage{amsmath}

\DeclareMathOperator*{\argmin}{argmin}

\newcommand{\E}{{\tt \mathbb{E}}}
\newcommand{\pr}{{\tt \mathrm{Pr}}}
\newcommand{\sys}{{\textsc{Capuchin}}}

\newcommand{\mc}[1]{\mathcal{#1}}

\newcommand{\ignore}[1]{}

\newcommand*{\rom}[1]{\expandafter\@slowromancap\romannumeral #1@}

\newcommand{\babak}[1]{{\texttt{\color{red} Babak: [{#1}]}}}
\newcommand{\bill}[1]{{\texttt{\color{olive} Bill: [{#1}]}}}

\newcommand{\indep}{\mbox{$\perp\!\!\!\perp$}}

\newcommand{\rep}{D'}
\newcommand{\icq}{Q_\varphi}
\newcommand{\hb}{D^*}

\newcommand{\att}{\mathbf{A}}

\newcommand\mvd{\twoheadrightarrow}

\newcommand{\fd}{\rightarrow}

\newcommand{\RNum}[1]{\uppercase\expandafter{\romannumeral #1\relax}}
\newcommand{\ie}{{\em i.e.}} 
\newcommand{\eg}{{\em e.g.}} 

\newcommand{\mmb}{{\bf MB}}

\newcommand{\mb}[1]{{\mathbf{#1}}}
\newtheorem{defn}{Definition}[section]

\newtheorem{example}[defn]{Example}
\newenvironment{myproof}[2] {\paragraph{Proof of {#1} {#2} :}}{\hfill$\square$}
\newtheorem{corollary}[defn]{Corollary}
\newtheorem{lemma}[defn]{Lemma}
\newtheorem{theorem}[defn]{Theorem}
\newtheorem{definition}[defn]{Definition}
\newtheorem{prop}[defn]{Proposition}

\newcommand{\proj}[1]{{\Pi}}
\newcommand{\sel}[1]{{\sigma}}

\newcommand{\cut}[1]{}
\newcommand{\eat}[1]{}

\newcommand{\defeq}{\stackrel{\text{def}}{=}}

\newcommand{\fairapp}{fairness application~}

%% file: intro.tex
\vspace*{-0.1cm}
\section{Introduction}
\label{sec:intro}

In 2014, a team of machine learning experts from Amazon Inc. began
work on an automated system to review job applicants' resumes.
According to a recent Reuters article~\cite{amazonhire2018}, the experimental system
gave job candidates scores ranging from one to five and was trained
on 10 years of recruiting data from Amazon.  However, by 2015
the team realized that the system showed a significant gender bias
towards male over female candidates because of
historical discrimination in the training data.  Amazon edited the system to
make it  gender agnostic, but there was no guarantee that discrimination did not occur through other means, and the project was
totally abandoned in 2017.

Fairness is increasingly recognized as a critical component of machine
learning (ML) systems, which make daily decisions that affect people's lives~\cite{courtland2018bias}.
The data on which these systems are trained reflect institutionalized
discrimination that can be reinforced and legitimized through automation.  A naive (and ineffective) approach sometimes used in practice is to
simply omit the protected attribute (say, race or gender) when
training the classifier. However, since the protected attribute is
frequently represented implicitly by some combination of proxy variables, the
classifier still learns the discrimination reflected in training data.
For example, zip code tends to predict race
due to a history of segregation~\cite{amazonrace2016,selbst2017disparate}; answers to personality
tests identify people with disabilities~\cite{bodie2017law,personalitywsj2017}; and keywords can reveal gender on a resume~\cite{amazonhire2018}.
As a result, a classifier trained
without regard to the protected attribute not only fails to remove discrimination, but it can complicate the detection and mitigation of
discrimination downstream via in-processing or post-processing
techniques~\cite{russell2017worlds,galhotra2017fairness,corbett2017algorithmic,chouldechova2017fair,kusner2017counterfactual,kilbertus2017avoiding,nabi2018fair,Veale:2018:FAD:3173574.3174014}, which we next describe.

%


The two main approaches to reduce or eliminate sources of discrimination are summarized in Fig.~\ref{tbl:famla}.
The most popular is the in-processing, where the ML algorithm itself is
modified; this approach must be reimplemented for every ML application.
The alternative is to process either the training data (pre-processing) or the output of the classifier itself (post-processing).
We advocate for the pre-processing strategy, which is agnostic to the choice of ML algorithm and instead interprets the problem as a database repair task.


%
One needs a quantitative measure of discrimination  in order to remove  it.
A large number of fairness definitions have been proposed (see Verma and Rubin for a recent discussion \cite{Verma:2018:FDE:3194770.3194776}),
which we broadly categorize in Fig.~\ref{tbl:famla}. The best-known measures are based on \emph{associative} relationships between the protected attribute
and the outcome.
For example, Equalized Odds requires that both protected and privileged groups have the same true positive (TP) and false positive (FP) rates.  However, it has been shown that associative definitions of fairness can be mutually exclusive~\cite{chouldechova2017fair} and fail to distinguish between discriminatory, non-discriminatory and spurious correlations between a protected attribute and the outcome of an algorithm~\cite{kilbertus2017avoiding,nabi2018fair,dwork2012fairness}.



\begin{example} \em \label{ex:berkeley} In a well-studied case, UC
	Berkeley was sued in 1973 for discrimination against females in
	graduate school admissions when it was found that 34.6\% of females
	were admitted in 1973 as opposed to 44.3\% of males.  It turned out
	that females tended to apply to departments with lower overall
	acceptance rates ~\cite{salimi2018bias}.  When broken down by department,  a slight bias toward female applicant
	was observed, \ignore{the admission
	rates for men and women were approximately
	equal} a result that did not constitute evidence for gender-based discrimination.
\end{example}

Such situations have recently motivated a search for a more principled measure
of fairness and discrimination based on \emph{causality}
\cite{kilbertus2017avoiding,nabi2018fair,kusner2017counterfactual,galhotra2017fairness,russell2017worlds}. These approaches measure the discriminatory causal influence of the protected attribute on the outcome of an algorithm.
However, they typically assume access to background information regarding the underlying causal model, which is unrealistic in practice.
For example, Kilbertus et al. assume the underlying casual model is provided as a structural equation model~\cite{kilbertus2017avoiding}.
Moreover, no existing proposals describe comprehensive systems for pre-processing data to mitigate causal discrimination.


This paper describes a new approach to removing discrimination by {\em
	repairing the training data} in order to remove the effect of any inappropriate and discriminatory
causal relationship between the protected attribute and classifier predictions, without assuming adherence to an underlying causal models.

\begin{figure*} \centering
	\begin{tabular}{ r|c|c| }
		\multicolumn{1}{r}{}
		&  \multicolumn{1}{c}{Associational}
		& \multicolumn{1}{c}{Causal} \\
		\cline{2-3}
		In-processing& \cite{kamishima2012fairness,pmlr-v54-zafar17a,calders2010three,kilbertus2017avoiding} & \cite{nabi2018fair,kilbertus2017avoiding,russell2017worlds} \\
		(Modify the ML Algorithm) & & \\ \cline{2-3}
		Pre/post-processing  &  \cite{feldman2015certifying,NIPS2017_6988,hardt2016equality,pmlr-v65-woodworth17a} & \sys \\
		(Modify the input/output Data) & & (this paper)\\ 	\cline{2-3}
	\end{tabular}
\caption{       \textmd{ Different categories of fairness aware machine-learning methods, based on whether they work with associations/causal definition and  whether they modifying algorithm/data to enforce fairness.}}
	\label{tbl:famla}
\end{figure*}

Our system, \sys, accepts a dataset consisting of a protected
attribute (e.g., gender, race, etc.), an outcome attribute (e.g.,
college admissions, loan application, or hiring decisions), and a set
of \emph{admissible variables}  through which it is
permissible for the protected attribute to influence the outcome.
For example, the applicant's choice of department in
Example~\ref{ex:berkeley} is considered admissible despite being correlated
with gender.  The system repairs the input data by inserting or removing tuples,
changing the empirical probability distribution to remove the
influence of the protected attribute on the outcome through any
causal pathway that includes inadmissible attributes.
That is, the repaired training data can be seen as a \emph{sample
	from a hypothetical fair world}. We make this notion more
precise in Section \ref{sec:fairdef}.



Unlike previous measures of fairness based on causality~\cite{nabi2018fair,kilbertus2017avoiding,russell2017worlds}, which require the
presence of the underlying causal model, our definition is based solely on
the notion of \emph{intervention}~\cite{pearl2009causality} and can be guaranteed even in the absence of causal models.
The user need only distinguish admissible and inadmissible attributes; we prove that this information is sufficient
to support the causal inferences needed to mitigate discrimination.

We use this \emph{interventional} approach to derive in
Sec.~\ref{sec:fairdef} a new fairness definition, called {\em
	justifiable fairness}. Justifiable fairness subsumes and improves on several
previous definitions and can correctly distinguish fairness violations
and non-violations that would otherwise be hidden by statistical
coincidences, such as Simpson's paradox.  We prove next, in
Sec.~\ref{sec:jfc}, that, if the training data satisfies a simple
saturated conditional independence, then any reasonable algorithm
trained on it will be fair.

Our core technical contribution, then, consists of a new approach to repair
training data in order to enforce the saturated conditional independence that guarantees fairness.
The database repair problem has been extensively studied in the
literature~\cite{DBLP:series/synthesis/2011Bertossi}, but in terms of database
constraints, not conditional independence. In Sec.~\ref{sec:repdata} we first define
the problem formally and then present a new technique to reduce it to a
multivalued functional dependency MVD~\cite{AbiteboulHVBook}.
Finally, we introduce new techniques to repair a dataset for an MVD
by reduction to the MaxSAT and Matrix Factorization problems.

We evaluate our approach in Sec~\ref{sec:exp} on two real datasets commonly studied
in the fairness literature, the adult dataset \cite{adult} and the COMPAS
recidivism dataset~\cite{valentino2012websites}. We find that our algorithms
not only capture fairness situations other approaches
cannot, but that they outperform the existing state-of-the-art
pre-processing approaches \emph{even on other fairness metrics for
	which they were not necessarily designed}. 
Surprisingly, our results show that our repair algorithms can mitigate
discrimination as well as prohibitively aggressive approaches, such as
dropping all inadmissible variables from the training set, while maintaining
high accuracy.  For example, our most flexible algorithm, which involves
a reduction to MaxSAT, can remove almost 50\% of the discrimination while
decreasing accuracy by only 1\% on adult data.

We make the following contributions:
\begin{itemize}
	\item We develop a new framework for causal fairness that does not require a complete causal model.
	\item We prove sufficient conditions for a fair classifier based on this framework.
	\item We reduce fairness to a database repair problem by linking causal inference to multivalued dependencies (MVDs).
	\item We develop a set of algorithms for the repair problem for MVDs.
	\item We evaluate our algorithms on real data and show that they meet our goals and outperform competitive methods on multiple metrics.
\end{itemize}


Section \ref{sec:peri} presents background on fairness and causality, while Section \ref{sec:counter_fair} describes sufficient conditions for
a fair classifier and derives the database repair problem.
In Section \ref{sec:repdata}, we present algorithms for solving the database repair problem and show, in Section \ref{sec:exp}, experimental evidence that our algorithms outperform the
state-of-the-art on multiple fairness metrics while preserving high utility.



%% file: preli.tex
\begin{table} \centering
	\centering
	\begin{tabular}{|l|l|} \hline
		\textbf{    Symbol} & \textbf{Meaning} \\ \hline
		$X, Y, Z$ & attributes (variables)\\
		$\mb{X}, \mb{Y}, \mb{Z}$ & sets of attributes \\
		$Dom(X), Dom(\mb{X})$ & their domains \\
		$x \in Dom(X), \mb{x} \in Dom(\mb{X})$ & a single value, a tuple of values \\
		$D$ & the database instance \\
		$\mb{V}$ & the attributes of the database $D$ \\
$\cm = \langle \mb U, \mb V , \mb F, \pr \rangle$ & Probabilistic Causal Model (PGM) \\
		$\cg$ & causal DAG \\
		$X \rightarrow Y$ & an edge in $\cg$ \\
		$\mb{Pa}(X)$ & the parents of $X$ in $\cg$ \\
		$\mb P$ & a path in $\cg$ \\
		$X \stackrel{*}{\rightarrow} Y$ & a directed path in $\cg$ \\
		$\mb Z \mvd \mb X$ & multivalued dependency (MVD) \\
		$\mb X \indep_{\pr} \mb Y | \mb Z$ or $\mb X \indep \mb Y | \mb Z$ & conditional independence \\
		$(\mb{X} \indep \mb{Y} |_d \ \mb{Z})$ & d-Separation in $\cg$. \\
		$\mmb(X)$ & The Markov boundary of $X$ \\
				$I$ & Inadmissible attributes \\
								$A$ & Admissible attributes \\
		\hline
	\end{tabular}
	\caption{      \textmd{ Notation used in the paper.}}
	\label{tab:notations}
\end{table}
{
	\section{Preliminaries}
	\label{sec:peri}
}


We review in this section the basic background on database repair, algorithmic
fairness and models of causality,  the
building blocks of our paper.

The notation used is summarized in Table~\ref{tab:notations}.  We
denote variables (i.e., dataset attributes) by uppercase letters,
$X, Y, Z, $ $V$; their values with lower case letters, $x,y,z, v$; and
denote sets of variables or values using boldface ($\mb X$ or
$\mb x$).  The domain of a variable $X$ is $Dom(X)$, and the domain of
a set of variables is $Dom(\mb X) = \prod_{Y\in \mb X} Dom(Y)$. In
this paper, all domains are discrete and finite; continuous domains
are assumed to be binned, as is typical.  A {\em database instance} $D$
is a relation whose attributes we denote as $\mb V$.  We assume set
semantics (\ie,\ no duplicates) unless otherwise stated, and we denote
the cardinality of $D$ as $n = |D|$.  Given a partition
$\mb X \cup \mb Y \cup \mb Z= \mb V$, we say that $D$ satisfies the
{\em multivalued dependency} (MVD) $\mb Z \mvd \mb X $ if
$D= \Pi_{\mb X \mb Z}(D) \Join \Pi_{\mb Z \mb Y}(D)$.

Typically, training data for ML is a bag $B$.  We convert it into a
set $D$ (by eliminating duplicates) and a probability distribution
$\pr$, which accounts for
multiplicies;\footnote{$\pr(\mb v) \defeq \frac{1}{|B|}\sum_{t \in
    B}1_{t=\mb v}$.} We call $D$ the support of $\pr$.  We say that
$\pr$ is {\em uniform} if all tuples have the same probability.  We
say $\mb X$ and $\mb Y$ are {\em conditionally independent (CI)} given
$\mb Z$, written $(\mb X \indep_\pr \mb Y | \mb Z)$, or just
$(\mb X \indep \mb Y | \mb Z)$
if $\pr(\mb x| \mb y, \mb z) =\pr(\mb x| \mb z)$ whenever
$\pr(\mb y, \mb z) > 0 $.  Conditional independences satisfy the
Graphoid axioms~\cite{pearl1985graphoids}, which are reviewed in
Appendix~\ref{app:addback} and are used in proofs.  When
$\mb V = \mb X \mb Y \mb Z$, then the CI is said to be {\em saturated}.
A uniform $\pr$ satisfies a saturated CI iff its support
$D$ satisfies the MVD $\mb Z \mvd \mb X$.  Training data usually does
not have a uniform $\pr$, and in that case the equivalence between the
CI and MVD fails~\cite{wong2000implication}; we address this issue in
Sec.~\ref{sec:repdata}.

The {\em database repair problem} is the following: we are given a set
of constraints $\Gamma$ and a database instance $D$, and we need to
perform a minimal set of updates on $D$ such that the new database
$\rep$ satisfies $\Gamma$~\cite{DBLP:series/synthesis/2011Bertossi}.
The problem has been studied extensively in database theory for
various classes of constraints $\Gamma$.  It is NP-hard even
when $D$ consists of a single relation (as it does in our paper)
and $\Gamma$ consists of functional
dependencies~\cite{DBLP:conf/pods/LivshitsKR18}.  In our setting,
$\Gamma$ consists of conditional independence statements, and it
remains NP-hard, as we show in Sec.~\ref{sec:repdata}.

\vspace*{0.1cm}
\subsection{Background on  Algorithmic Fairness}
\label{sec:cvs}

\begin{figure*}  \centering
	\begin{tabular}{|l|l|} \hline
		\textbf{     Fairness Metric}  & \textbf{Description}  \\ \hline
		Demographic Parity (DP) \cite{calders2009building}  & $S \indep O$  \\
		a.k.a. Statistical Parity \cite{dwork2012fairness} & \\
		or Benchmarking \cite{simoiu2017problem} & \\ \hline
		Conditional Statistical parity \cite{corbett2017algorithmic} & $S \indep O |\mb A$ \\ \hline
		%
		Equalized Odds (EO) \cite{hardt2016equality} \footnotemark[2] & $S \indep O| Y $  \\
		a.k.a.  Disparate Mistreatment  \cite{zafar2017fairness} & \\ \hline
		Predictive Parity (PP)\cite{chouldechova2017fair} \footnotemark[3]  & $S \indep Y| O $  \\
		a.k.a. Outcome Test  \cite{simoiu2017problem} & \\
		or Test-fairness  \cite{chouldechova2017fair} & \\
		or Calibration \cite{chouldechova2017fair}, & \\
		or Matching Conditional Frequencies \cite{hardt2016equality} & \\ \hline
	\end{tabular}  
	\caption{      \textmd{ Common associational definitions of fairness and their conditional independence statement counterparts.}} \label{tbl:asdef}
\end{figure*}

Algorithmic fairness considers a {\em protected attribute} $S$, the
{\em response variable} $Y$, and a prediction algorithm
$A : Dom(\mb X) \rightarrow Dom(O)$, where $\mb X \subseteq \mb V$,
whose prediction is denoted $O$ (some references denote it $\tilde Y$)
and called\ignore{ {\em output} or} {\em outcome}.  For simplicity, we
assume $S$ classifies the population into protected $S=1$ and
privileged $S=0$, for example, female and male.  Fairness definitions
can be classified as associational or causal.


\paragraph{\bf Associational fairness} is based on statistical measures on the
variables of interest; a summary is shown in Fig.~\ref{tbl:asdef}.
\textit{Demographic Parity} (DP)
\cite{calders2010three,kamiran2009classifying,zemel2013learning,simoiu2017problem,dwork2012fairness}, requires an algorithm to classify both
the protected and the privileged group with the same probability,
$\pr(O=1|S=1)=\pr(O=1|S=0)$. As we saw
in Example~\ref{ex:berkeley}, the lack of statistical parity cannot be
considered as evidence for gender-based discrimination; this
has motivated the introduction of \textit{Conditional Statistical Parity} (CSP)
\cite{corbett2017algorithmic}, which controls for a set of
admissible factors  $\mb A$, i.e.,
$\pr(O=1|S=1,\mb A= \mb a)=\pr(O=1|S=0,\mb A=\mb a)$.  
 Another popular measure used for predictive
classification algorithms is  \textit{Equalized Odds} (EO), which
requires that both protected and privileged groups to have the same false
positive (FP) rate, $\pr(O=1|S=1,Y=0)=\pr(O=1|S=0,Y=0)$ , and the same
false negative (FN) rate, $\pr(O=0|S=1,Y=1)=\pr(O=0|S=0,Y=1)$ \ignore{, or,
equivalently, $(O \indep S|Y)$.} Finally, \textit{Predictive Parity} (PP)
requires that both protected and unprotected groups have the same
predicted positive value (PPV),
$\pr(Y=1|O=i,S=0)=\pr(Y=1|O=i,S=1) \ \text{for} \ i,=\{1,0\} $. It has been shown that these measures can be mutually exclusive \cite{chouldechova2017fair} (see Appendix \ref{app:addback}).

\paragraph{\bf  Causal fairness \cite{kusner2017counterfactual,kilbertus2017avoiding,nabi2018fair,russell2017worlds,galhotra2017fairness}}
was motivated by the need to address difficulties generated by
associational fairness and assumes an underlying causal model.  We
first discuss causal DAGs before reviewing causal fairness.

\subsection{Background on Causal DAGs }

\label{sec:causal:dag}

We now review causal directed acyclic graphs (DAGs) and refer the reader
to Appendix \ref{app:addback} and \cite{pearl2009causality} for more details.

\paragraph*{\bf Causal DAG}
A \textit{causal DAG} $\cg$ over set of variables $\mb V$ is a directed acyclic graph that
models the functional interaction between variables in $\mb V$.  Each node $X$ represents
a variable in $\mb V$ that is functionally determined by: (a) its parents $\mb{Pa}(X)$
in the DAG, and (b) some set of \emph{exogenous} factors that need not appear in the DAG,
as long as they are mutually independent.
This functional interpretation leads to the same decomposition of the  joint probability distribution of $\mb V$ that characterizes Bayesian networks \cite{pearl2009causality}:
\begin{align}
\pr(\mb V) = & \prod_{X \in \mb V} \pr(X | \mb{Pa}(X)) \label{eq:bayesian1}
\end{align}

\paragraph*{\bf $d$-Separation and Faithfulness}
A common inference question in a causal DAG is how to determine whether a
CI $(\mb X \indep \mb Y | \mb Z)$ holds.  A sufficient criterion is
given by the notion of d-separation,  a syntactic condition
$(\mb X \indep \mb Y |_d \mb Z)$ that can be checked directly on the
graph.  $\pr$ and $\cg$ are
called {\em Markov compatible} if $(\mb X \indep \mb Y |_d \mb Z)$
implies $(\mb X \indep_\pr \mb Y | \mb Z)$; if the converse
implication holds, then we say that $\pr$ is {\em faithful} to $\cg$.
The following is known:

\begin{prop} \label{prop:d:separation}  If $\cg$ is a causal DAG and
  $\Pr$ is given by Eq.(\ref{eq:bayesian1}), then they are Markov
  compatible.
\end{prop}

\begin{figure*} 
	\hspace*{.8cm}	\begin{subfigure}{0.4\textwidth} 
	\hspace*{.8cm}	\includegraphics[scale=0.15]{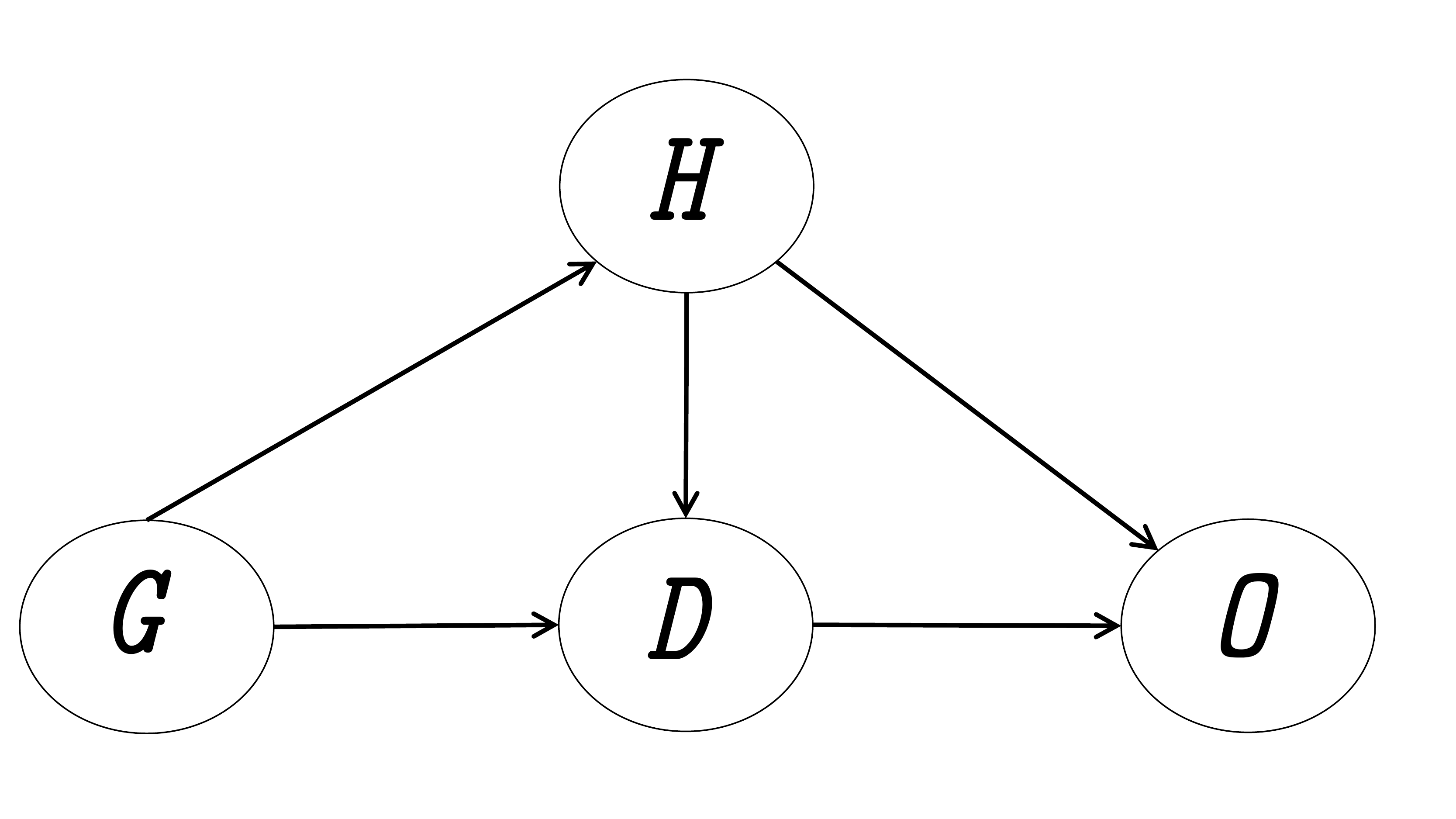}
	\caption{}
	\end{subfigure}
	\begin{subfigure}{0.4\textwidth} 
	\hspace*{.8cm}	\includegraphics[scale=0.15]{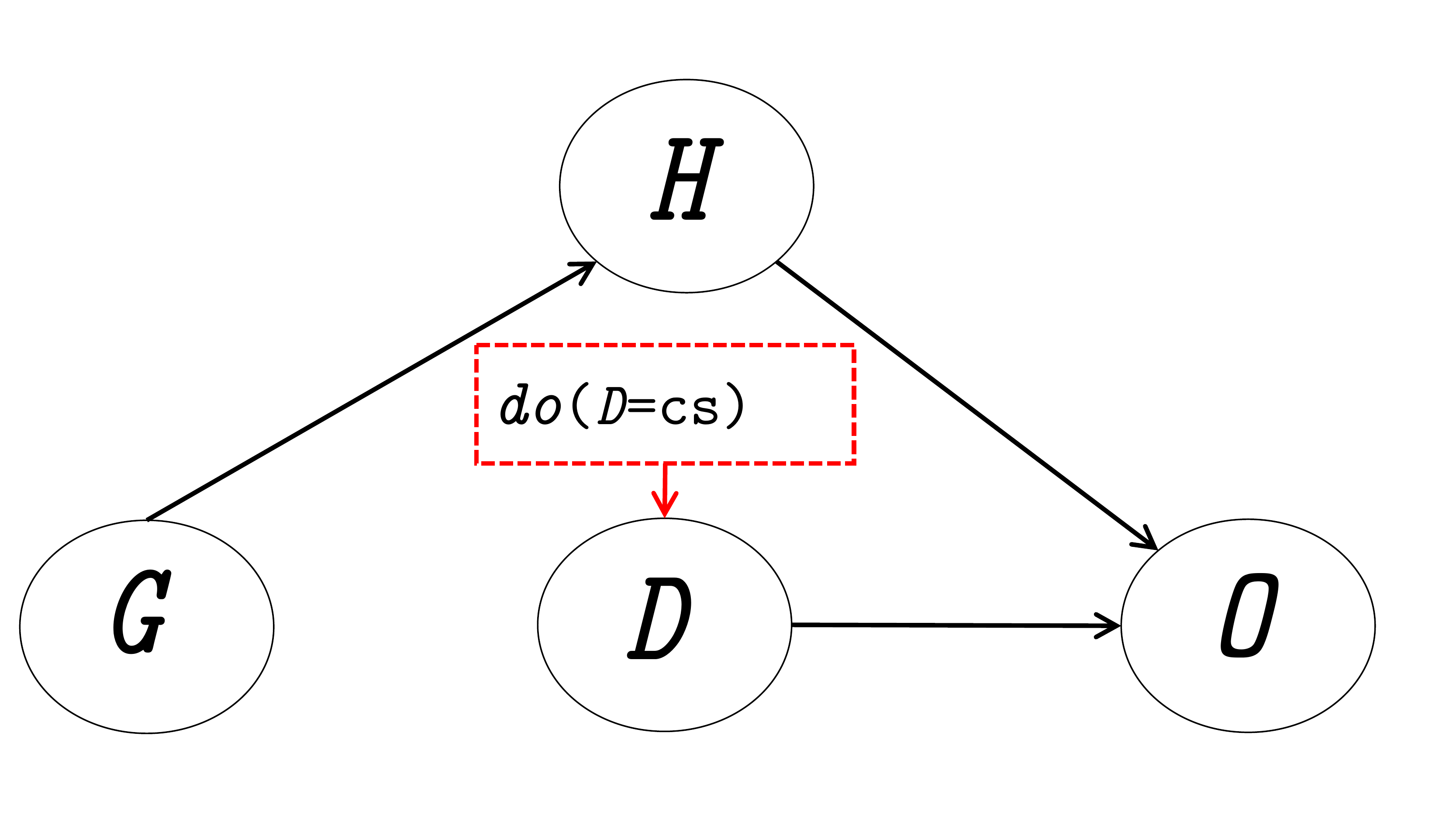}
		\caption{ }
	\end{subfigure}
	\caption{(a) Represents a causal DAG with $A=$ admission outcome, $G=$ applicant's gender,
		$H=$ applicant's hobbies and $D=$ applicant's choice of department (cf. Ex.~\ref{ex:cmexm}).
		(b) Represents  the causal DAG obtained after external interventions (cf. Ex.~\ref{ex:cmexm}).}
	\label{fig:cgex}
\end{figure*}

\paragraph*{\bf Counterfactuals and \texttt{do} Operator}
A \textit{counterfactual} is an intervention where we actively modify
the state of a set of variables $\mb X$ in the real world to some
value $\mb X= \mb x$ and observe the effect on some output $Y$.
Pearl~\cite{pearl2009causality} described the $do$ operator that
allows this effect to be computed on a causal DAG, denoted
$\pr(Y|do(X=x))$.  To compute this value, we assume that $X$ is
determined by a constant function $X=x$ instead of a function provided
by the causal DAG.  This assumption corresponds to a modified graph
with all edges into $\mb X$ removed, and values of these variables are
set to $\mb x$.  The Bayesian rule Eq.(\ref{eq:bayesian1}) for the
modified graph defines $\pr(Y|do(\mb X=\mb x))$; the exact expression
is in \cite[Theorem 3.2.2]{pearl2009causality}. We give an
alternative and, to our best knowledge, new formula expressed by
introducing some compensating factors; the proof is in
Appendix~\ref{app:proof}: 

\begin{theorem}  \label{theo:af} Given a causal DAG $\cg$ and a set of
  variables $\mb X \subseteq \mb V$, suppose
  $\mb X=\{X_0,X_1 \ldots X_m \}$ are ordered such that $X_i$ is a
  non-descendant of $X_{i+1}$ in $\cg$. The effect of a set of
  interventions $do(\mb X= \mb x )$ is given by the following {\em
    extended adjustment formula}:
{
	\begin{align}
\pr(y|do(\mb X= \mb x )) &=& \nonumber \\
    &&  \hspace*{-2cm}\sum_{ \mb z \in Dom(\mb Z)} \pr( y | \mb x, \mb z )  \bigg( \prod_{i=0}^{m} \pr\big(\mb{pa}(X_i)\bigg|  \bigcup_{j=0}^{i-1} \mb{pa}(X_j ), \bigcup_{j=0}^{i-1}x_j  \big) \bigg) \label{eq:af}
	\end{align}
}
	\noindent where $\mb Z= \bigcup_{X \in \mb X} \mb{Pa}(X)$ and
	$j\geq 0$.
\end{theorem}
In particular, if $\mb X$ has no parents, then intervention coincides
with conditioning, $\pr(y|do(\mb X = \mb x)) = \pr(y|\mb X = \mb x)$.


\begin{example} \em \label{ex:cmexm} Continuing Example~\ref{ex:berkeley},
	Fig.~\ref{fig:cgex}(a) shows a small fragment of the causal DAG of
	the admission process in a college.  Admissions decisions are made
	independently by each department and are based on a rich collection
	of information about the candidates, such as test scores, grades,
resumes, statement of purpose, etc.  These characteristics affect not
	only the admission decisions, but also which department the
	candidate chooses to apply to.  We
	show only a tiny fragment of the causal graph, where $O=$ admission
	outcome, $D=$ department, $G=$ candidate's gender, and $H=$
	hobbies, which can be influenced by gender.
	\footnote{In the Amazon hiring
		example~\cite{amazonhire2018}, hobbies correlated with gender, \eg,\ {\em Captain of the women's chess team}.}
	The admissions office anonymizes gender, but it does consider extracurricular activities
	such as hobbies, so we include an edge $H \rightarrow O$.  Since different
	genders apply to departments at different rates, there is an edge $G \rightarrow D$.
	Some departments may tend to attract applicants with certain hobbies (e.g.,
	the math department may attract applicants who play chess),
	so we also include an edge $H \rightarrow D$.  The joint distribution is given by
	%
{	\begin{align}
	\pr(g,h,d,o)=\pr(g)\pr(h|g)\pr(d|g,h)\pr(o|h,d) \label{exjd1}
	\end{align}}
      Consider the counterfactual: {\em update the
        applicant's department to $cs$}.  We compare
      the marginal probability of $O$, the conditional probability,
      and the intervention:
      {
	\begin{align}
	\pr(o|D=\text{cs}) = & \sum_{g,h} \pr(g) \pr(h|g) \pr(D=\text{cs}|g,h) \pr(o|D=\text{cs},h)\nonumber \\
	\pr(o|do(D=\text{cs})) = & \sum_{g,h} \pr(g)\pr(h|g)  \pr(o|D=\text{cs},h) \label{exjd2}
	\end{align}
}
The expression for intervention (\ref{exjd2}), based on
\cite[Theorem 3.2.2]{pearl2009causality} is obtained from the
conditional probability by removing the term $\pr(D=\text{cs}|g,h)$,
or equivalently deleting the edge $G\rightarrow D$ from the
graph in Fig.~\ref{fig:cgex}(b).  Alternatively, we can express the
intervention using Eq.(\ref{eq:af}) (notice that
$\mb{Pa}(D) = \set{G,H}$):
{	\begin{align}
	\pr(o|do(D=\text{cs})) = & \sum_{g,h} \pr(o|g,h,D=\text{cs})\pr(h|g)  \pr(g) \label{eq:babak:do}
	\end{align}}
	%
The reader may check that Eq.(\ref{exjd2}) and (\ref{eq:babak:do}) are equivalent.

\end{example}


\vspace*{-0.2cm}



\subsection{Causal Fairness}
\label{sec:cf}
  \paragraph*{\bf Counterfactual Fairness}
  Kusner et
  al.~\cite{kusner2017counterfactual,DBLP:journals/corr/KusnerLRS17}
  (see also the discussion in~\cite{loftus2018causal}) defined a
  classifier as {\em counterfactually fair} if the protected attribute
  of an individual is not a cause of the outcome of the classifier for
  that individual, i.e., had the protected attributes of the
  individual been different, and other things being equal, the outcome
  of the predictor would have remained the same. However, the
  definition of counterfactual fairness in
  \cite{kusner2017counterfactual} captures individual-level fairness
  only under certain assumptions (see Appendix~\ref{app:addback}).
  \ignore{In terms of causal DAGs, a classifier is counterfactually
    fair if there is no directed path from $S$ to $\tilde{Y}$.}
  Indeed, it is known in statistics that individual-level
  counterfactuals can not be estimated from data \cite{rubin1970thesis,rubin1986statistics,rubin2008comment}.
%

%

  \paragraph*{\bf Proxy Fairness} To avoid individual-level
  counterfactuals, a common is to study
  population-level counterfactuals or interventional distributions
  that capture the effect of interventions at population level rather
  than individual level
  \cite{pearl2009causal,rubin1970thesis,rubin1986statistics}.
  Kilbertus et. al. \cite{kilbertus2017avoiding} defined proxy
  fairness as follows:
  {
    \begin{align}
      P(\tilde{Y}=1| do(\mb P=\mb p ))=P(\tilde{Y}=1| do(\mb  P=\mb p')) \label{eq:pfair}
    \end{align}
  }
  \noindent for any $\mb p, \mb p' \in Dom(\mb P)$, where $\mb P$
  consists of proxies to a sensitive variable $S$ (and might include
  $S$).  Intuitively, a classifier satisfies proxy fairness in
  Eq~\ref{eq:pfair}, if the distribution of $\tilde{Y}$ under two
  interventional regimes in which $\mb P$ set to $\mb p$ and $\mb p'$
  is the same. Thus, proxy fairness is not an individual-level notion.
  Next example shows proxy fairness fails to capture group-level
  discrimination in general.  
%

%
  \begin{example} \label{ex:obsvsprox} \em To illustrate the difference
    between counterfactual and proxy fairness, consider the college
    admission example.  Both departments make decisions based on
    students' gender and qualifications, $O=f(G,D,Q)$, for a binary
    $G$ and $Q$. The causal DAG is
    $G \rightarrow O, D \rightarrow O, Q \rightarrow O$. Let $D=U_D$
    and $Q=U_Q$, where $U_D$ and $U_Q$ are exogenous factors that are
    independent and that are uniformly distributed, e.g.,
    $P(U_Q=1)=P(U_Q=0)=\frac{1}{2}$. Further suppose
    $f(G, \text{'A'},Q)= G \land Q$ and
    $f(G, \text{'B'},Q)= (1-G) \land Q$, i.e., dep.~A admits only
    qualified males and dep.~B admits only qualified females.  This
    admission process is proxy-fair\footnote{Here $D$ is not a proxy
      to $G$, because $D \indep G$ by assumption.}, because
    $P(O=1| do(G=1))=P(O=1| do(G=0))=\frac{1}{2}$.  On the other hand,
    it is clearly individually-unfair, in fact it is group-level
    unfair (for all applicants to the same department).  To capture
    individual fairness, counterfactual
    fairness~\cite{kusner2017counterfactual,DBLP:journals/corr/KusnerLRS17}
    is a non-standard definition that does both conditioning {\em and}
    intervention on the sensitive attribute.
    \ignore{($P(-|G=g,\texttt{do}(G=g))$.}  Conditioning ``extracts
    information from the individual to learn the background
    variables''~\cite[pp.11, footnote 1]{loftus2018causal}.
  \end{example}

\paragraph*{\bf Path-specific fairness} These definitions are based on
graph properties of the causal graph, \eg, prohibiting specific paths from the
sensitive attribute to the outcome~\cite{nabi2018fair,loftus2018causal}; however,
identifying path-specific causality from data requires very strong assumptions and is often impractical~\cite{avin2005identifiability}.

%% file: transpar.tex
\vspace*{-0.2cm}
\section{Defining and Enforcing Algorithmic Fairness}
\label{sec:counter_fair}

\revb{In this section we introduce a new definition of fairness,
  which, unlike proxy fairness~\cite{kilbertus2017avoiding}, captures
  correctly group-level fairness, and, unlike counterfactual
  fairness~\cite{kusner2017counterfactual,DBLP:journals/corr/KusnerLRS17}
  is based on the standard notion of intervention and, hence, it is
  testable from the data.  In the next section we will describe how to
  repair an unfair training dataset to enforce fairness.
}


\subsection{{Interventional Fairness}}
\label{sec:fairdef}

This section assumes that the causal graph is given. The
algorithm computes an output variable $O$ from input variables $\mb X$
(Sec.~\ref{sec:cvs}). We begin with a definition describing when an
outcome $O$ is causally independent of the protected attribute $S$ for
any possible configuration of a given set of variables $\mb K$.

\begin{definition}[$\mb K$-fair]   \label{def:cnf} Fix a set of
  attributes $\mb K \subseteq \mb V-\set{S,O}$.  We say that an
  algorithm $\mc A : Dom(\mb X) \rightarrow Dom(O)$ is $\mb K$-fair
  w.r.t. a protected attribute $S$ if, for any context $\mb K = \mb k$
  and every outcome $O=o$, the following holds:
	{ 
	\begin{eqnarray}
	\pr(O=o| do(S=0),do(\mb K = \mb k))= \pr(O=o| do(S=1), do(\mb K = \mb k)) \label{eq:cfair}
	\end{eqnarray}
}
\end{definition}



\revb{We call an algorithm \reve{{\em interventionally fair}} if it is
  $\mb K$-fair for every set $\mb K$.  Unlike proxy fairness, this
  notion captures correctly group-level fairness, because it ensures
  that $S$ does not affect $O$ in \emph{any configuration} of the
  system obtained by fixing other variables at some arbitrary values.
  Unlike counterfactual fairness, it does not attempt to capture
  fairness at the individual level, and therefore it uses the standard
  definition of intervention (the \texttt{do}-operator).  In fact, we
  argue that interventional fairness is the strongest notion of
  fairness that is testable from data, yet captures correctly
  group-level fairness. We illustrate with an example (see also
  Ex~\ref{ex:ap_str}).}



\begin{example} \label{ex:intsprox} \em  In contrast to proxy
    fairness, interventional fairness correctly identifies the
    admission process in Ex.  \ref{ex:obsvsprox} as unfair at
    department-level. This is because the admission process fails to
    satisfy $\set{D}$-fairness since,
    $P(O=1| do(G=0), do(D=\text{'A'}))=0$ but
    $P(O=1| do(G=1), do(D=\text{'A'}))=\frac{1}{2}$. Therefore,
    interventional fairness is a more fine-grained notion than proxy
    fairness.
  \reve{We note however that, interventional fairness does not
    guarantee individual faintness in general. To see this suppose the
    admission decisions in both departments are based on student's
    gender and an unobserved exogenous factor $U_O$ that is uniformly
    distributed, i.e., $O=f(G,U_O)$, such that $f(G, 0)= G$ and
    $f(G,1)=1-G$. Hence, the causal DAG is $G \rightarrow O$.  Then
    the admission process is $\emptyset$-fair because,
    $P(O=1| do(G=1))=P(O=1| do(G=0))=\frac{1}{2}$. Therefore, it is
    interventionally fair (since $\mb V-\set{O,G} = \emptyset$).
    However, it is clearly unfair at individual level.  If the
    variable $U_0$ were endogenous (i.e. known to the algorithm), then
    the admission process is no longer interventionally fair, because
    it is not $\set{U_o}$-fair:
    $P(O=1| do(G=1), do(U_o=1))= P(O=1| G=1, U_o=1)=0$, while
    $P(O=1| do(G=1), do(U_o=1))=P(O=1| G=0, U_o=1)=1$. Under the same setting counterfactual fairnesses \cite{kusner2017counterfactual,DBLP:journals/corr/KusnerLRS17} fails to capture individual-level discrimination in this example (see Appendix~\ref{app:addback}).
}
\end{example}


In practice, interventional fairness is too restrictive, as we show
below. To make it practical, we allow the user to classify variables
into {\em admissible} and {\em inadmissible}.  The former variables
through which it is permissible for the protected attribute to
influence the outcome.  In Example~\ref{ex:berkeley}, the user would
label department as admissible since it is considered a fair use in
admissions decisions, and would (implicitly) label all other variables
as inadmissible, for example, hobby.  Only users can identify this
classification, and therefore admissible variables are part of the
problem definition:


\begin{definition}[Fairness application]  \label{def:app} A \fairapp
  over a domain $\mb V$ is a tuple $(\mc A, S, \mb A, \mb I)$, where
  $\mc A$ is an algorithm $Dom(\mb X) \rightarrow Dom(O)$;
  $\mb X \subseteq \mb V$ are its input variables; $S,O \in \mb V$ are
  the protected attribute and outcome, and
  $\mb A \cup \mb I = \mb V - \set{S,O}$ is a partition of the
  variables into admissible and inadmissible.
\end{definition}

We can now introduce our definition of fairness:

\begin{definition}[Justifiable fairness]  \label{def:jf} A \fairapp 
  $(\mc A,$ $S, \mb A, \mb I)$ is {\em justifiability fair} if it is
  $\mb K$-fair w.r.t. all supersets $\mb K \supseteq \mb A$.
\end{definition}


Notice that
interventional fairness corresponds to the case where no variable is
admissible, i.e., $\mb A = \emptyset$.


We give next a characterization of justifiable fairness in terms of
the structure of the causal DAG:
\vspace*{-0.1cm}
\begin{theorem} \label{th:jfcd}  If all directed paths from $S$ to $O$
  go through an admissible attribute in $\mb A$, then the algorithm is
  justifiably fair.  If the probability distribution is faithful to
  the causal DAG, then the converse also holds.
\end{theorem}

To  ensure interventional fairness,  a sufficient
condition is that there exists no path from $S$ to $O$ in the causal
graph (because $\mb A = \emptyset$).  \reve{(Hence, under faithfulness, interventional fairness implies fairness at individual-level, i.e., intervening on the sensitive attribute does not change the counterfactual outcome of individuals.}) Since this is too strong in most
scenarios, we adopt justifiable fairness instead.
\begin{figure*} \centering
	\begin{subfigure}{0.3\textwidth} 
		\hspace*{-0.8cm}
		\includegraphics[scale=0.15]{Fig2_a.pdf}
		\caption{College~\RNum{1}}
	\end{subfigure}
	\begin{subfigure}{0.3\textwidth} 
		\includegraphics[scale=0.15]{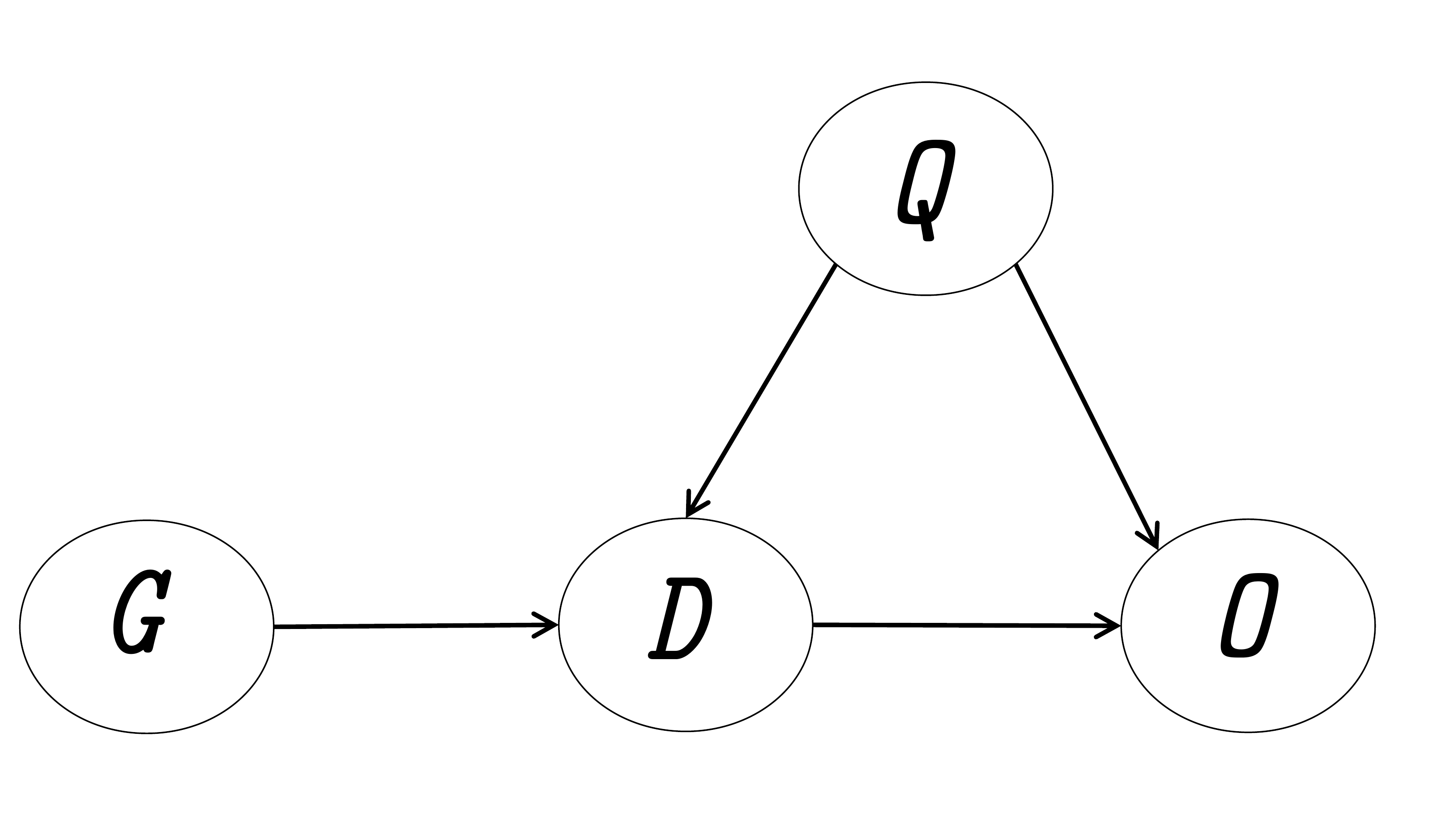}
		\caption{ College~\RNum{2}}
	\end{subfigure}

	{
		\begin{tabular}{c@{\qquad}ccc@{\qquad}ccc}
			\toprule
			\multirow{2}{*}{\raisebox{-\heavyrulewidth}{{\bf College~\RNum{1}}}} & \multicolumn{2}{c}{Dept.~A} & \multicolumn{2}{c}{Dept.~B} & \multicolumn{2}{c}{Total}  \\
			\cmidrule{2-7}
			& Admitted & Applied & Admitted &Applied & Admitted & Applied \\
			\midrule
			Male& 16 &  20 & 16 & 80 & 32 & 100 \\
			Female& 16 & 80 & 16 & 20 & 32 & 100 \\
			\bottomrule
		\end{tabular}
\hspace*{-0.2cm}			\begin{tabular}{c@{\qquad}ccc@{\qquad}ccc}
			\toprule
			\multirow{2}{*}{\raisebox{-\heavyrulewidth}{\bf College~~\RNum{2}}} & \multicolumn{2}{c}{Dept.~A} & \multicolumn{2}{c}{Dept.~B} & \multicolumn{2}{c}{Total}  \\
			\cmidrule{2-7}
			& Admitted & Applied & Admitted &Applied & Admitted & Applied \\
			\midrule
						Male& 10 & 10 & 40 & 90 & 50 & 100 \\
			Female& 40 &  50 & 10 & 50 & 50 & 100 \\

			\bottomrule
		\end{tabular}
	}
	\caption{      \textmd{ Admission process representation in two colleges where  the associational notions of fairness fail (see Ex.\ref{ex:ap_str}).}}
	\label{adm_rate}
\end{figure*}
We illustrate with an example.
%
%

\begin{example} \label{ex:ap_str} \em Fig~\ref{adm_rate} shows how fair or
  unfair situations may be hidden by coincidences but exposed through
  causal analysis.  In both examples, the protected attribute is
  gender $G$, and the admissible attribute is department $D$. Suppose
  both departments in College~\RNum{1} are admitting only on the basis
  of  their applicants' hobbies.  Clearly, the admission process is
  discriminatory in this college because department~A admits 80\% of
  its male applicants and 20\% of the female applicants, while
  department~B admits 20\% of male and 80\% of female applicants.  On
  the other hand, the admission rate for the entire college is the
  same 32\% for both male and female applicants, falsely suggesting
  that the college is fair.  Suppose $H$ is a proxy to $G$ such that $H=G$ ($G$ and $H$ are the same), then proxy
  fairness classifies this example as
  fair: indeed, since Gender has no parents in the causal graph,
  intervention is the same as conditioning, hence
  $\pr(O=1|do(G=\text{i}))=\pr(O=1| G = \text{i})$ for $i=0,1$.  Of the previous methods, only conditional
  statistical parity correctly indicates discrimination.  We
  illustrate how our definition correctly classifies this examples as
  unfair.  Indeed, assuming the user labels the department $D$ as
  admissible, $\set{D}$-fairness fails because, by Eq.(\ref{eq:af}),
  $\pr(O=1|do(G=1),do(D=\text{'A'}))=\sum_h
  \pr(O=1|G=1,D=\text{'A'},h)\pr(h|G=1) =
  \pr(O=1|G=1,D=\text{'A'})=0.8$,
%
%
%
%
  and, similarly $\pr(O=1|do(G=0),do(D=\text{'A'}))=0.2$.
  Therefore, the admission process is not justifiably fair.

  Now, consider the second table for College~\RNum{2}, where both
  departments~A and B admit only on the basis of student
  qualifications $Q$.  A superficial examination of the data suggests
  that the admission is unfair: department A admits 80\% of all
  females, and 100\% of all male applicants; department B admits 20\%
  and 44.4\% respectively.  Upon deeper examination of the causal DAG,
  we can see that the admission process is justifiably fair because
  the only path from Gender to the Outcome goes through department,
  which is an admissible attribute.  To understand how the data could
  have resulted from this causal graph, suppose 50\% of each gender
  have high qualifications and are admitted, while others are
  rejected, and that 50\% of females apply to each department but
  more qualified females apply to department A than to B (80\%
  v.s. 20\%). Further, suppose fewer males apply to department A, but
  all of them are qualified.  The algorithm satisfies demographic
  parity and proxy fairness but fails to satisfy conditional
  statistical parity since $\pr(A=1|G=1,D=\text{A})=0.8$ but
  $\pr(A=1|G=0,D=\text{A})=0.2)$.  Thus, conditioning on $D$ falsely
  indicates discrimination in College~\RNum{2}. One can check that the
  algorithm is justifiably fair, and thus our definition also
  correctly classifies this example; for example, $\set{D}$-fairness
  follows from Eq.(\ref{eq:af}):
  $\pr(O=1|do(\text{G}=\text{i}),do(D=d))= \sum_{q}\pr(O=1|G=i,d,q))
  \nonumber \pr(q|G=i)= \frac{1}{2}$.
  To summarize, unlike previous definitions of fairness, justifiable
  fairness correctly identifies College~\RNum{1} as discriminatory and
  College~\RNum{2} as fair.

\end{example}

\ignore{
	\begin{figure}
		\begin{subfigure}{0.2\textwidth} \centering
			\hspace*{-0.8cm}
			\includegraphics[scale=0.1]{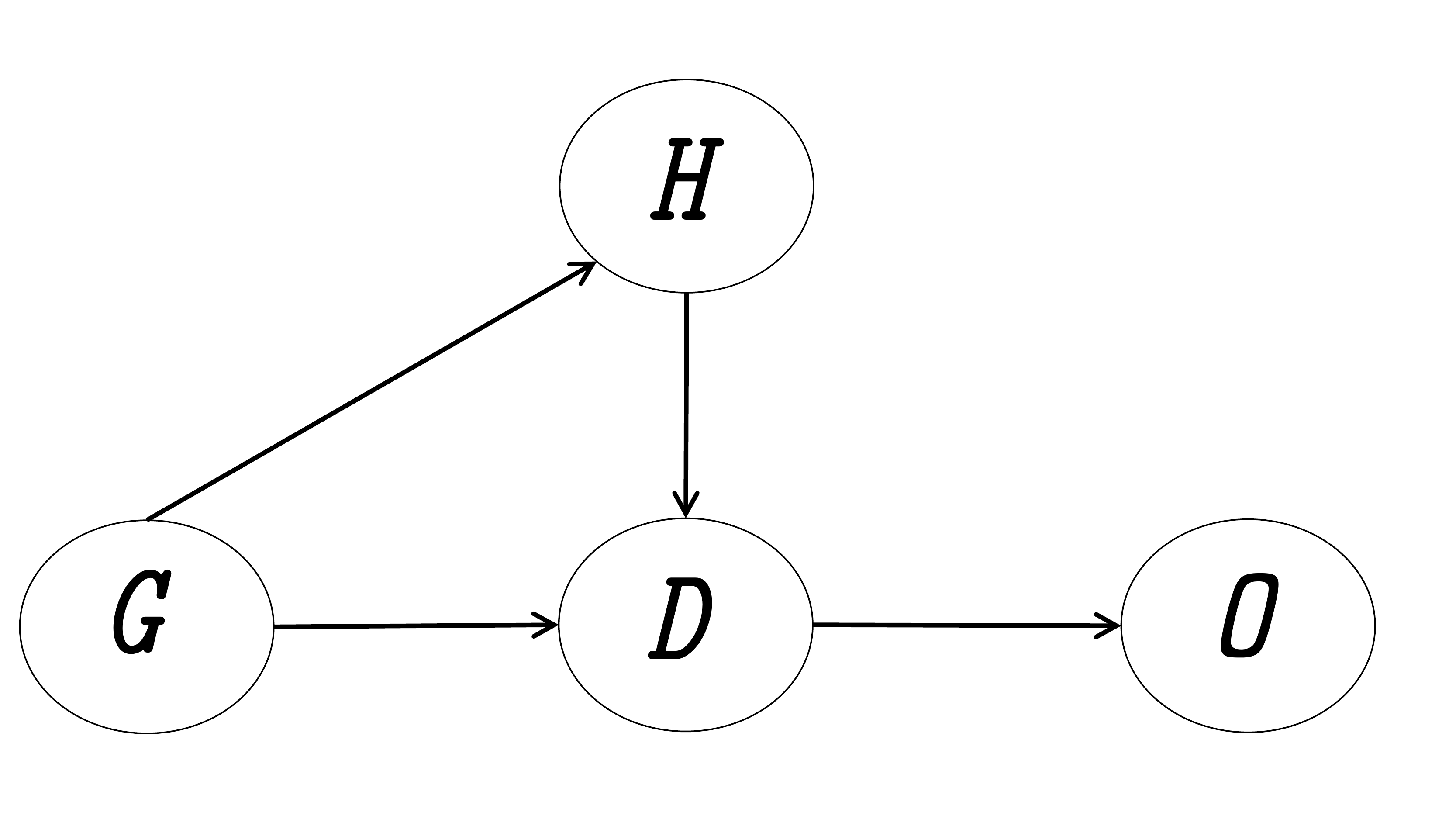}
			\caption{ }
		\end{subfigure}
		\begin{subfigure}{0.2\textwidth} \centering
			\includegraphics[scale=0.1]{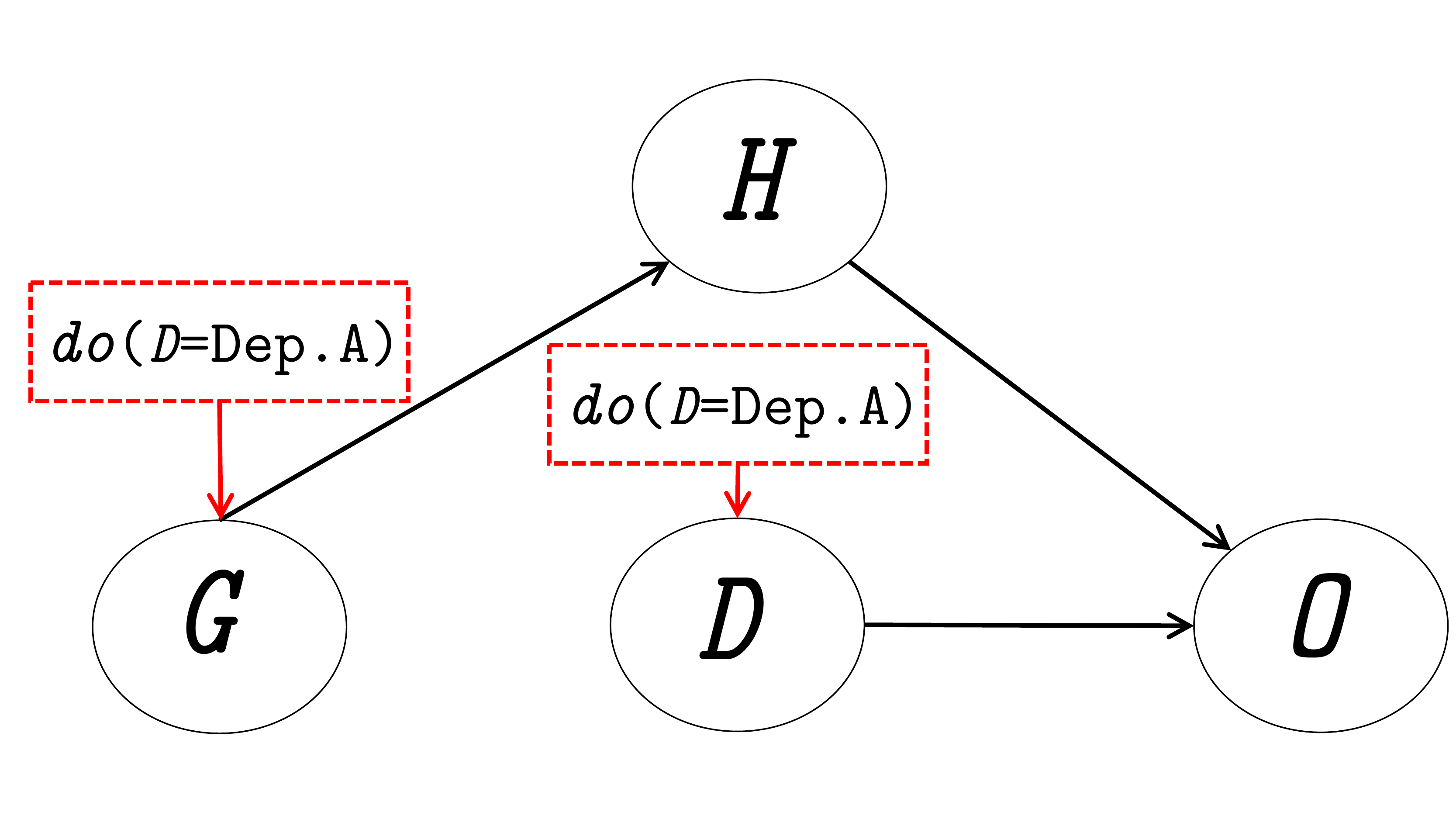}
			\caption{ }
		\end{subfigure}

		\ignore{
			\vspace*{2mm}	\begin{subfigure}{0.2\textwidth} \centering
				\hspace*{-0.8cm}
				\includegraphics[scale=0.4]{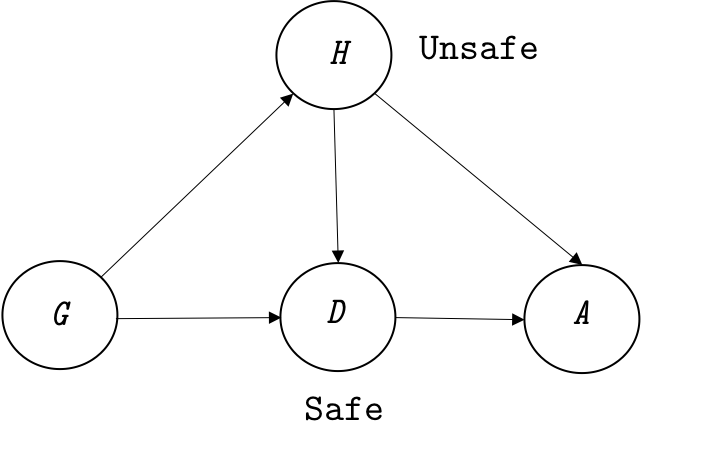}
				\caption{ }
			\end{subfigure}
			\begin{subfigure}{0.2\textwidth} \centering
				\hspace*{0.5cm}
				\includegraphics[scale=0.4]{fig/admi.png}
				\caption{ }

			\end{subfigure}
		}
		\caption{Represent a PCM in which the admission process is contextually fair w.r.t. department.}
		\label{fig:cgex2}
	\end{figure}
}

\ignore{
	\begin{figure}

		\begin{subfigure}{0.2\textwidth} \centering
			\includegraphics[scale=0.15]{fig/gDAG.png}
		\end{subfigure}
		\begin{subfigure}{0.2\textwidth} \centering

			\begin{align}
			O &: &  \text{Outcome} \nonumber \\
			\mb N &: &  \text{Inadmissibles} \nonumber \\
			\mb D &: &  \text{Admissibles} \nonumber \\
			S &: &  \text{Sensitive} \nonumber \\
			\mb D' &: &  \text{Other Admissibles} \nonumber
			\end{align}
		\end{subfigure}
		\caption{A generic causal DAG that illustrates an algorithm that is not Justifiably fair.}
		\label{fig:dic}
	\end{figure}
}


\subsection{Testing Fairness on the Training Data}
\label{sec:jfc}

In this section we introduce a sufficient condition for testing
justifiable fairness, which uses only the training data $D, \pr$
(Sec.~\ref{sec:peri}) and does not require access to the causal graph
$\cg$.  We  assume only that $\cg$ and $\pr$ are Markov compatible
(Sec.~\ref{sec:causal:dag}). The training data has an additional
response variable $Y$.  As before, we assume a \fairapp $(\mc A,$
$S, \mb A, \mb I)$ is  given and  that the algorithm is a
good prediction of the response variable, \ie\,
%
%
%
$\pr(Y=1| \mb X=\mb x) \approx \pr(O=1| \mb X=\mb x)$; we call the
algorithm a \emph{reasonable} classifier to indicate that it satisfies
this condition. \reve{ Note that this is a typical assumption in pre-processing approachs, see e.g., \cite{NIPS2017_6988} and needed to decouple the
	the issues of model accuracy and fairness. If the distributions of $\pr(Y=1| \mb X=\mb x)$ and $\pr(O=1| \mb X=\mb x)$ could be arbitrarily far apart, no fairness claims can be made about a classifier that, for example, imposes
	a pre-determined distribution on the outcome predictions rather than learning
	an approximation of $\pr(Y=1| \mb X=\mb x)$ from the training data.}


We first establish a technical condition for fairness based on the
Markov boundary, and then we simplify it.  Recall that, give a probability
distribution $\Pr$, the {\em Markov boundary} of a variable
$Y \in \mb V$, denoted $\mmb(Y)$, is a minimal subset of
$\mb V-\set{Y}$ that satisfies the saturated conditional independence
$(Y \indep_\pr \mb V-(\mmb(Y) \cup \set{Y}) | \mmb(Y))$. Intuitively,
$\mmb(Y)$ shields $Y$ from the influence of other variables.  It is
usually assumed that the Markov boundary of a variable is unique (see
Appendix~\ref{app:addback}).  We prove:


\begin{theorem} \label{theo:suff_jf}  A sufficient condition for a
  \fairapp $(\mc A, S, \mb A, \mb I)$ to be justifiably fair is
  $\mmb(O) \subseteq \mb A$.
\end{theorem}

If $\pr$ is faithful to the causal graph, then the theorem follows
immediately from Theorem~\ref{th:jfcd}; but we prove it without
assuming faithfulness in Appendix~\ref{app:proof}.  The condition in
Theorem~\ref{theo:suff_jf} can be checked without knowing the causal
DAG, but requires the computation of the Markov boundary;
moreover, it is expressed in terms of the outcome $O$ of the
algorithm.  We derive from here a sufficient condition without
reference to the Markov boundary, which refers only to the response
variable $Y$ present in the training data.



\begin{corollary} \label{col:fair_learn} Fix a training data $D, \pr$,
  where $Y \in \mb V$ is the training label, and $\mb A, \mb I$ are
  admissible and inadmissible attributes.  Then any reasonable
  classifier trained on a set of variables $\mb X \subseteq \mb V$ is
  justifiably fair w.r.t. a protected attribute $S$, if any of the following hold:
	\begin{itemize}
		\item[(a)] $\pr$ satisfies the CI $(Y \indep \mb X \cap \mb I  | \mb X \cap \mb{A} )$, or
		\item[(b)] $\mb X \supseteq \mb A$ and $\pr$ satisfies the saturated CI $(Y \indep \mb \mb I |\mb{A})$.
	\end{itemize}
\end{corollary}

The proof is the Appendix.
While condition (a) is the weaker
assumption, condition (b) has the advantage that the CI is saturated.
Our method for building a fair classifier is to repair the training
data in order to enforce (b).


\subsection{Building Fair Classifiers}
\label{sec:build}

This leads us to the following methods for building justifiably fair classifiers.

\paragraph*{\bf Dropping Inadmissible Attributes} A naive way to
satisfy Corollary~\ref{col:fair_learn}(a) is to set $\mb X = \mb A$,
in other words  to
train the classifier only
on  admissible attributes
This method guarantees fairness, but, as we will show in
Sec.~\ref{sec:experi}, dropping even one inadmissible variable can
negatively affect the accuracy of the classifier.  Moreover, this
approach cannot be used in data release situations, where all
variables must be included. Releasing data that reflect  discrimination can
 unintentionally reinforce and amplify discrimination in other
 contexts that data is used.

\paragraph*{\bf Repairing Training Data}
Instead, our approach is to repair the training data to enforce the
condition in Corollary~\ref{col:fair_learn}(b).
We consider the saturated CI
$(Y \indep \mb I |\mb{A})$ as an {\em integrity constraint} that
should always hold in training data $D, \pr$.  \sys\ performs a
sequence of database updates (viz., insertions and deletions of
tuples) to obtain another training database $D'$ to satisfy
$(Y \indep \mb I |\mb{A})$.
We describe this repair problem in Sec.~\ref{sec:repdata}.
To the causal DAG, this approach can be seen as
modifying the underlying causal model to enforce the fairness
constraint.  However, instead of intervening on the causal DAG, which
we do not know and over which we have no control, we intervene on
the training data to ensure fairness. 
\reva{Note that minimal repairs are crucial for preserving the utility of data.  Specifically, we need to ensure that the joint distribution of training and repaired data are close. Since there is no general metric to measure the distance between  two distributions that works well for all datasets and applications, in Sec~\ref{sec:repdata} we propose several minimal repair methods. We empirically show that these methods behave differently for different datasets and ML algorithms.} \revb{We also note that the negative effect of repair on utility depends on several factors such as the size of data, sparsity of data,  repair method, ML algorithm, strongness of dependency that the repair method enforces; hence accuracy should be trade with fairness at training time. }



%% file: repair.tex
\vspace*{-0.2cm}
\section{Repairing Training Data to Ensure Fairness}
\label{sec:repdata}

We have shown in Corollary~\ref{col:fair_learn} that, if the training
data $D$ satisfies a certain saturated conditional independence (CI),
then a classification algorithm trained on $D, \pr$ is justifiably
fair.  We show here how to modify (repair) the training data to
enforce the CI and thus ensure that any reasonable classifier trained
on it will be justifiably fair.




	\vspace*{-0.3cm}
\subsection{Minimal Repair for MVD and CI}
\label{sec:prob_stat}

We first consider repairing an MVD.  Fix an MVD $\mb Z \mvd \mb X$ and
a database $D$ that does not satisfy it.  The minimal database repair
problem is this: find another database $D'$ that satisfies the MVD
such that the distance between $D$ and $D'$ is minimized.  In this section,
 we restrict the distance function to the symmetric difference, i.e, $|\Delta(D, D')|$.

\begin{figure} 
  \centering
  \begin{minipage}[b]{0.32\hsize}\centering
    \[
      \begin{tabular}{c|ccc|c}
	\cline{2-4}
        D: &	X & Y & Z & $\pr$ \\
	\cline{2-4}
        $t_1$&	$a$ & $a$ & $c$ & $3/8$ \\
        $t_2$&	$a$ & $b$ & $c$ & $2/8$ \\
        $t_3$&	$b$ & $a$ & $c$ & $2/8$ \\
        $t_4$&	$b$ & $b$ & $d$ & $1/8$ \\
	\cline{2-4}
      \end{tabular}
    \]
  \end{minipage}
  \begin{minipage}[b]{0.32\hsize}\centering
    \[
      \begin{tabular}{c|ccc|}
	\cline{2-4}
        $D_1:$ &	X & Y & Z \\
	\cline{2-4}
        $t_1$&	$a$ & $a$ & $c$ \\
        $t_2$&$a$ & $b$& $c$\\
        $t_3$&$b$ & $a$ & $c$\\
        $t_4$&$b$ & $b$ & $c$ \\
        $t_5$& $b$ & $b$ & $d$\\
	\cline{2-4}
      \end{tabular}
    \]
  \end{minipage}
  \begin{minipage}[b]{0.32\hsize}\centering
    \[
      \begin{tabular}{c|ccc|}
	\cline{2-4}
        $D_2:$	&X & Y & Z \\
	\cline{2-4}
        $t_1$&	$a$ & $a$ & $c$\\
        $t_2$& $a$ & $b$ & $c$ \\
        $t_4$& $b$ & $b$ & $d$ \\
	\cline{2-4}
      \end{tabular}
    \]
\end{minipage}
    \caption{      \textmd{ A simple database repair: $D$ does not satisfy the MVD $Z \mvd X$. In $D_1$, we inserted
the tuple $(b,b,c)$ to satisfy the MVD, and in $D_2$ we deleted the tuple $(b,a,c)$ to satisfy the MVD.}}
    \label{fig:simple:repair}
  \end{figure}
\begin{example} \label{ex:rep1} \em
Consider the database $D$ in Fig.~\ref{fig:simple:repair} (ignoring the
probabilities for the moment), and the MVD $Z \mvd X$.  $D$ does not
satisfy the MVD.  The figure shows two minimal repairs, $D_1, D_2$,
one obtained by inserting a tuple, and the other by deleting a tuple.
\end{example}




However, our problem is to repair for a saturated CI, not an MVD,
since that is what is required in Corollary~\ref{col:fair_learn}.  The repair
problem for a database constraint is well-studied in the literature,
but here we need to repair to satisfy a CI, which is not a database
constraint.  We first  formally define the repair problem for a CI
then show how to reduce it to the repair for an MVD.  More precisely,
our input is a database $D$ and a probability distribution $\Pr$, and
the goal is to define a ``repair'' $D',\pr'$ that satisfies the given
CI.


We assume that all probabilities are rational numbers.  Let the {\em
  bag associated} to $D, \pr$ to be the smallest bag $B$ such that
$\pr$ is the empirical distribution on $B$.  In other words, $B$ is
obtained by replicating each tuple $t \in D$ a number of times
proportional to $\pr(t)$. \footnote{Equivalently, if the tuples have
  probabilities $p_1/q, p_2/q, \ldots$ (same denominator), then each
  tuple $t_i$ occurs exactly $p_i$ times in $B$.}  If $\pr$
is uniform, then $B=D$.
   
\begin{definition} 
  The minimal repair of $D, \pr$ for a saturated CI
  $(\mb X; \mb Y|\mb Z)$ is a pair $D',\pr'$ such that $\pr'$
  satisfies the CI and $|\Delta(B,B')|$ is minimized, where $B$ and
  $B'$ are the bags associated to $D,\pr$ and $D',\pr'$, respectively.
\end{definition}

Recall that $\mb V$ denotes the set of attributes of $D$.  Let $\pr$
be any probability distribution on the variables $\set{K} \cup \mb V$,
where $K$ is a fresh variable not in $\mb V$.
   
\begin{lemma} \label{lemma:k:v} If $\pr$ satisfies
  $(K\mb X; \mb Y | \mb Z)$, then it also satisfies
  $(\mb X; \mb Y | \mb Z)$.
\end{lemma}

The lemma follows immediately from the Decomposition axiom in Graphoid (see Appendix~\ref{app:addback}).


%
%

We now describe our method for computing a minimal repair of $D, \pr$
for some saturated CI.  First, we compute the bag $B$ associated to
$D, \pr$.  Next, we add the new attribute $K$ to the tuples in $B$ and
assign distinct values to $t.K$ to all duplicate tuples $t$, thus
converting $B$ into a set $D_B$ with attributes $K \cup \mb V$.
Importantly, we use as few distinct values for $K$ as possible,
i.e., we enumerate the instances of each unique tuple.  More precisely, we
define:
\begin{align}
  D_B =  \set{ (i,t)| t \in B, i = 1,\ldots,|t_B|} \label{eq:bad-set}
\end{align}
were $|t_B|$ denotes the number of occurrences (or multiplicity) of a
tuple $t$ in the bag $B$.
Then, we repair $D_B$ w.r.t. to the MVD
$\mb Z \mvd K\mb X$, obtaining a repaired database $D_B'$.  Finally,
we construct a new training set $D' = \Pi_{\mb V}(D_B')$, with the
probability distribution obtained by marginalizing the empirical
distribution on $D_B'$ to the variables $\mb V$.  We prove the
following:




\begin{theorem} \label{th:bag_set}  Let $D$ be a database and $\pr$ a
  probability distribution on its tuples, and let $B$ be the
  associated bag (with attributes $\set{K} \cup \mb V$).  Fix a
  saturated CI $(\mb X; \mb Y|\mb Z)$, and let $B'$ be a minimal
  repair for the MVD $\mb Z \mvd K\mb X$.  Then, $D', \pr'$ is a
  minimal repair of $D, \pr$ for the CI, where $D'$ is $B'$ with
  duplicates removed, and $\pr'$ is the empirical distribution on
  $B'$.
\end{theorem}


We illustrate with an example.

\begin{figure}

\begin{minipage}[t]{0.22\hsize}\centering
	\[
	\begin{tabular}{c|ccc|}
	\cline{2-4}
	B: &	X & Y & Z \\
	\cline{2-4}
	&	$a$ & $a$ & $c$ \\
	&	$a$ & $a$ & $c$ \\
	&	$a$ & $a$ & $c$ \\
	&	$a$ & $b$ & $c$\\
	&	$a$ & $b$ & $c$\\
	&	$b$ & $a$ & $c$\\
	&	$b$ & $a$ & $c$\\
	&	$b$& $b$ & $d$ \\
	\cline{2-4}
	\end{tabular}
	\]
\end{minipage}
\begin{minipage}[t]{0.23\hsize}
	\[
	\begin{tabular}{c|cccc|}
	\cline{2-5}
	$D_B:$ & K &X & Y & Z \\
	\cline{2-5}
&1&		$a$ & $a$ & $c$ \\
&2&	$a$ & $a$ & $c$ \\
&3&	$a$ & $a$ & $c$ \\
&1&	$a$ & $b$ & $c$\\
&2&	$a$ & $b$ & $c$\\
&1&	$b$ & $a$ & $c$\\
&2&	$b$ & $a$ & $c$\\
&1&	$b$& $b$ & $d$ \\
	\cline{2-5}
	\end{tabular}
	\]
\end{minipage} 
\begin{minipage}[t]{0.23\hsize}\centering 
	\[
\begin{tabular}{c|cccc|}
\cline{2-5}
$D'_B:$ & K &X & Y & Z \\
\cline{2-5}
&1&	$a$ & $a$ & $c$ \\
&2&	$a$ & $a$ & $c$ \\
&1&	$a$ & $b$ & $c$\\
&2&	$a$ & $b$ & $c$\\
&1&	$b$ & $a$ & $c$\\
&1&     $b$ & $b$ & $c$\\
&1&	$b$& $b$ & $d$ \\
\cline{2-5}
\end{tabular}
\]
\end{minipage}
\begin{minipage}[t]{0.23\hsize}\centering 
\[
\begin{tabular}{c|ccc|c}
\cline{2-4}
$D':$ &X & Y & Z & $\pr'$  \\
\cline{2-4}
&	$a$ & $a$ & $c$ & $2/7$ \\
&	$a$ & $b$ & $c$ & $2/7$ \\
&	$b$ & $a$ & $c$ & $1/7$ \\
&	$b$ & $b$ & $c$ & $1/7$ \\
&	$b$ & $b$ & $d$ & $1/7$ \\
\cline{2-4}
\end{tabular}
\]
\end{minipage}
  \caption{      \textmd{ Repairing a conditional independence (CI).}}
  \label{fig:repair:ci}
\end{figure}
   
\begin{example} \label{ex:rep2} \em In Example~\ref{ex:rep1} we showed two
  repairs $D_1, D_2$ of the database $D$ in
  Fig~\ref{fig:simple:repair} for the MVD $Z \mvd X$.  Consider now
  the probability distribution, $\pr$ shown in the figure.  Suppose
  we want to repair it for the CI $(X; Y | Z)$.  Clearly, both $D_1$
  and $D_2$, when endowed with the empirical distribution {\em do}
  satisfy this CI, but they are very poor repairs because they
  completely ignore the probabilities in the original training data,
  which are important signals for learning.  Our definition captures
  this by insisting that the repaired bag $B'$ be close to the bag $B$
  associated to $D,\pr$ (see $B$ in Fig.~\ref{fig:repair:ci}), but the
  sets $D_1$ and $D_2$ are rather far from $B$.  Instead, our method
  first converts $B$ into a set $D_B$ by adding a new attribute $K$
  (see Fig.~\ref{fig:repair:ci}) then, it repairs $D_B$ for the MVD
  $Z \mvd KX$, obtaining $D_B'$.  The final repair $D',\pr'$ consists
  of the empirical distribution on $D_B'$, but with the attribute $K$
  and duplicates removed.
\end{example}

We note that, in order for Theorem~\ref{th:bag_set} to hold, it is
critical that we use minimum distinct values for the
attribute $K$ in $D_B$; otherwise minimal repairs of $D_B$ are no
longer minimal repairs of the original data $D, \pr'$.  For example,
if we use distinct values for $K$, thus making $K$ a key, then only
subset of $D_B$ that satisfies the MVD $\mb Z \mvd K\mb X$ is the
empty set.

\vspace*{-0.2cm}
\subsection{\bf Reducing Minimal Repair to 3SAT}
\label{sec:prov}

Corollary~\ref{col:fair_learn} requires us to repair the training data
$D$ to satisfy a CI.  We have shown how to convert this problem into
the problem of repairing a derived data $D_B$ to satisfy an MVD.  In
this section we describe how to find a minimal repair for an MVD by
reduction to the weighted MaxSAT problem.

We denote the database by $D$, the MVD by $\varphi: \mb Z \mvd \mb X$,
and assume that $D$'s attributes are
$\mb V = \mb X \cup \mb Y \cup \mb Z$.  Recall that $D$ satisfies the
MVD iff $D = \Pi_{\mb X \mb Z}(D) \Join \Pi_{\mb Y \mb Z}(D)$.  Since
we want to allow repairs that include both insertions and deletions,
we start by finding an upper bound on the set of tuples that we may
want to insert in the database. For example, one can
restrict the set of tuples to those that have only constants that
already occurring in the database, i.e., an upper bound is $ADom^k$, where
$ADom$ is the active domain of $D$, and $k$ is the arity of $D$.
However, this set is too large in practice.  Instead, we prove that it
suffices to consider candidate tuples in a much smaller set, given by:
$\hb \defeq   \Pi_{\mb X \mb Z}(D) \Join \Pi_{\mb Z \mb Y}(D)$.
\ignore{ {  
\begin{align*}
\hb \defeq &  \Pi_{\mb X \mb Z}(D) \Join \Pi_{\mb Z \mb Y}(D)
\end{align*}
}}

\begin{prop} \label{prop:min_rep_lin}  Any minimal repair $\rep$ of $D$
  for an MVD satisfies $\rep \subseteq \hb$.
\end{prop}
Next, we associate the following Boolean Conjunctive query to the MVD $\varphi$:
{   
\begin{align}
\icq \leftarrow \ D( \mb X_1, \mb Y_1, \mb Z),  D( \mb X_2, \mb Y_2, \mb Z),  \neg
D( \mb X_1, \mb Y_2, \mb Z)  \label{eq:cq_ci}
\end{align} }
It follows immediately that $D \not \models \varphi$ iff
$D \models \icq$, and therefore the repair problem becomes: modify the
database $D$ to make $\icq$ false.  For that purpose, we use the
lineage of the query $\icq$.  By the previous proposition, we know
that we need to consider as candidates for insertions only those tuples in  $\hb$; hence we compute the lineage over the set of
possible tuples $\hb$.  We briefly review here the construction of the
lineage and refer the reader to~\cite{xu2018provenance} and the
references there for more detail.  We associate a distinct Boolean
variable $X_t$ to each tuple $t \in \hb$, and consider all mappings
$\theta : Var(\icq) \rightarrow ADom(D)$ such that each of the three
tuples--
$\hb(\theta(X_1),\theta(Y_1),\theta(Z)),
\hb(\theta(X_2),\theta(Y_2),\theta(Z))$, and
$\hb(\theta(X_1),$ $\theta(Y_2),\theta(Z))$-- are in $\hb$.  Then, the
lineage and its negation are:

{ 
\begin{align}
\Phi_\varphi = &  \bigvee_{\theta }
\left(X_{\hb(\theta(X_1),\theta(Y_1),\theta(Z))} \land   X_{\hb(\theta(X_2),\theta(Y_2),\theta(Z))} \land  \neg X_{\hb(\theta(X_1),\theta(Y_2),\theta(Z))} \right) \label{eq:le}\\
\neg \Phi_\varphi = &  \bigwedge_{\theta} \left(\neg X_{\hb(\theta(X_1),\theta(Y_1),\theta(Z))} \lor \neg  X_{\hb(\theta(X_2),\theta(Y_2),\theta(Z))} \lor X_{\hb(\theta(X_1),\theta(Y_2),\theta(Z))}\right) \label{eq:3cnf}
\end{align}
}

Recall that an {\em assignment} is a mapping from Boolean variables
$X_t$ to $\set{0,1}$.  Thus, our goal is to find an assignment
satisfying the 3CNF $\neg \Phi_\varphi$, which is as close as possible
to the initial assignment $X_t = 1$ for $t \in D$, $X_t = 0$ for
$t \in \hb - D$.

We briefly review the weighted MaxSAT problem here.  Its input is a
3CNF $\mb F$ whose clauses are partitioned into
$\mb F = (\mb F_h, \mb F_s, \mc C)$, where $\mb F_h$ are called the
hard clauses, and $\mb F_s$ are the soft clauses, and a function
$\mb C : \mb F_s \rightarrow R^+$  associates a non-negative cost
with each  soft clause.  A solution to the problem finds an
assignment that satisfies all  hard constraints, and maximizes the
weight of the satisfied soft constraints.

To ensure ``closeness'' to the initial assignment, we add to the
Boolean formula a clause $X_t$ for every $t \in D$, and a clause
$\neg X_t$ for every $t \in \hb - D$.  The final 3CNF formula is:

{  
\begin{align*}
  \Psi = & \underbrace{(\neg \Phi_\varphi)}_{\mbox{hard clauses}}
 \wedge
\underbrace{\bigwedge_{t \in D} X_t \wedge \bigwedge_{t \in \hb-D} (\neg X_t)}_{\mbox{soft clauses}}
\end{align*}}

The algorithm constructing $\Psi$ is shown in
Algorithm~\ref{algo:maxsat}.

%

\begin{algorithm} 
	\DontPrintSemicolon
	\KwIn{A database $D$ with vairables $\mb X \cup \mb Y  \cup \mb Z$ and a saturated CI $\varphi: (\mb X \indep \mb Y |\mb Z )$}
	\KwOut{ A 3CNF $\Psi$ consisting of hard and soft clauses.}
	Compute $\hb(\mb X_1,\mb Y_2, \mb Z) = D(\mb X_1,\mb Y_1,\mb Z) \land D(\mb X_2,\mb Y_2,\mb Z)$ \\
		\For{$t \in \hb$}{If $t \in D$, add the soft clause $X_t$ to $\Psi$ \\
		If $t \in \hb-D$ add the soft clause $(\neg X_t)$ to  $\Psi$}
	Compute $C(\mb X_1,\mb Y_1,\mb X_2,\mb Y_2,  \mb Z) = \hb(\mb X_1,\mb Y_1,\mb Z) \land \hb (\mb X_2,\mb Y_2,\mb Z)$\\
\For{$t \in C$}{
	$t_1 \gets t[\mb X_1,\mb Y_1,\mb Z]$; $t_2 \gets t[\mb X_2,\mb Y_2,\mb Z]$;	$t_3 \gets t[\mb X_1,\mb Y_2,\mb Z]$\\
	Add the hard clause $(\neg X_{t_1} \lor \neg X_{t_2} \lor X_{t_3})$ to $\Psi$\\}

	\caption{Converts the problem of finding a database repair w.r.t. a CI statement
		into solving a general CNF formula.} \label{algo:maxsat}
\end{algorithm}

%
%
%
 
\begin{example} \em Continuing Ex.~\ref{ex:rep1}, we observe that
  $\hb=D_1$; hence, there are 5 possible tuples.  The lineage expression for $\varphi$  and it negation are:
  {  
	\begin{align}
	\Phi_\varphi  &=&  (X_{t_1} \land  X_{t_4}  \land  \neg X_{t_2}) \lor  ( X_{t_2}  \land  X_{t_3}  \land \neg X_{t_1})  \ \lor  \nonumber \\ &&
	( X_{t_3} \land  X_{t_2}  \land \neg X_{t_4}) \lor  ( X_{t_4} \land  X_{t_1}  \land \neg X_{t_3}) \nonumber
	\end{align} }
	Hence, {  
		\begin{align}
	\neg \Phi_\varphi  &=&  (\neg X_{t_1} \lor   \neg X_{t_4}  \lor   X_{t_2}) \land  ( \neg X_{t_2}  \lor \neg  X_{t_3}  \lor  X_{t_1})  \ \land  \nonumber \\ &&
	( \neg X_{t_3} \lor  \neg X_{t_2}  \lor   X_{t_4}) \land  ( \neg X_{t_4} \lor  \neg X_{t_1}  \lor  X_{t_3}) \nonumber
	\end{align}}
	The reader can check that the repairs $D_1$ and $ D_2$ in
        Ex.~\ref{ex:rep1} are corresponded to some satisfying
        assignment of $\neg \Phi_\varphi $, e.g., $D_2$ obtained from
        the truth assignment $\sigma(X_{t_1})=\sigma(X_{t_2})=1$,
        $\sigma(X_{t_3})=\sigma(X_{t_5})=0$; both satisfy all clauses
        in $\neg \Phi_\varphi$.  The formula $\Psi$ that we give as
        input to the weighted MaxSAT consists of $\neg \Phi_\varphi$
        plus these five clauses:
        $X_{t_1} \wedge X_{t_2} \wedge X_{t_3} \wedge X_{t_4} \wedge
        \neg X_{t_5}$, each with cost 1.  MaxSAT will attempt to
        satisfy as many as possible, thus finding a repair that is
        close to the initial database $D$.
\end{example}

Note that repairing a database w.r.t. a CI $(\mb X \indep \mb Y |\mb Z )$ can be reduced to repairing subsets $\sigma_{\mb Z=\mb z}(D)$ for $\mb z \in Dom(Z)$ w.r.t. the marginal independence $(\mb X \indep \mb Y)$.  Therefore, the problem is highly parallelizable.  \sys\ partition  subsets $\Pi_{\mb Z}(D)$ into chunks of even size (if possible) and repairs them in parallel (see Sec~\ref{sec:rep_metho_com}).

\begin{algorithm}  
	\DontPrintSemicolon
	\KwIn{ A bag $B$ with attributes $\mb V= \mb X \mb Y \mb Z $ a CI statment $(\mb X  \indep \mb Y| \mb Z)$.}
	\KwOut{$B'$ a repair of  $B$}
	\For{$\mb z \in Dom(\mb Z)$} {
		$ M^{B'}_{\mb X}, M^{B'}_{\mb Y} \gets  \mb{Factorize}(M^{B_{\mb z}}_{\mb X, \mb Y})$\\
		$ M^{B'{\mb z}}_{\mb X, \mb Y} \gets \frac{1}{|B_{\mb Z}|}{M^{B'}_{\mb X}}^\intercal M^{B'}_{\mb Y}$\\
	}

	$ \Return  \ \text{$B'$ associated with} \ \mb M^{B'}_{\mb X, \mb Y,\mb Z}=\{M^{B'{\mb z}}_{\mb X, \mb Y} \}$

	\caption{Repair using Matrix Factorization.} \label{algo:smi}
\end{algorithm}

\vspace*{-0.25cm}
\subsection{\bf Repair via Matrix Factorization}
	\label{sec:mf}
In this section, we use matrix factorization to repair a bag w.r.t. a CI statement. We are given a bag $B$ to which we associate the empirical distribution $\pr(\mb v) = \frac{1}{|B|}\sum_{t \in B}1_{t=\mb v}$, and a CI statement $\varphi: (\mb X \indep \mb Y| \mb Z)$ such that $B$ is inconsistent with $\varphi$, meaning  that $\varphi$ does not hold in $\pr$.  Our goal is to find a repair of $B$, i.e., a bag $B'$ that is close to $B$ such that $(\mb X \indep \mb Y|_{\pr'} \mb Z)$, where
$\pr'$ is the empirical distribution associated to $B'$.

First, we review the problem of non-negative rank-one matrix factorization. Given a matrix $\mb M \in \mathbb{R}^{n \times m}$, the problem of  {\em rank-one nonnegative matrix factorization (NMF)} is the  minimization problem:  $\argmin_{\mb U \in \mathbb{R}_+^{n \times 1}, \mb V  \in \mathbb{R}_+^{1 \times m}} \rVert \mb M-\mb U\mb V \rVert_F $, where  $\mathbb{R}_+ $ stands for  non-negative real numbers and   $\rVert . \rVert_F$ is the Euclidean norm  of a matrix.\footnote{Recall that a matrix is of rank-one if and only if it can be represented by the outer product of two vectors.}

We express the connection between our repair problem and the NMF problem using  contingency matrices. Given three disjoint subsets of attributes $\mb X, \mb Y, \mb Z \subseteq \mb V$,  let  $m=|Dom(\mb X)|$, $n=|Dom(\mb Y)|$,  $k=| Dom(\mb Z)|$ and  $B_{\mb z}=\sigma_{\mb Z=\mb z}(B)$.   A {\em multiway-contingency matrix} over  $\mb X$,  $\mb Y$ and $\mb Z$ consists of $k$  $n \times m$ matrices $\mb M^{B}_{\mb X, \mb Y,\mb Z}=\{ \mb M^{B_{\mb z}}_{\mb X, \mb Y} | \mb z \in Dom(\mb Z) \}$ where, $\mb M^{B_{\mb z}}_{\mb X, \mb Y}(ij)= \sum_{t \in B} 1_{t[\mb X \mb Y]=  ij}$.  Intuitively, $\mb M^{B_{\mb z}}_{\mb X, \mb Y}(ij)$ represents the  joint frequency of $\mb X$ and $\mb Y$ in a subset of bag with $\mb Z=\mb z$.

The following  obtained immediately from the connection between independence and rank of a contingency matrix.



\begin{prop} \label{theo:md}  Let $B$ be a bag and $\pr$ be the
empirical  distribution associated to $B$. It holds that $(\mb X \indep \mb Y|_\pr \mb Z)$ iff each contingency matrix $\mb M \in \mb M^{B}_{\mb X, \mb Y,\mb Z} $ is of rank-one.
\end{prop}

We illustrate with an example.
\begin{example}  \em  Let $\mb M_1=\begin{bmatrix}
	1      &1  \\
	1      &0  \\
\end{bmatrix}$, $\mb M_2=\begin{bmatrix}
0      &0  \\
0      &1  \\
\end{bmatrix}$,  $\mb M_3=\begin{bmatrix}
1      &1  \\
1    &1  \\
\end{bmatrix}$, $\mb M_4=\begin{bmatrix}
1      &1  \\
0      &0  \\
\end{bmatrix}$. The following contingency matrices are associated to  $D$, $D_1$ and $D_2$ in  Ex.~\ref{ex:rep1}:
$\mb M^{D}_{\mb X, \mb Y,\mb Z}=\{\ \mb M_1, \mb M_2 \}$,
$\mb M^{D_1}_{\mb X, \mb Y,\mb Z}=\{\ \mb M_3, \mb M_2 \}$ and
$\mb M^{D_2}_{\mb X, \mb Y,\mb Z}=\{\  \mb M_4, \mb M_2 \}$. The reader can verify that $\mb M_2, \mb M_3$ and $ \mb M_4$ are of rank-one but $ \mb M_1$ is not. It is clear that, $D$ is inconsistent with  $\varphi$ but $  D_1$ and $  D_2$ are consistent with $\varphi$.
\end{example}

The following implied  from  NP-hardness of
NMF \cite{vavasis2009complexity}.

\begin{prop}  \label{theo:md2} The problem of repairing a database w.r.t. a single CI is NP-hard in general.
\end{prop}

Based on Prop~\ref{theo:md}, we propose Algorithm \ref{algo:smi} for
repairing a bag w.r.t. a single CI
$ \varphi: (\mb X \indep \mb Y| \mb Z$).  The algorithm works as
follows: for each $\mb z \in Dom(\mb Z)$, it uses the
$\mb{Factorize}$ subroutine to factorize the $n \times m$ contingency
matrix $\mb M^{B_{\mb z}}_{\mb X, \mb Y}$ into a $1 \times n$ matrix
$\mb M^{B'}_{\mb X}$ and a $1 \times m$ matrix
$\mb M^{B'}_{\mb Y}$. Then, it uses the product of
${\mb M^{B'}_{\mb X}}^\intercal$ and $\mb M^{B'}_{\mb X}$ to
construct a new bag $B'$. It is clear that
${\mb M^{B'}_{\mb X}}^\intercal \mb M^{B'}_{\mb Y}$ is of rank-one
by construction; thus, the algorithm always returns a bag $B'$
that is consistent with $\varphi$. Note that any off-the-shelf NMF
algorithm (such as \cite{fevotte2011algorithms}) can be used in
Algorithm \ref{algo:smi}, to minimize the Euclidean distance between $\pr$ and $\pr'$, the empirical distributions associated to $B$ and $B'$, respectively. \ignore{(These approached can be adopted for computing approximate repairs.)}  In addition, we use the simple
factorization of $\mb M^{B_{\mb z}}_{\mb X, \mb Y}$ into
$\mb M^{B_{\mb z}}_{\mb X}$ and $\mb M^{B_{\mb z}}_{ \mb Y}$, i.e.,
the marginal frequencies of $\mb X$ and $\mb Y$ in $B_{\mb z}$.  We
refer to this simple factorization as {\em Independent Coupling
  (IC)}. It is easy to see that KL-divergence between $\pr$ and $\pr'$ is
bounded by conditional mutual information
$I(\mb X \indep \mb Y| \mb Z)$.


%% file: discussion.tex
\vspace*{-0.25cm}
\section{Discussion}
\label{sec:diss} 
\reva{
  \paragraph*{\bf Generalizability to Unseen Test Data} In the
  following we briefly discuss the generalizability of the proposed
  repair algorithm to unseen test data.  Recall that the bag $B$
  represents the training data, $B'$ its repair, and let $T$ be the
  unseen test data.  We prove the following in Appendix~\ref{app:proof}:

  \begin{lemma} \label{prop:gen}  If the repaired data satisfies
    $(Y \indep S,\mb I|_{\pr_{B'}} \mb A)$ and the unseen test data
    satisfies $\pr_T(s,\mb i|\mb a) = \pr_{B'}(s,\mb i|\mb a)$, then
    the unseen test data also satisfies
    $(Y \indep S,\mb I|_{\Pr_T} \mb A)$
  \end{lemma}


  The goal of  repair is precisely to satisfy
  $(Y \indep S,\mb I|_{\pr_{B'}} \mb A)$, hence the classifier trained
  on the repaired data $B'$ will be justifiable fair on the test data
  $T$ provided that $\pr_T(s,\mb i|\mb a) = \pr_{B'}(s,\mb i|\mb a)$.
  It is generally assumed that the test and training data are drawn
  from the same distribution $\pr$.  By the law of large numbers, the
  empirical distribution of i.i.d samples of size
  $N\rightarrow \infty$ converges to $\pr$, hence $\pr_T=\pr_B$ in the
  limit.  Therefore, the algorithm will be justifiable fair on the
  test data, provided that the repair is done such that
  $\pr_B(s,\mb i| \mb a) = \pr_{B'}(s,\mb i |\mb a)$. This condition
  is satisfied by the IC repair method which simply repair data by coupling marginal distributions, because it holds by
  construction that
  $\pr_{B'}(y,s,\mb i,\mb a) = \pr_{B}(y,\mb a)\pr_{B}(s,\mb i,\mb
  a)/\pr_{B}(\mb a)$.   In contrast, the condition is only
  approximately satisfied by the MaxSAT and MF approaches, translating
  to slightly weaker fairness guarantees on unseen test data.  Nevertheless, we
  empirically show in Sec~\ref{sec:exp} that MaxSAT and MF approaches
  maintain a significantly better balance between accuracy and fairness. We note that our repair methods can be naturally extended to repair both training and test data for stricter fairness grantees, we consider this extension as future work.
\ignore{Specifically, IC radically changes the empirical distribution for sparse data, which negatively impacts the accuracy of the ML algorithms and usability of data.}
\ignore{We note that the condition established in Lemma\ref{prop:gen} can be fed into the MaxSAT
and MF approaches as a constraint, which we defer to future work.
We also note that our repair methods can be naturally extended to repair both training and test data, we consider this extension as future work.}
\vspace*{-.2cm}	\ }

\revb{
  \paragraph*{\bf  Scalability}  As shown in Sec~\ref{sec:repdata}, repairing data w.r.t. a single CI  is an NP-complete problem.
	Therefore, the scalability of our proposed repair methods is equal to that of MaxSAT solvers and approximation algorithms for matrix factorization.
	However, our repair problem is embarrassingly parallel and can be scaled to large datasets by partitioning data into small chunks formed by the conditioning set (see Sec~\ref{sec:experi}).
	In this paper we focused on a single CI, which suffices for many real world fairness applications. We leave the natural extension to future work.
}



%% file: experiment.tex




\begin{table}[] \centering
	\begin{small} 
		\begin{tabular}{@{}lrrrrrrrrr@{}}\toprule
			{Dataset} & {Att. [$\#$]} & {Rows[$\#$]} &  IC & MF &  MS(H.) &  MS(S.)   \\ \midrule
			\textbf{Adult}~\cite{adult} & 10 & 48k & 12  & 20& 40 & 30 \\ \hdashline
			\textbf{\nipsadult}~\cite{NIPS2017_6988} & 4 & 48k & 2  & 3& 20  & NA \\ \hdashline
			\textbf{COMPAS}~\cite{larson2016we}& 7  & 7k & 2 &3 &7 & 8\\ \hdashline
			\textbf{\nipscompas}~\cite{NIPS2017_6988}& 5  & 7k & 2 &3 &9 & NA\\ \hdashline
		\end{tabular}
	\end{small}
	\caption{      \textmd{ Runtime in seconds for  experiments in Sec.~\ref{sec:extendtoend}.}}
	\label{tbl:data}
\end{table}

\vspace{-0.2cm}
\section{Experimental Results}
\label{sec:exp}

\label{sec:experi}
This section presents experiments that evaluate the feasibility and
efficacy of \sys.  We aim to address the following questions.
{\bf Q1}: What is the end-to-end performance of \sys\ in terms of utility and fairness, with respect to our different algorithms?
{\bf Q2}: To what extent are the repaired datasets modified by the repair process of \sys?
{\bf Q3}: How does \sys\ compare to state-of-the-art pre-processing methods for enforcing fairness in predictive classification algorithms? \revd{Table~\ref{tbl:data} reports the running time of the repair algorithms.}


\subsection{Degree of Discrimination}
\label{sec:rod}


To assess the effectiveness of the proposed approaches, we next propose a metric that quantifies the degree of discrimination of a classification algorithm.

If we have access to the causal DAG, we could directly compute the {\em degree of interventional discrimination}
of an algorithm: given admissible variables $\mb A$,
for each $\mb K \supseteq \mb A$, compute the ratio of the LHS and RHS of Eq.~\ref{eq:cfair}
using Theorem~\ref{theo:af}, and average the
results.
However,  in many practical
settings we must make judgments about the fairness of an algorithm whose inputs are unknown and
	that may access information that is not even available in the dataset.  We cannot assume access
to an underlying causal DAG in these situations.
For example, in
the case of COMPAS, the recidivism prediction tool, it has been
hypothesized that the algorithm is truly racially discriminatory
\cite{larson2016we}; \sys\ confirms this in Sec~\ref{sec:extendtoend}. 
However, the algorithm itself
is not available to determine which inputs were used and how they might relate.
Instead, we propose a new metric for discovering evidence of potential
discrimination from data that uses the causal framework we described but is
still applicable in situations where all we know is which attributes in the Markov boundary of $O$  are admissible.

\begin{definition} \em Given a\fairapp
	$(\mc A,$ $S, \mb A, \mb I)$, let $\mb A_b=\mmb(O)- \mb I$. We quantify the {\em  ratio of observational discrimination (ROD)} of  $\mc A$ against $S$  in a context $\mb A_b=\mb a_b$ as $\delta(S;O| \mb a_b) \defeq  \frac{\pr(O=1|S=0, \mb a_b) \pr(O=0|S=1 , \mb a_b)}{\pr(O=0|S=0, \mb a_b) \pr(O=1|S=1,  \mb  a_b)}$.
\end{definition}
%
%
Intuitively, ROD calculates the effect of membership in a protected group on the odds of the positive outcome of $\mc A$ for subjects that are similar on $\mb A_b=\mb a_b$ ($\mb A_b$ consists of admissible attributes in the Markov boundary of the outcome).
If $\delta(S;O| \mb a_b)=1$ , then there is no observational evidence that $\mc A$ is discriminatory toward subjects with similar characteristics $\mb a_b$. If  $\delta(S;O| \mb a_b)>1$, then the algorithm potentially discriminates against the protected group, and vice versa if $\delta(S;O| \mb a_b)<1$. ROD is sensitive to the choice of a context $\mb A_b=\mb a_b$ by design. The overall ROD denoted by $\delta(S,O|\mb A_b)$ can be computed by averaging $\delta(S,O|\mb a_b)$  for all $\mb a_b \in \mb A_b$. For categorical data, standard methods in {\em meta analysis} for computing {\em pooled odds ratio} and assessing statistical significance can be applied (see~\cite{chang2017meta,loux2017comparison}). It is easy to see for faithful distributions that ROD=1 coincides with justifiable fairness (see Prop~\ref{prop:deg} in the Appendix~\ref{app:proof}).

\vspace*{-0.3cm}
\subsection{Setup}
\label{sec:setup}
The datasets used for experiments are listed in Table~\ref{tbl:data}.
We implemented our MaxSAT encoding algorithm in Python. For every instance of the input
data, our algorithm constructed the appropriate data files in WCNF format. We used the
Open-WBO~\cite{martins2014open} solver to solve the weighted MaxSAT instances.
%


We report the empirical utility of each classifier using Accuracy (ACC) = $\frac{TP+TN}{TP+FP+FN+TN}$ via 5-fold cross-validation.  We evaluate using three classifiers: Linear Regression (LR), Multi-layer Perceptron (MLP), and Random Forest (RF).  We selected LR and RF for comparison with Calmon et al.~\cite{NIPS2017_6988}.  We added MLP because it was the highest accuracy method out of ten alternative methods tested on the original (unrepaired) data.  We do not report on these other methods for clarity.



We evaluated using the fairness metrics in Table~\ref{tab:expmetrics}. For computing these metrics, conditional expectations were estimated as prescribed in  \cite{salimi2018bias}. We used standard techniques in meta-analysis to compute the pooled odds ratio \cite{chang2017meta}, and its statistical significance, needed to compute ROD. Specifically, we reported the p-value of the ROD, where the null hypothesis was ROD=1; (low p-values suggest the observed ROD is not due to random variation). We combined the p-values from cross-validation test datasets using Hartung's method \cite{hartung1999note}; p-values were dependent due to the overlap in cross-validation tests.  We normalized ROD between 0 and 1, where 0 shows no observational discrimination. We reported the absolute value of the averages of all metrics computed from each test dataset, where the smaller the value, the less the discrimination exhibited by the classifier.

\begin{table*} \centering
\begin{tabular}{ll}
\toprule
\textbf{Metric}    & \textbf{Description} and \textbf{Definition}\\
\midrule
ROD      & \begin{tabular}[c]{@{}l@{}}Ratio of Observation Discrimination:\\ (See Sec.\ref{sec:rod})\end{tabular} \\
\hdashline
DP       & \begin{tabular}[c]{@{}l@{}}Demographic Parity: \\ $Pr(O=1 | S=1) - Pr(O=1 | S=0)$\end{tabular}                                          \\
\hdashline
TPB     & \begin{tabular}[c]{@{}l@{}}True Positive Rate Balance: \\ $Pr(O=1 | S=1,Y=1) - Pr(O=1 | S=0, Y=1)$\end{tabular}                         \\
\hdashline
TNB     & \begin{tabular}[c]{@{}l@{}}True Negative Rate Balance:\\ $Pr(O=0 | S=1, Y=0)-Pr(O=0 | S=0, Y=0)$\end{tabular}                           \\
\hdashline
CDP      & \begin{tabular}[c]{@{}l@{}}Conditional Statistical Parity:\\ $\E_{\mb a}[Pr(O = 1|S = 1, \mb a) - Pr(O = 1|S = 0, \mb a)]$\end{tabular}                \\
\hdashline
CTPB    & \begin{tabular}[c]{@{}l@{}}Conditional TPRB:\\ $\E_{\mb a}[Pr(O = 1|S = 1,Y = 1, \mb a) - Pr(O = 1|S =0, Y=1, \mb a)]$\end{tabular}                      \\
\hdashline
CTNB    & \begin{tabular}[c]{@{}l@{}}Conditional TNRB: \\ $\E_{\mb a} [\Pr(O=0|S=1,Y=0,\mb a)-\Pr(O=0|S=0, Y=0,\mb a)]$\end{tabular}              \\
\hdashline
\end{tabular}
\caption{\textmd{Fairness metrics used in our experiments.}}
\label{tab:expmetrics}
\end{table*}

\begin{figure}
	\begin{subfigure}[t]{1\textwidth} \centering
		\hspace*{-1cm}		\includegraphics[scale=0.15]{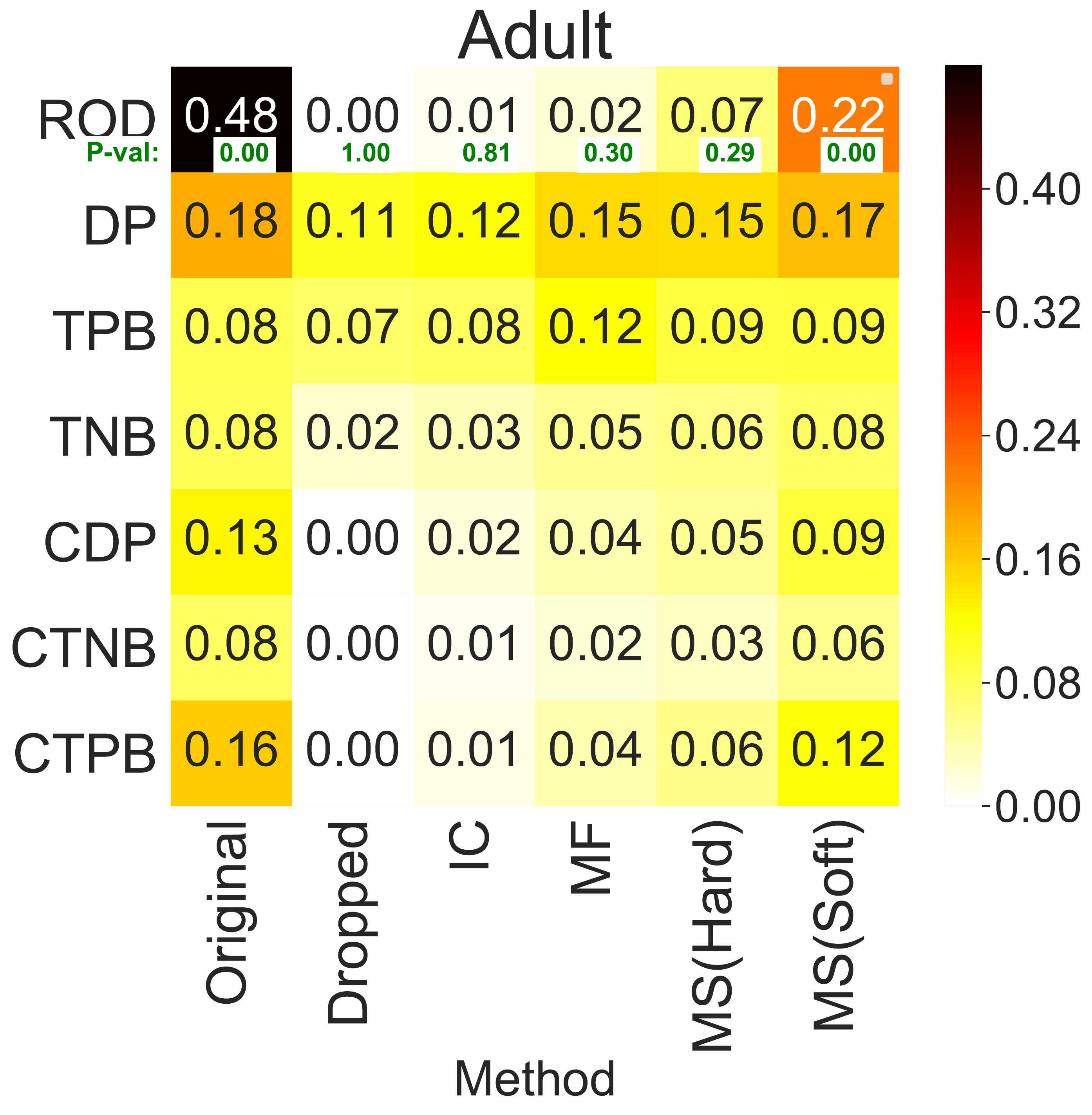}
		\includegraphics[scale=0.15]{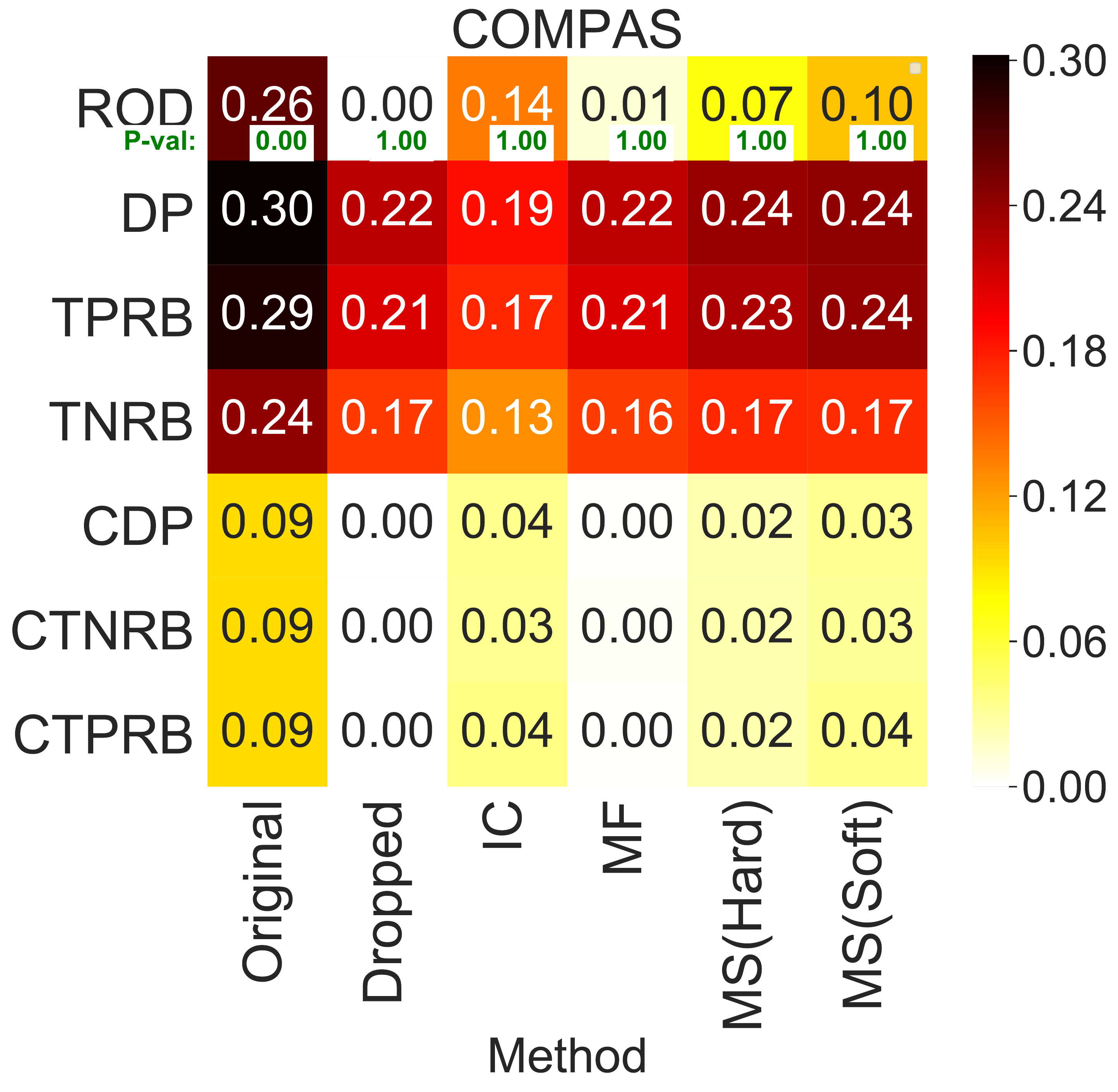}
		\includegraphics[scale=0.15]{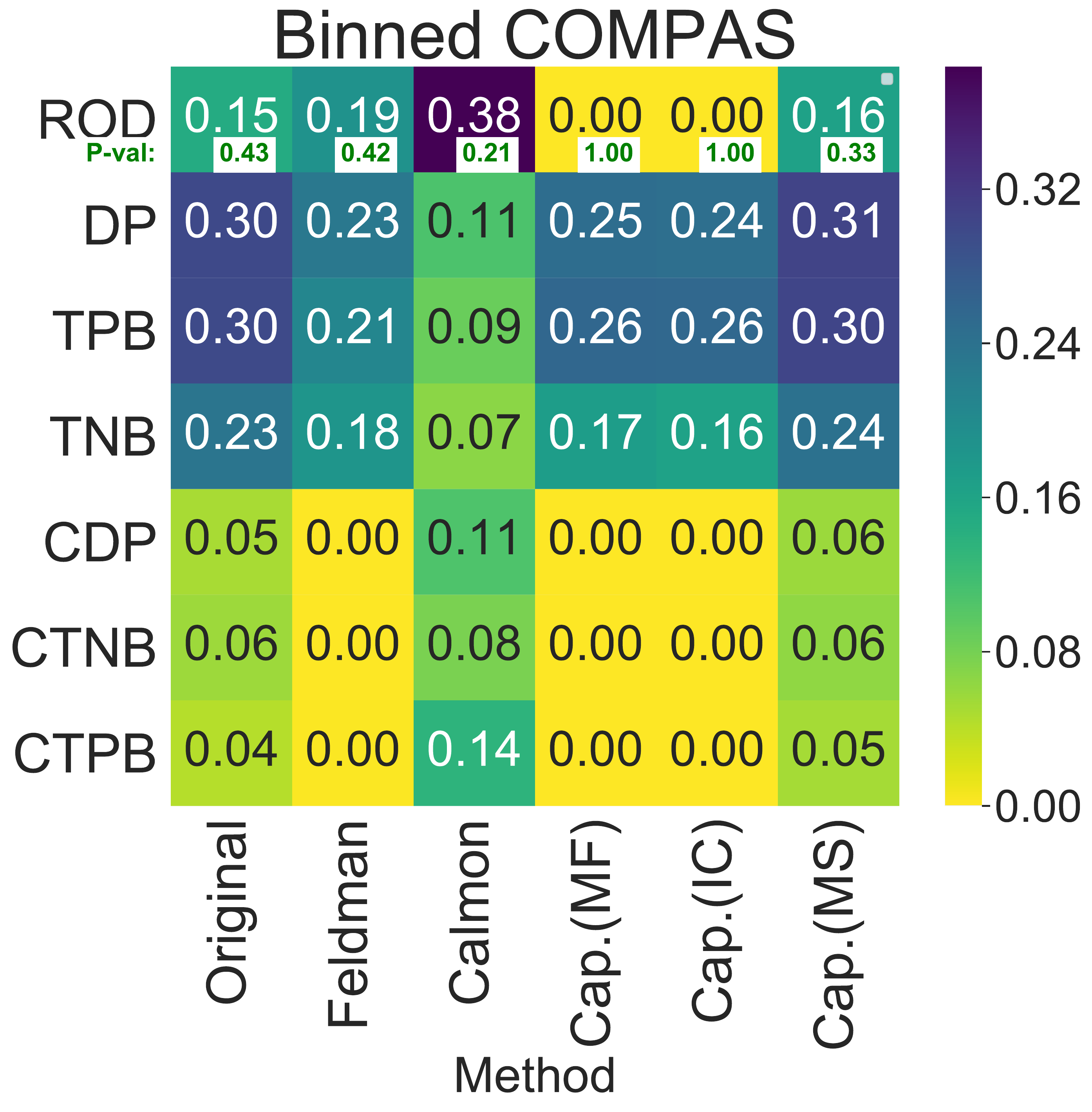}
		\includegraphics[scale=0.15]{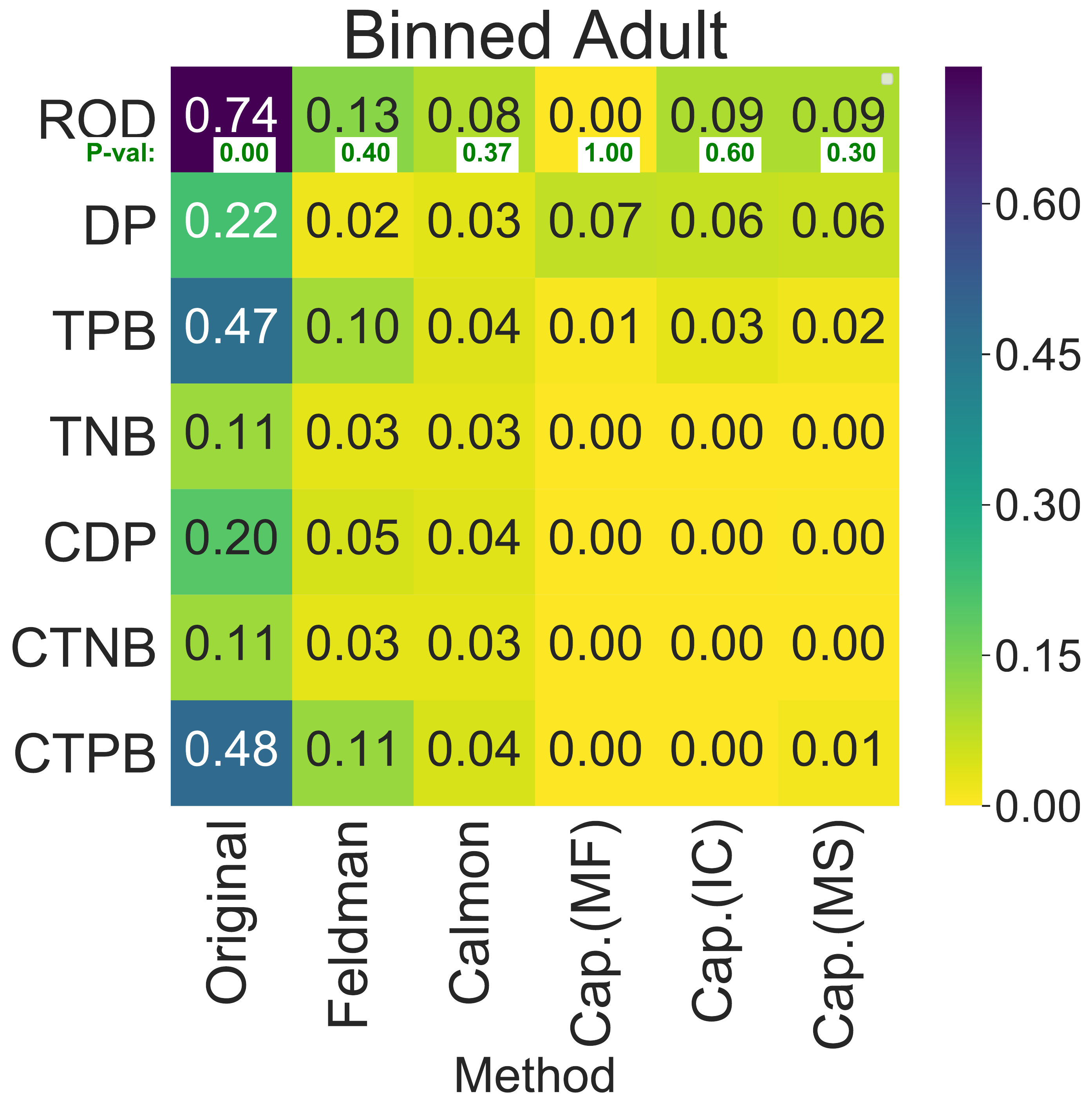}
\caption{ \textmd{Fairness Comparison of \sys\ for MLP classifier.}}
	\end{subfigure}
		\begin{subfigure}[t]{1\textwidth} \centering
		\hspace*{-1cm}		\includegraphics[scale=0.15]{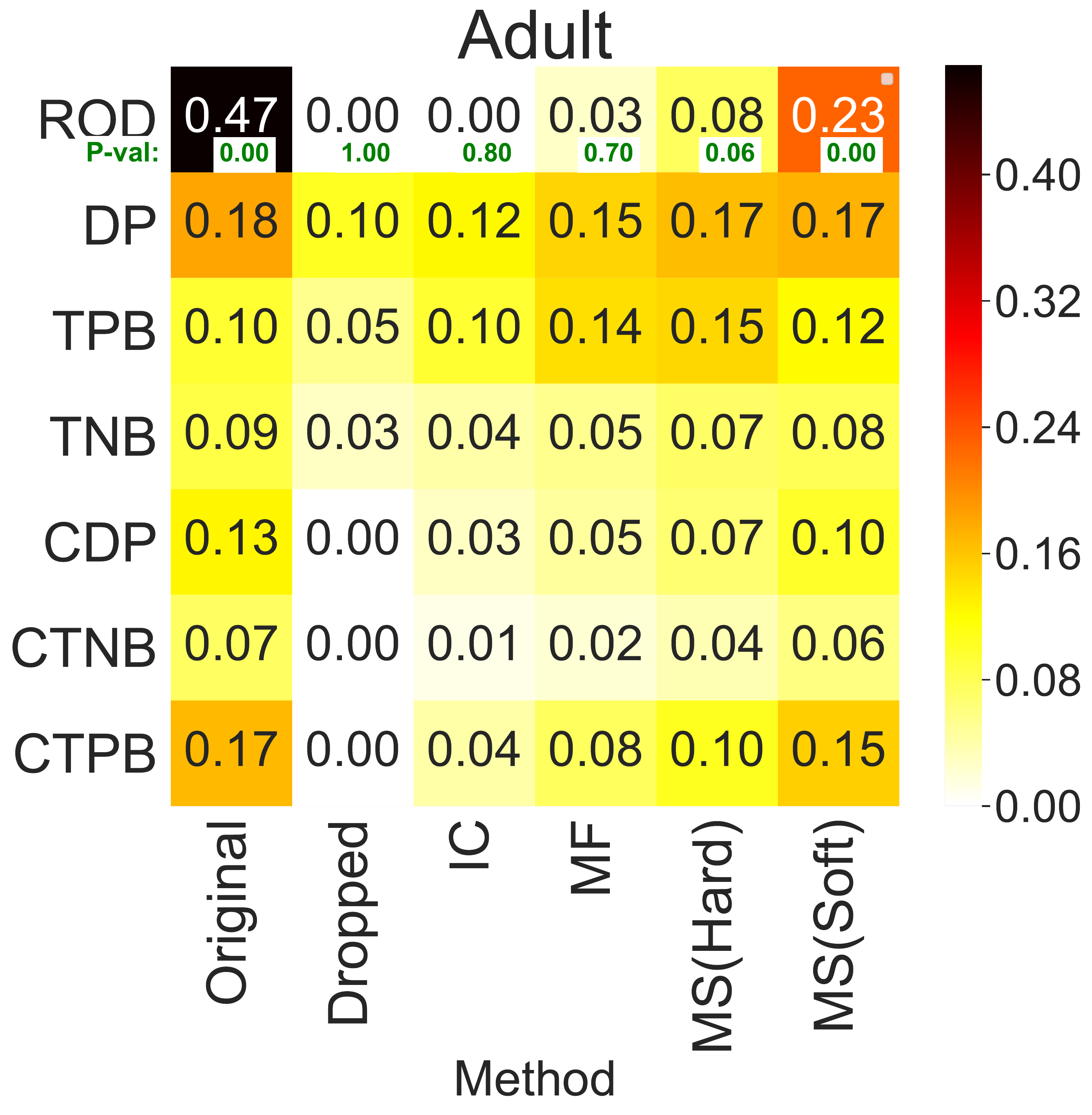}
		\includegraphics[scale=0.15]{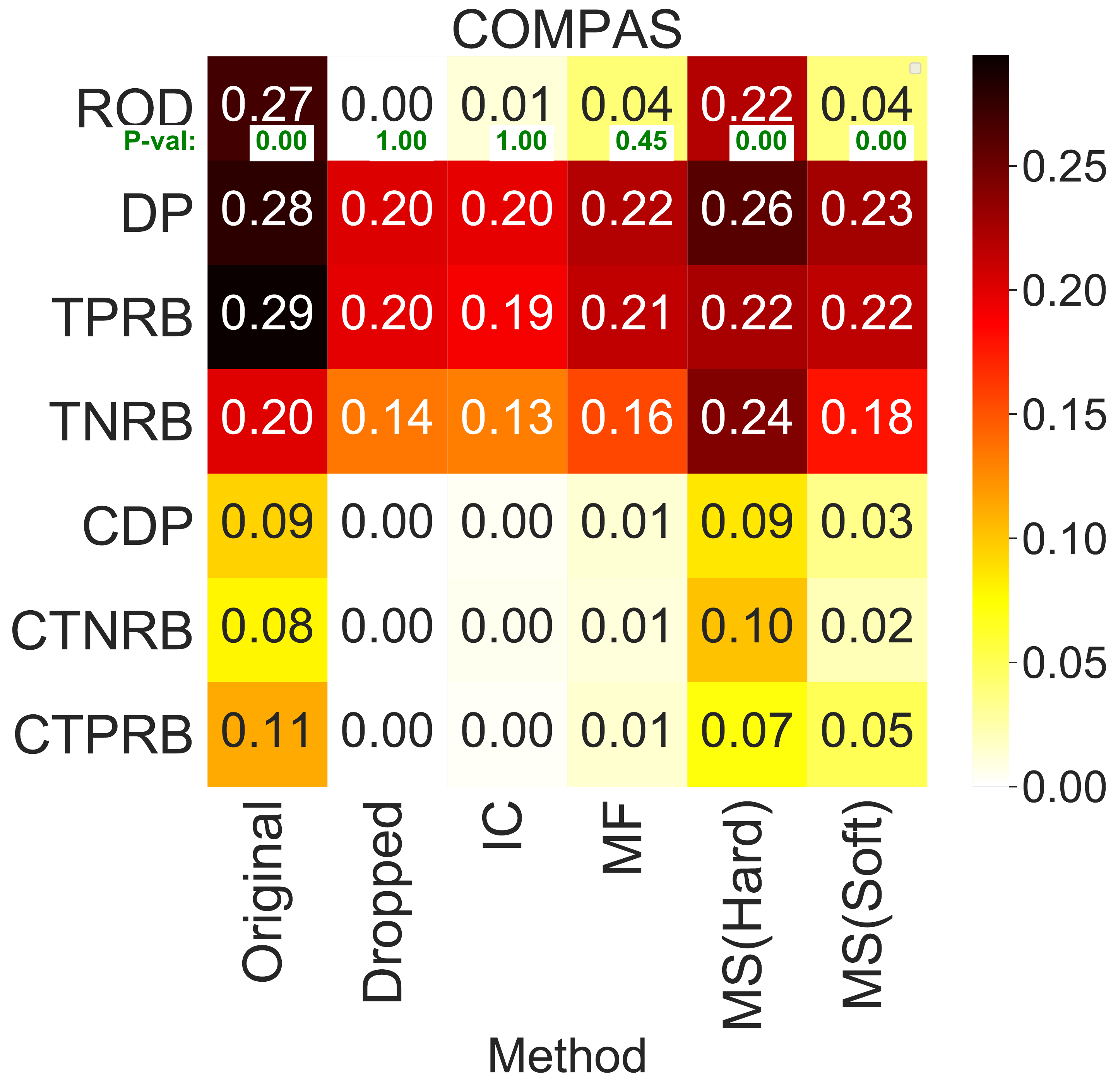}
		\includegraphics[scale=0.15]{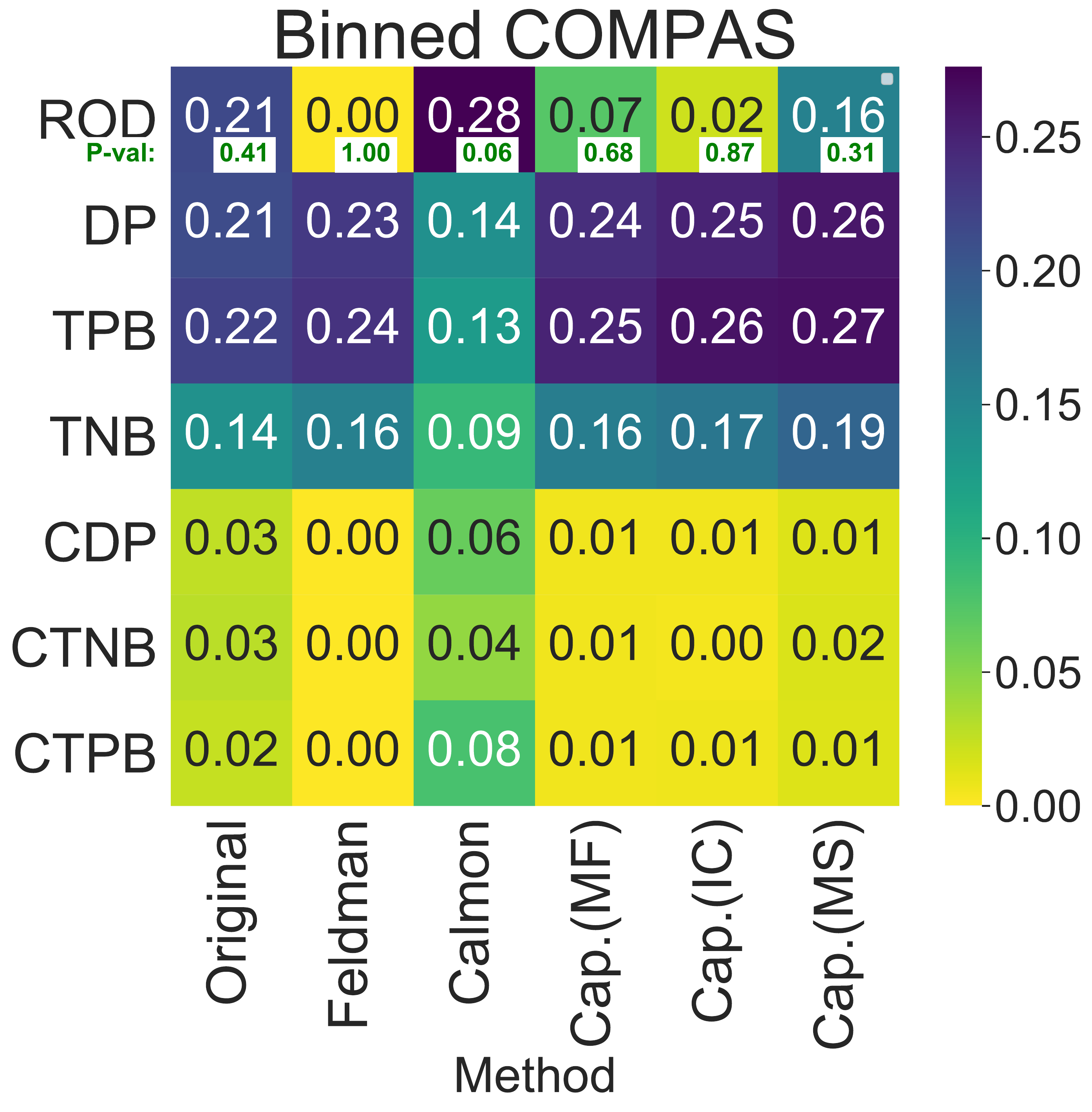}
		\includegraphics[scale=0.15]{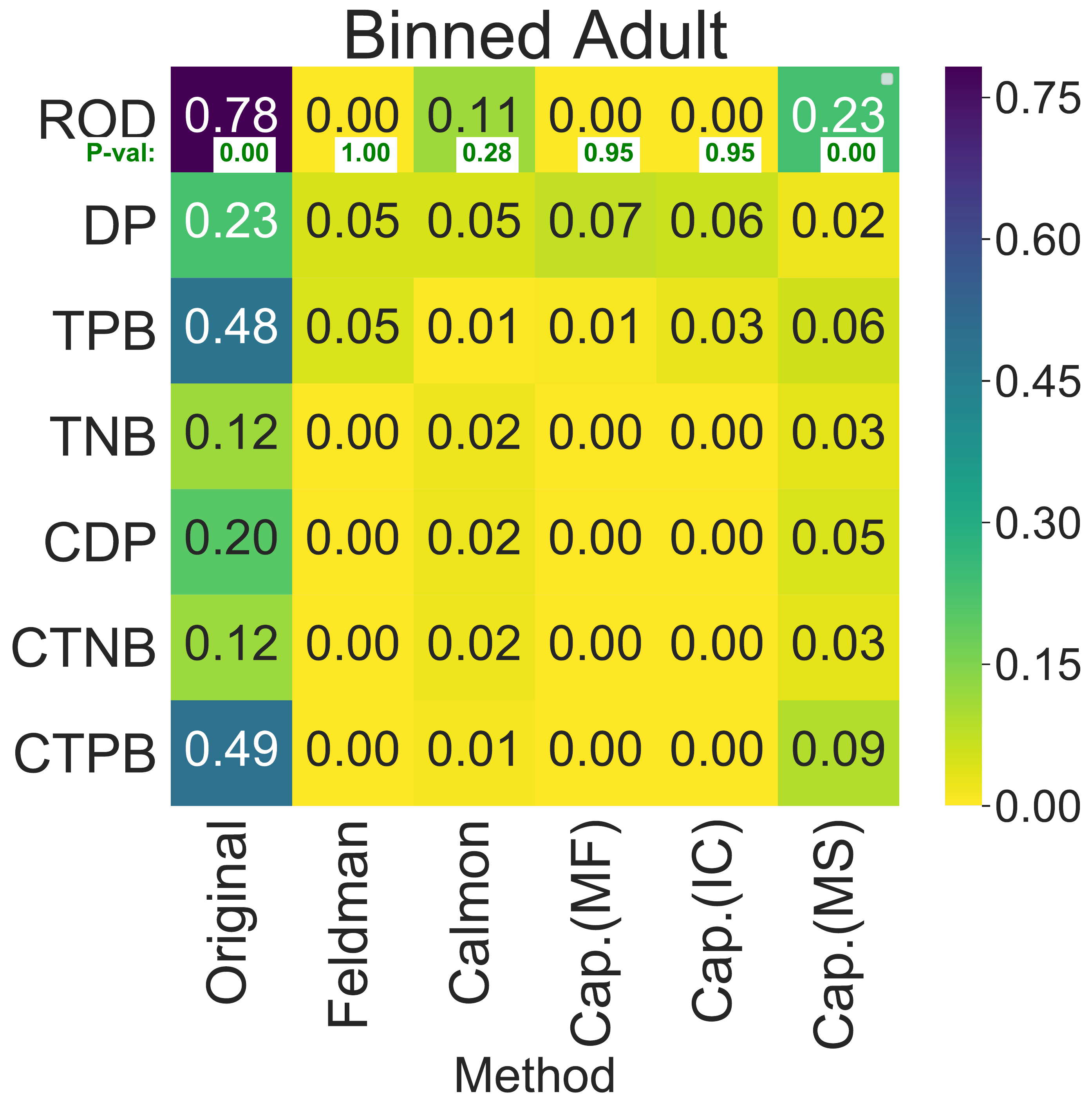}
		\caption{ \textmd{Fairness Comparison of \sys\ for RF classifier.}}
	\end{subfigure}

	\begin{subfigure}[t]{1\textwidth} \centering
		\hspace*{-1cm}		\includegraphics[scale=0.15]{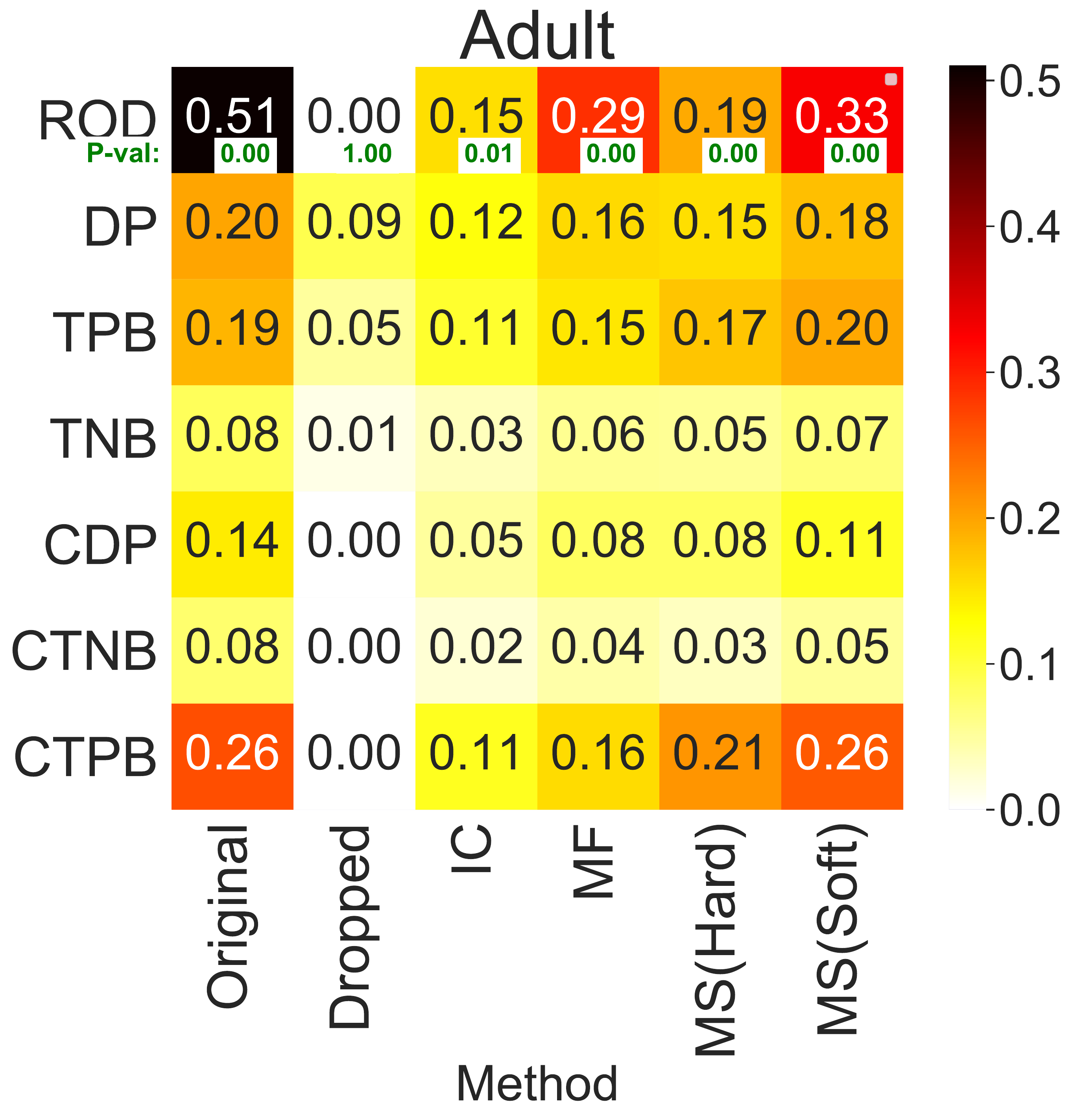}
		\includegraphics[scale=0.15]{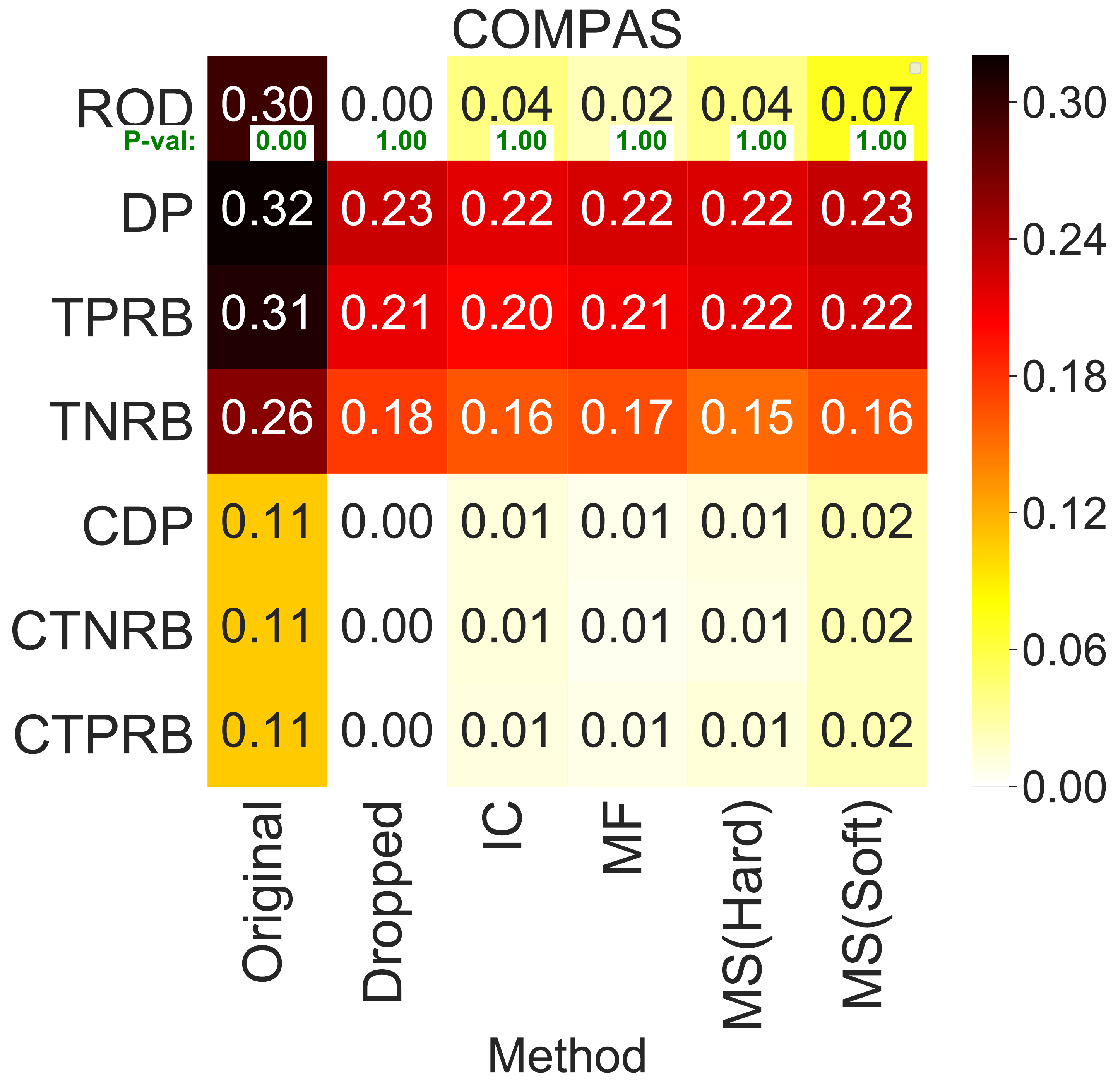}
		\includegraphics[scale=0.15]{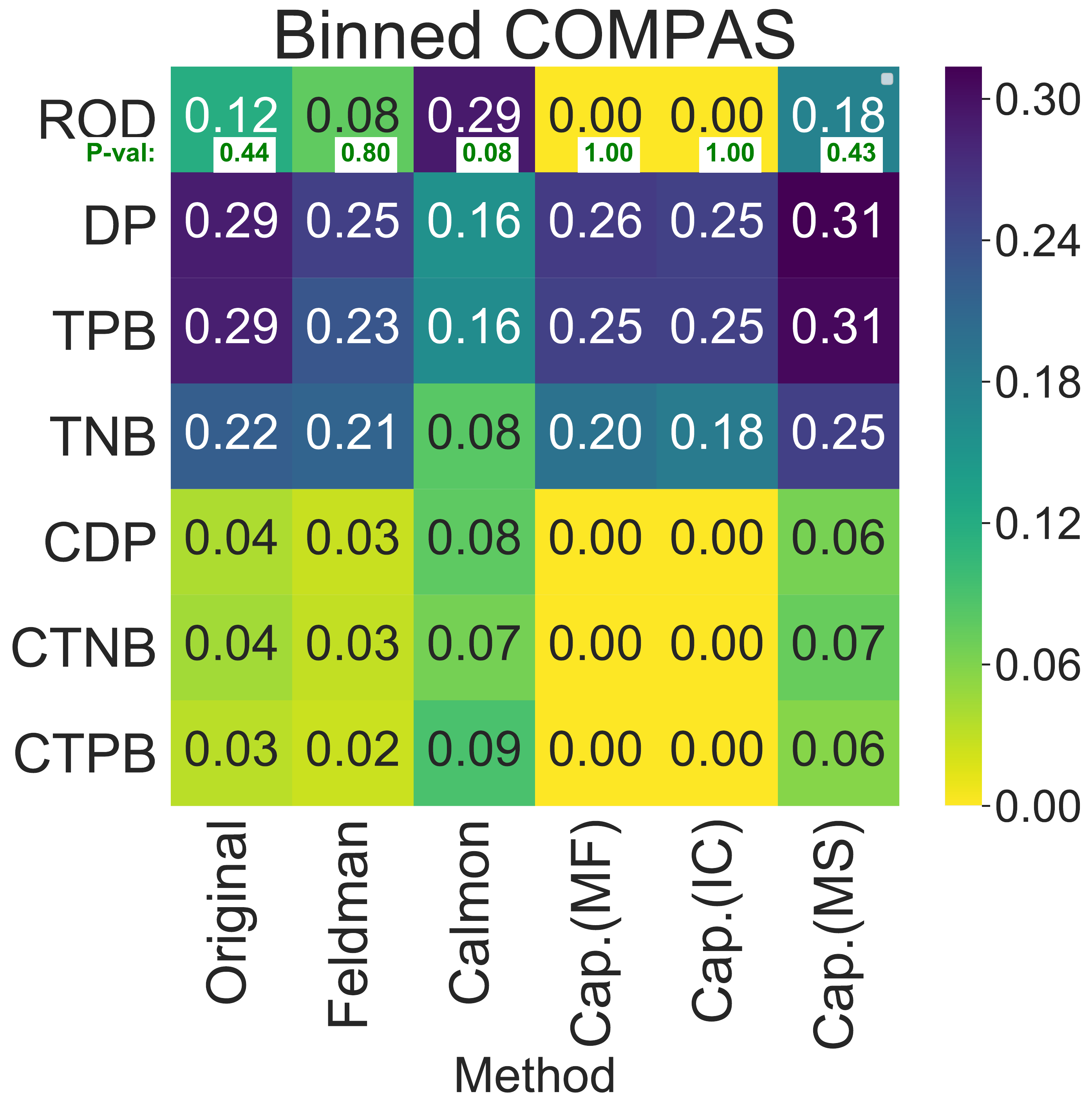}
		\includegraphics[scale=0.15]{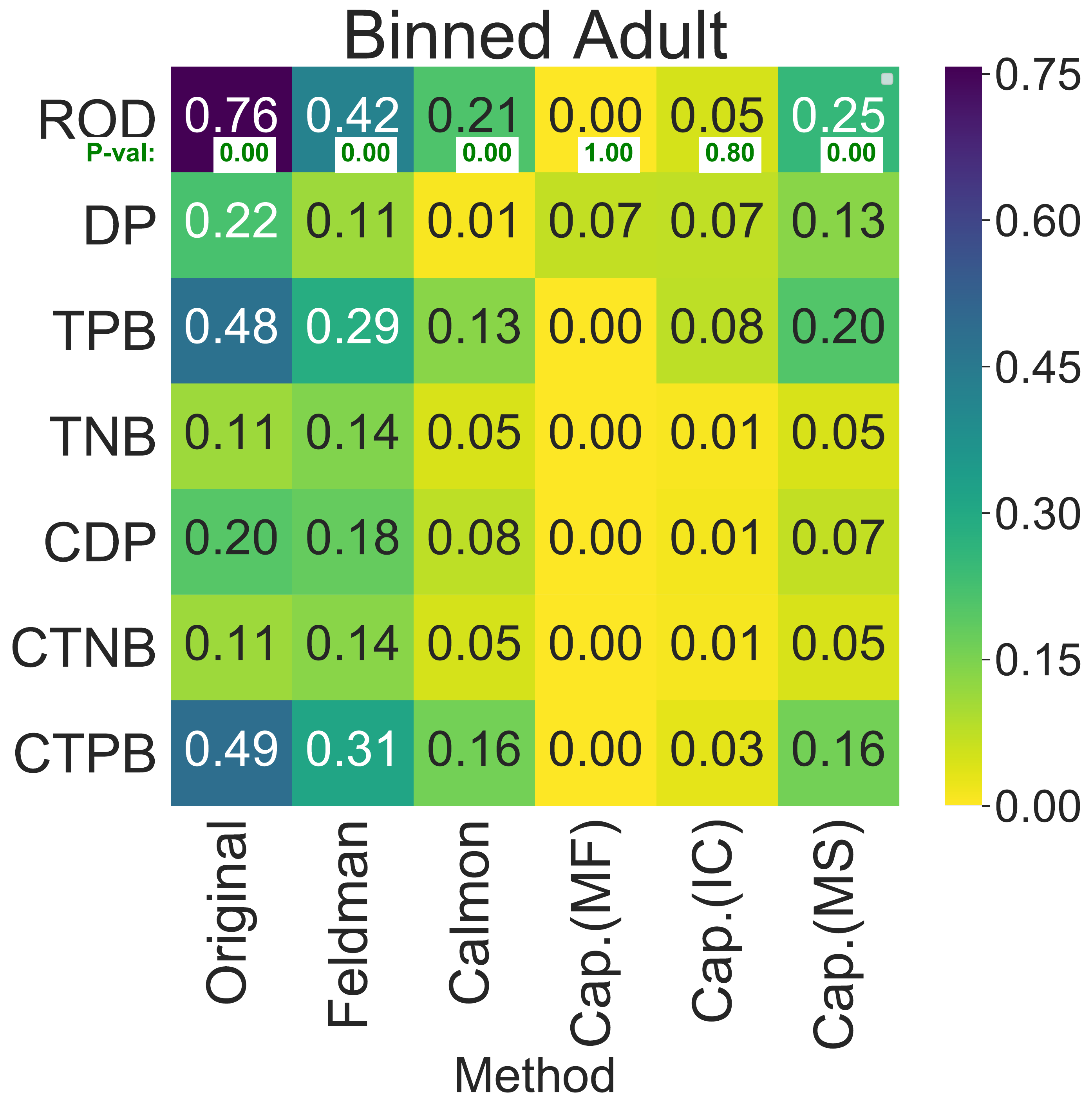}
		\caption{ \textmd{Fairness Comparison of \sys\ for LR classifier.}}
	\end{subfigure}
	\caption{ \textmd{Bias reduction performance  of \sys\ for MLP classifier.}}
		\label{fig:MLPbias}
\end{figure}

\subsection{End-To-End Results}

\label{sec:extendtoend}

In the following experiments, a fairness constraint was enforced on training data using \sys\ repair algorithms (cf. Sec~\ref{sec:repdata}).  Specifically, each dataset was split into five training and test datasets.  All training data were repaired separately using Matrix Factorization (MF), Independent Coupling (IC) and two versions of the MaxSAT approach: MS(Hard), which feeds all clauses of the lineage of a CI into MaxSAT, and MS(Soft), which only feeds small fraction of the clauses.  We tuned  MaxSAT to enforce CIs approximately. We then measured the utility and discrimination metrics for each repair method as explained in Sec~\ref{sec:setup}. For all datasets, the chosen training variables included the Markov boundary of the outcome variables, which were learned from data using the Grow-Shrink algorithm \cite{margaritis2003learning} and permutation  \cite{salimi2018bias}. 

{\bf Adult data.} Using this dataset, several prior efforts in algorithmic fairness have reported  gender discrimination based on a strong statistical dependency between income and gender in favor of males \cite{luong2011k,vzliobaite2011handling,tramer2017fairtest}. However, it has been shown that Adult data is inconsistent \cite{salimi2018bias} because its income attribute reports household income for married individuals, and there are more married males in data. Furthermore, data reflects the historical income inequality that can be reinforced by ML algorithms. 
We used \sys\ to remove the mentioned sources of discrimination from Adult data.  Specifically, we categorized the attributes in Adult data as follows: $(\mb S)$ sensitive attributes: gender (male, female); $(\mb A)$ admissible attributes: hours per week, occupation, age, education, etc.; $(\mb N)$ inadmissible attributes: marital status; $(Y)$ binary outcome: high income.  As is common in the literature, we assumed that the potential influence of gender on income through some or all of the admissible variables was fair; {for example, gender influences education and occupation, which in turn influence income, but, for this experiment, these effects were not considered discriminatory.} However, the direct influence of gender on income, as well as its indirect influence on income through marital status, were assumed to be  discriminatory. To remove the bias, we enforced the CI  $(Y \indep \mb S, \mb N| \mb D)$ on training datasets using the \sys\ repair algorithms. Then, we trained the classifiers on both original and repaired training datasets using the set of variables $ \mb A \cup \mb N \cup \mb S$. We also trained the classifiers on original data using only $\mb A$, i.e., we dropped the sensitive and inadmissible variables.

\begin{figure}[htbp]  \centering
	\begin{subfigure}{0.7\textwidth} 
	\hspace*{-.9cm}		\includegraphics[scale=0.5]{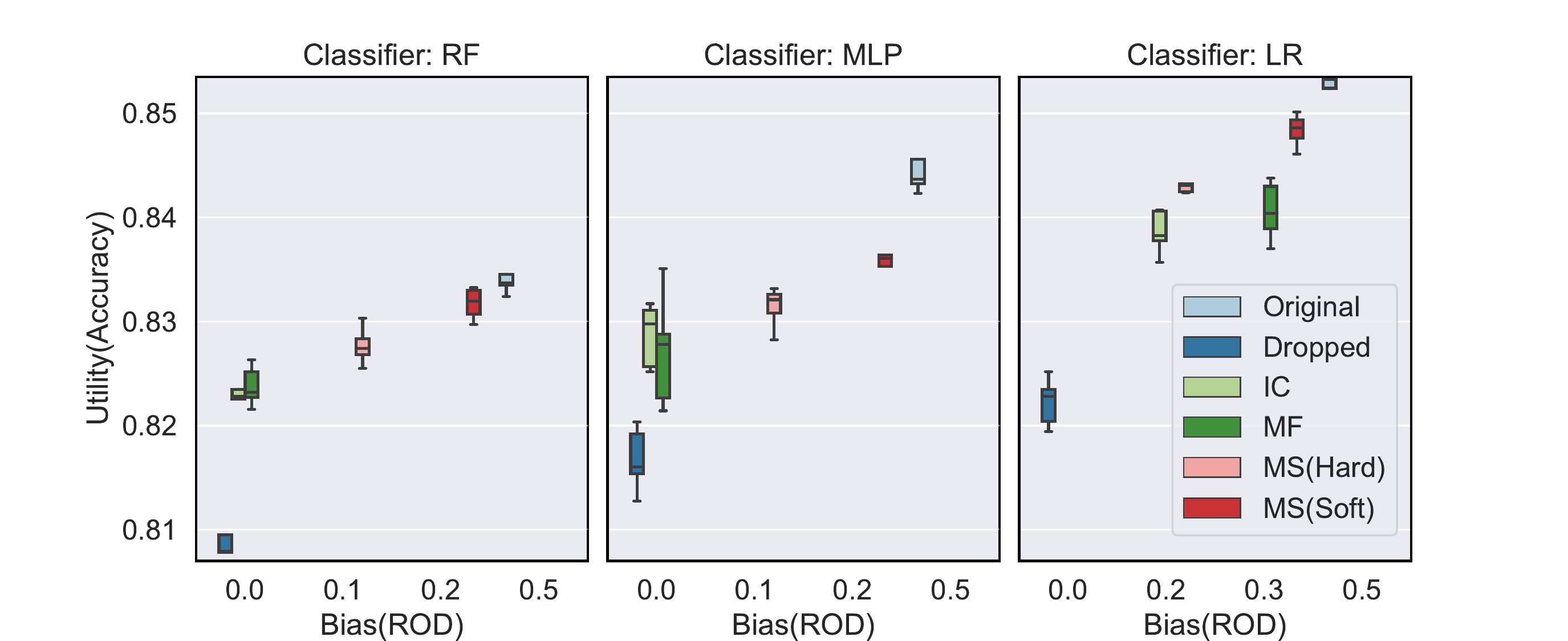}
	\end{subfigure}
	\caption{\textmd{Performance of \sys\ on Adult data.}}
				\label{fig:adult}
\end{figure}

Fig.~\ref{fig:adult} compares the  utility and bias of \sys\ repair methods on Adult data.  As shown, all repair methods successfully reduced
the ROD for all classifiers. As shown in Fig.~\ref{fig:MLPbias}, the repaired data also improved associational fairness measures: the \sys\ repair methods had an effect similar to dropping the sensitive and inadmissible variables completely, but they delivered much higher accuracy (because the CI was enforced approximately).   {The residual bias after repair was expected since: (1) the classifier was only an approximation, and (2) we did not repair the test data.   However, as shown in most cases, the residual bias indicated by ROD was not statistically significant. This shows that our methods are robust (by design) to the mismatch between the distribution of repaired data and test data.}  These repair methods delivered surprisingly good results: when partially repairing data using the MaxSAT approach, i.e, using MS(Soft), almost 50\% of the bias was removed while accuracy decreased by only 1\%. We also note that the residual bias generally favored the protected group (as opposed to the bias in the original data). 

\begin{figure}[htbp] \centering
	\begin{subfigure}{0.7\textwidth}
		\hspace*{-.9cm}	\includegraphics[scale=0.5]{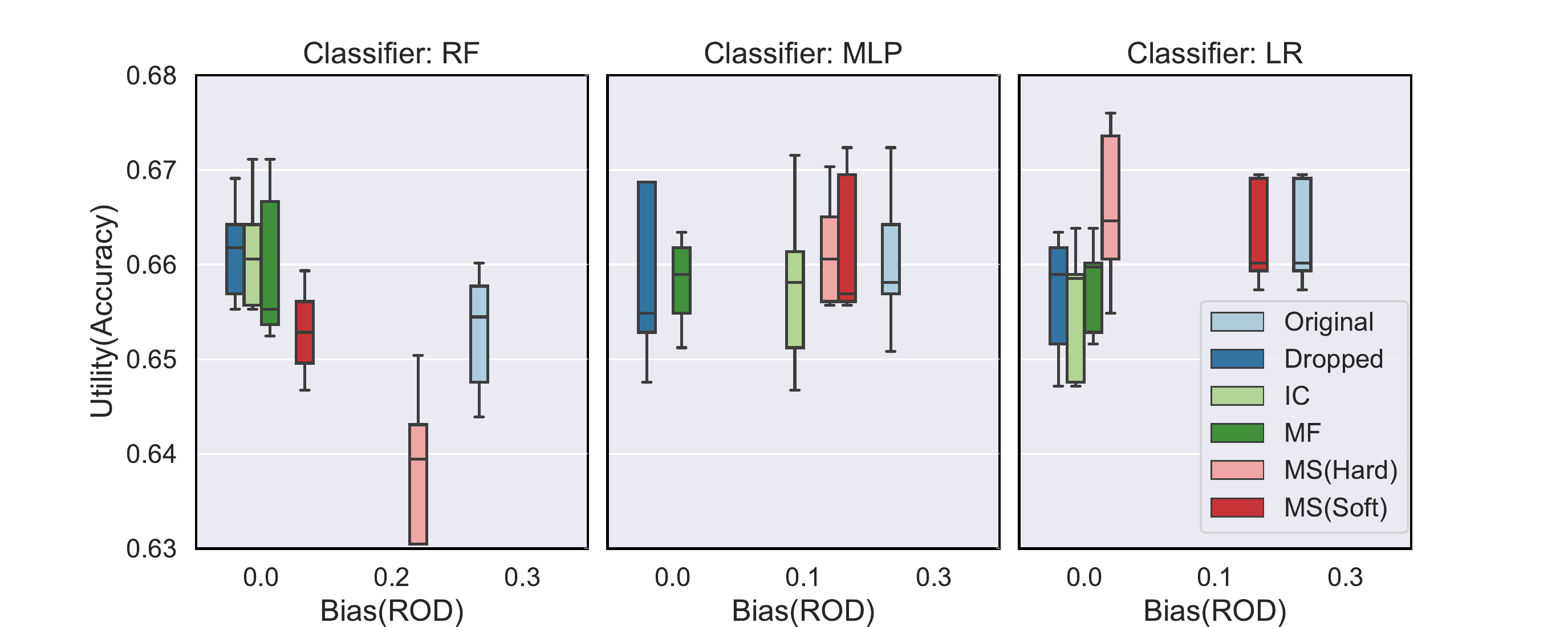}
	\end{subfigure}	\caption{\textmd{Performance of \sys\ on COMPAS data.}}
\label{fig:compas}
\end{figure}

{\bf COMPAS.}  For the second experiment, we used the ProPublica COMPAS dataset \cite{larson2016we}. This dataset contains records for all offenders in Broward County, Florida in 2013 and 2014. We categorized the attributes in COMPAS data as follows: $(\mb S)$ protected attributes: race (African American, Caucasian); $(\mb A)$ admissible attributes:  number of prior convictions, severity of charge degree, age; \textbf{(Y)} binary outcome: a binary indicator of whether the individual is a recidivist. As is common in the literature, we assumed that it was fair to use the admissible attributes to predict recidivism even though they can potentially be influenced by race, and our only goal in this experiment was to address the direct influence of race. We pursued the same steps as explained in the first experiment. \ignore{we used \sys\ to enforce the CI  $(Y \indep \mb S| \mb A)$. Then, we trained the classifiers on both original and repaired training datasets on $ \mb A \cup \mb S$. We also trained the classifiers on the original data but only on $\mb A$, i.e., we dropped the sensitive variable. }
Fig.~\ref{fig:compas} compares the bias and utility of \sys\ repair methods to original data.  As shown, all repair methods successfully reduced the ROD. However, we observed that MF and IC performed better than MS on COMPAS data (as opposed to Adult data); see~\ref{sec:rep_metho_com} for an explanation. \ignore{An interesting observation is that the significant DP, TPRB and TNRB values could be explained by the admissible variables. \bill{I didn't get the point of the last sentence} Also} {We observed that in some cases, repair improved the accuracy of the classifiers
on test data by preventing overfitting.}

In addition, we used \sys\ to compute the ROD  for the COMPAS score and compared it to the same quantity computed for the ground truth. While we obtained a 95\% confidence interval of $(0.7, 0.9)$ for ROD for ground truth, we obtained a 95\% confidence interval of (0.3,  0.5) for ROD for the COMPAS score (high, low). That is, for individuals with the same number of prior convictions and severity of charges, COMPAS overestimated the odds of recidivism by a factor close to 2. The fact that the admissible variables explain the majority of the association between race and recidivism --- but not the association between COMPAS scores and recidivism --- suggests COMPAS scores are highly racially biased.

\subsection{Comparing \sys\ Repair Methods}
\label{sec:rep_metho_com}
\begin{figure}[ht]  \centering
  \includegraphics[width=.33\linewidth]{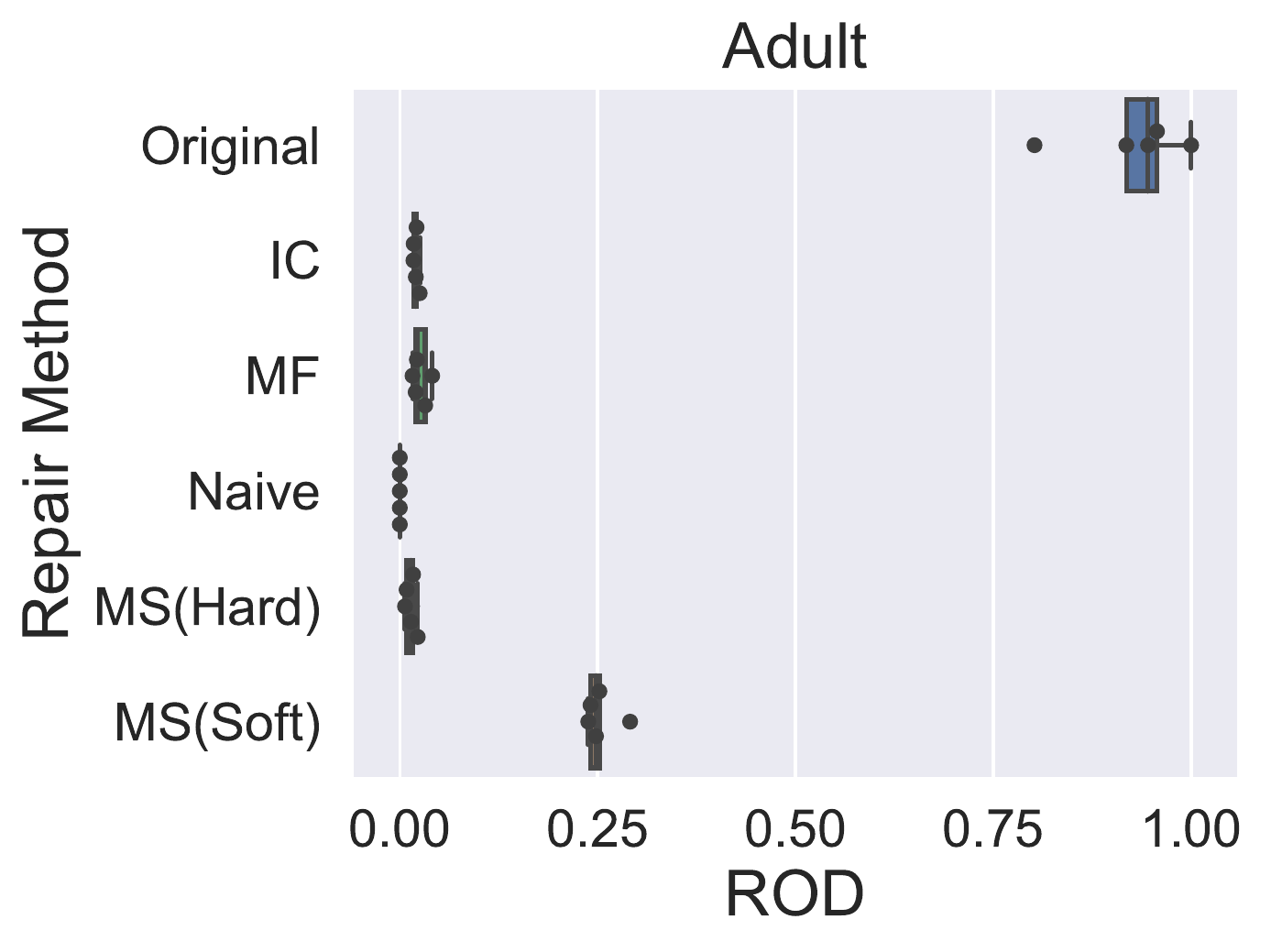}
\hspace*{0.8cm}    \includegraphics[width=.3\linewidth]{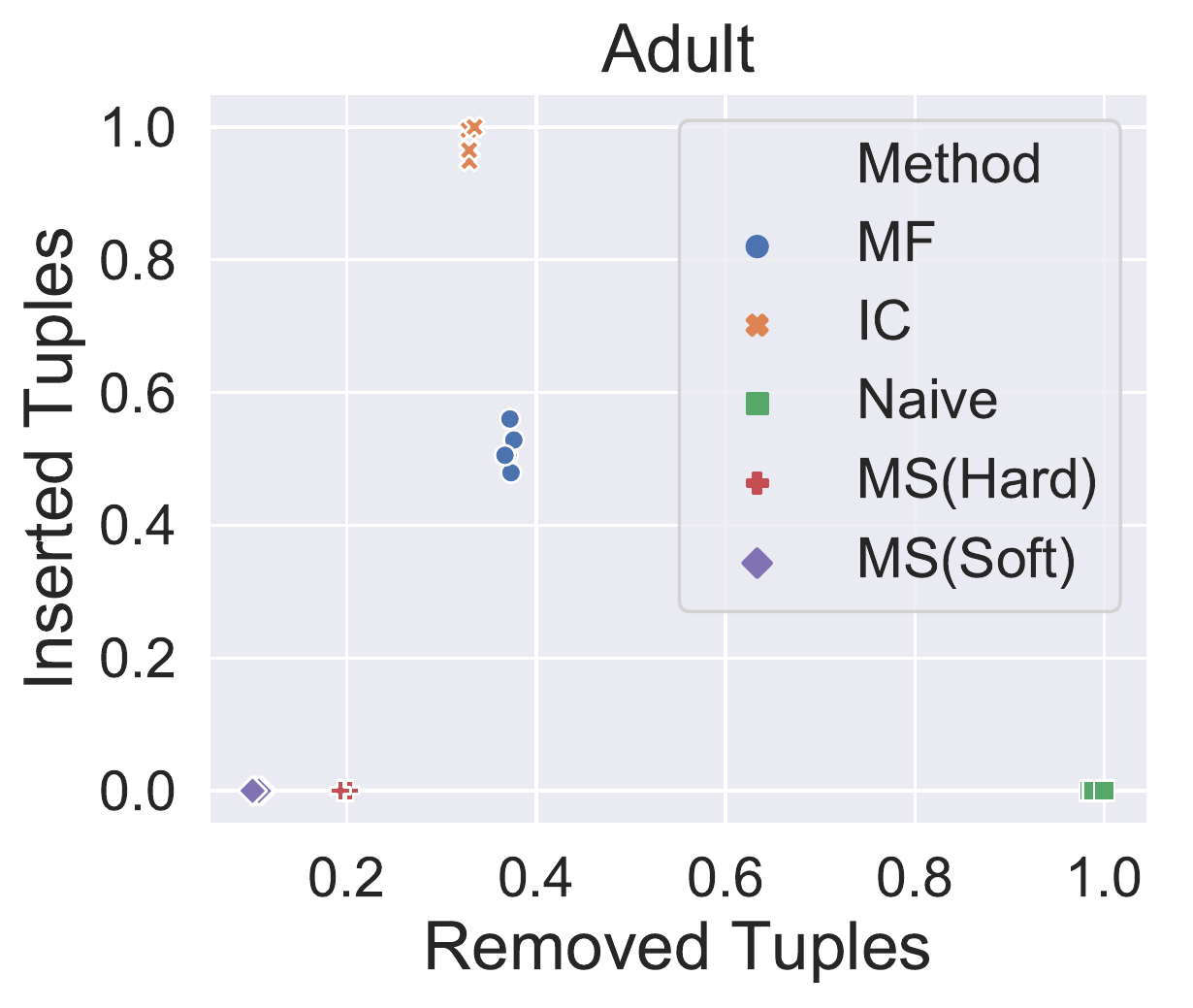}
    \\
      \includegraphics[width=.33\linewidth]{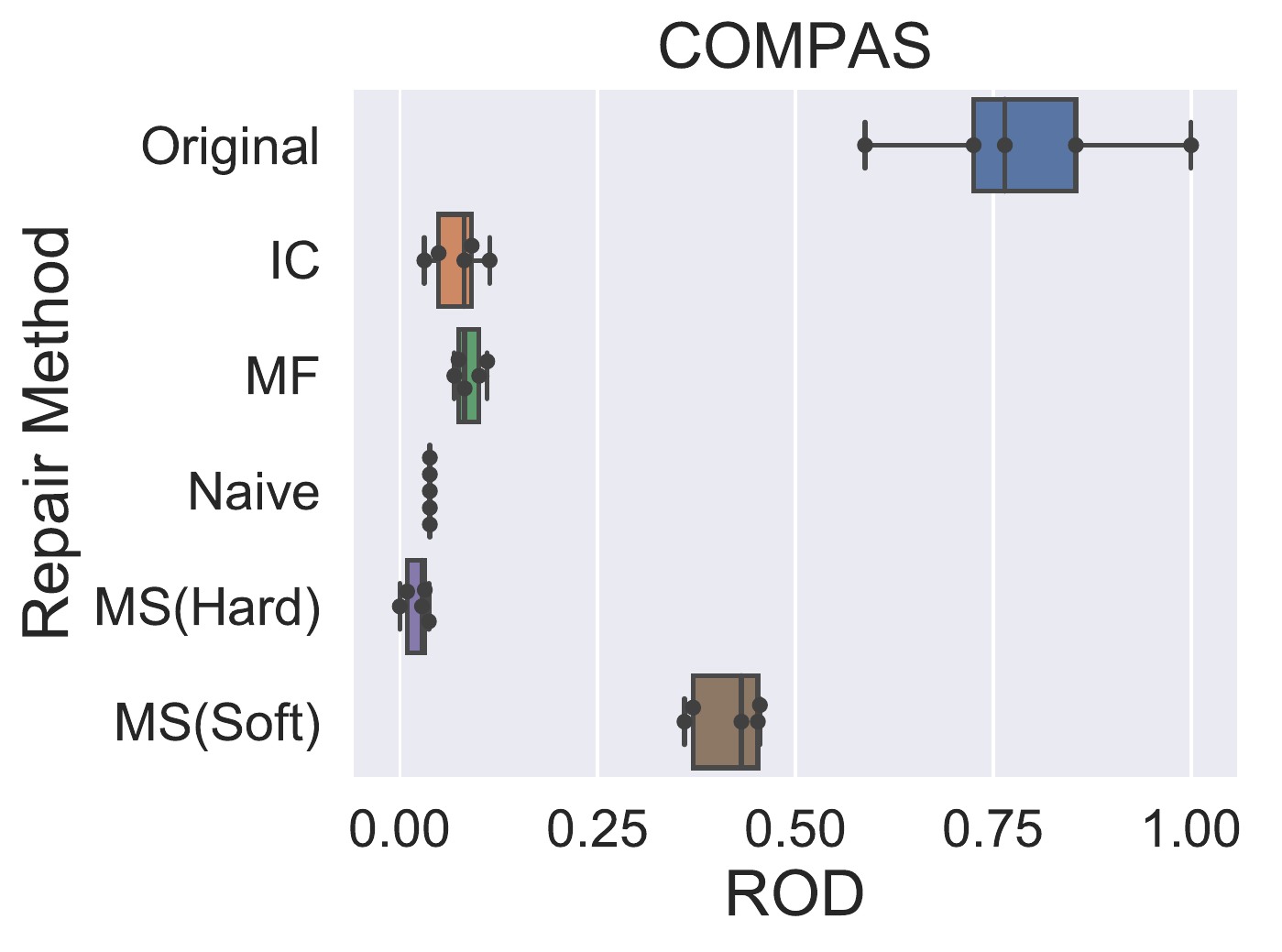}
  \hspace*{0.8cm}         \includegraphics[width=.3\linewidth]{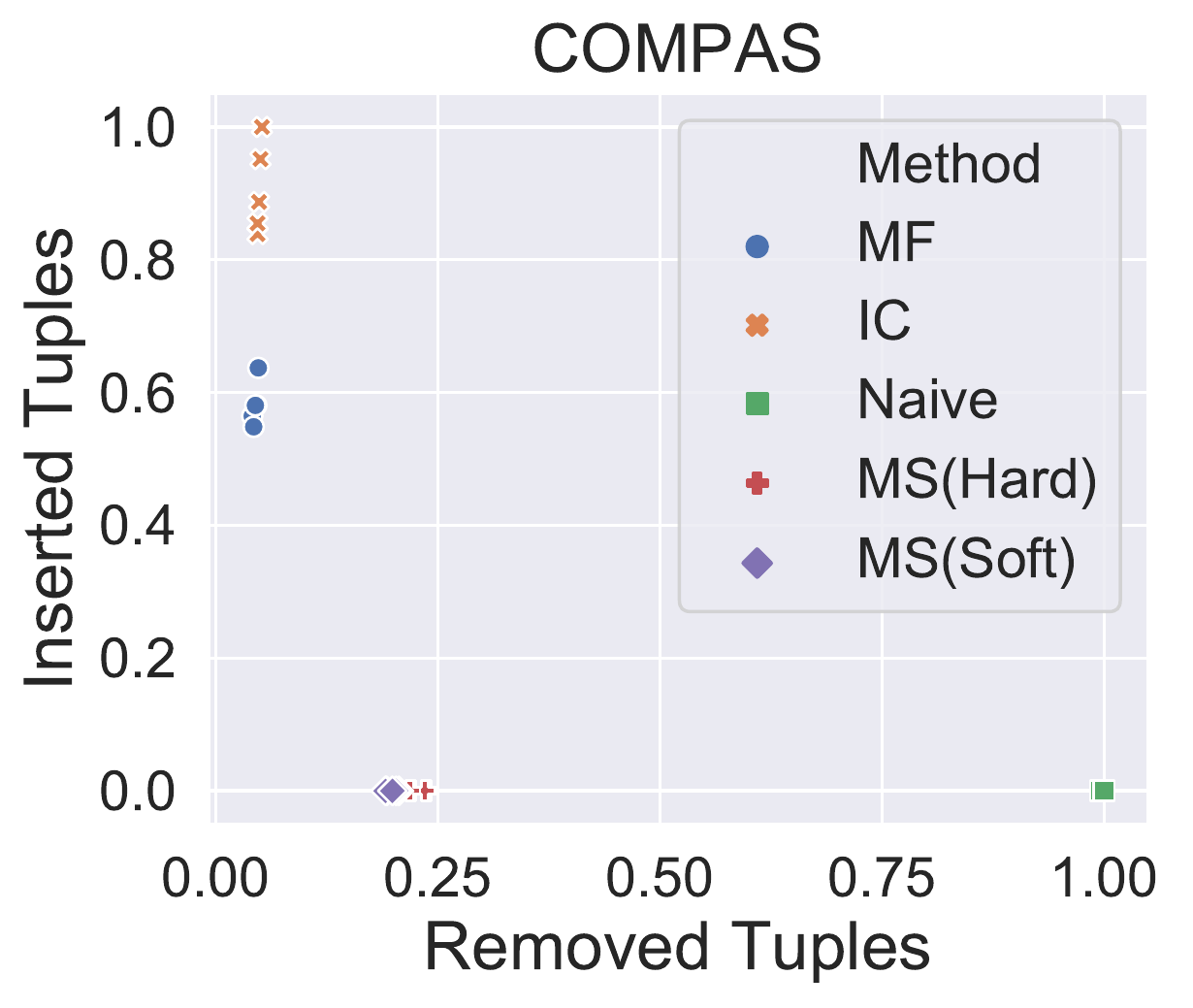}
        	\caption{\textmd{Comparison of different repair methods. All methods successfully enforced fairness constraint, but with different ratio of tuple insertion and deletion. On Adult data  MaxSAT outperformed the other methods due to sparsity.}}
	\label{fig:com_rep}
\end{figure}

To compare  \sys\ repair methods beyond the utility experiments in Sec~\ref{sec:extendtoend}, we compared the number of tuples added and deleted for each method, as well as the bias reduction on training data. Fig~\ref{fig:com_rep} reports these measures for the experiments in Sec~\ref{sec:extendtoend}. Note that all numbers were normalized between 0 and 1, where ROD=1 shows no discrimination.  For Adult data, we tuned the MS approach to repair data only by tuple deletion and compared it to a naive approach that repaired data using lineage expression but without using the MaxSat solver. As shown in Fig~\ref{fig:com_rep}, the MaxSat approach removed up to 80\% fewer tuples than the naive approach. 

\begin{figure}[htbp] \centering

	\begin{subfigure}{0.8\textwidth} 
\hspace{1.9cm}	\includegraphics[scale=0.4]{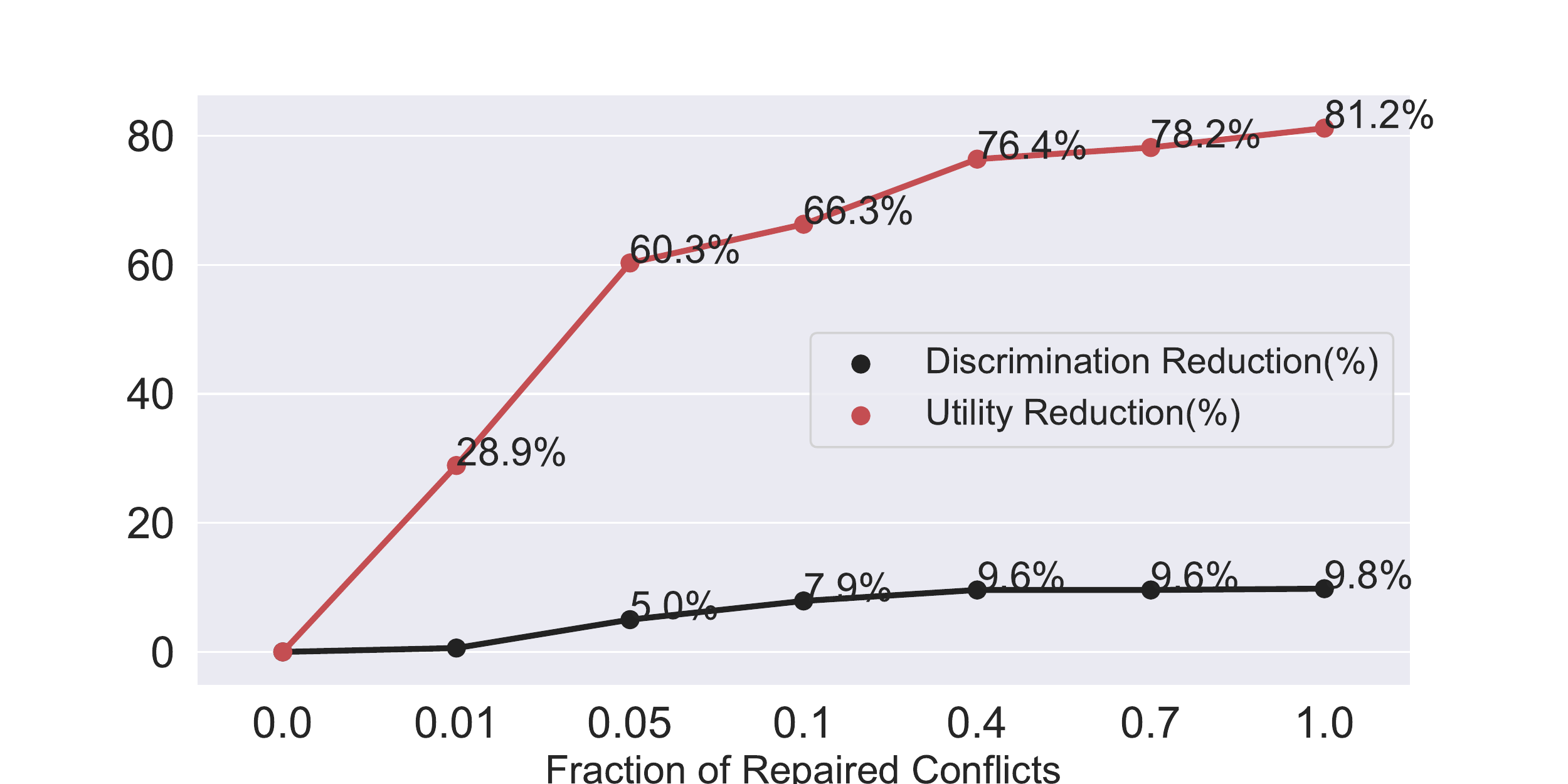}
	\end{subfigure}
	\caption{\textmd{Bias-utility trade off in MaxSAT approach.Repairing only small fraction of conflicts with the fairness constraint in Adult data significantly reduced discrimination.}}
	\label{fig:bias-accur}
\end{figure}

In general, the MaxSAT approach was the most flexible repair method (since it can be configured for partial repairs).  Further, it achieved better classification accuracy, and it balanced tuple insertion and deletion.  {Further, it could be extended naturally to multiple CIs, though we defer this extension for future work.}  In terms of the utility of classification, the MS approach performed better on sparse data in which the conditioning groups consisted of several attributes. Figure \ref{fig:bias-accur} shows that repairing a very small fraction of inconsistencies (i.e., clauses in the lineage expression of the associated CI) in the experiment conducted on Adult data  (Sec~\ref{sec:extendtoend})  led to a significant discrimination reduction. This optimization makes the MS approach more appealing in terms of balancing bias and utility.  However, for dense data, IC and MD performed better. This difference was because the size of the lineage expression grew very large when the conditioning sets of CIs consisted of only a few attributes.

\begin{figure} \centering
	\begin{subfigure}{0.3\textwidth} \centering
\hspace{-3.3cm}	\includegraphics[scale=0.5]{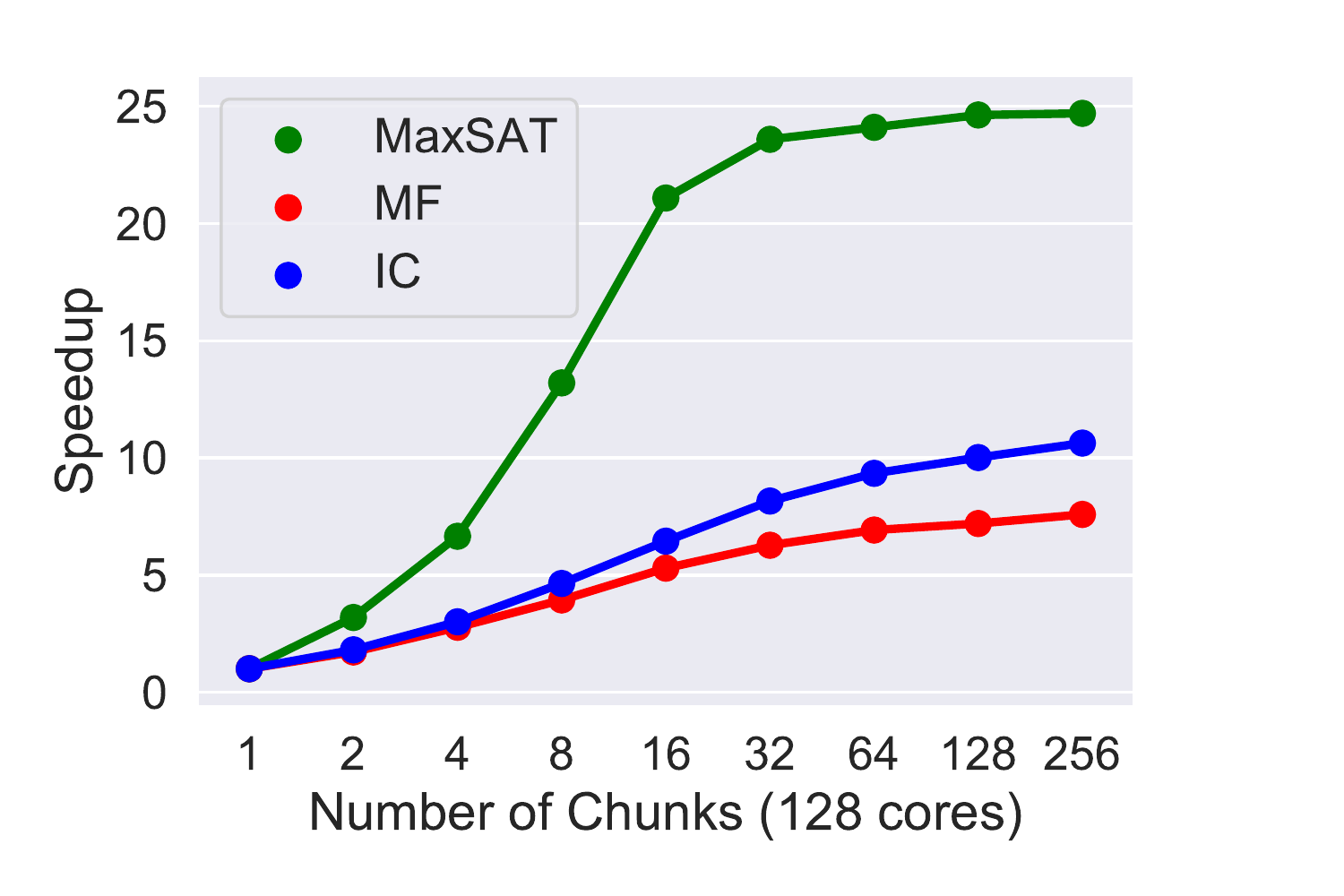}
		\caption*{(a)}
	\end{subfigure}
	\begin{subfigure}{0.3\textwidth} \centering
		\includegraphics[scale=0.5]{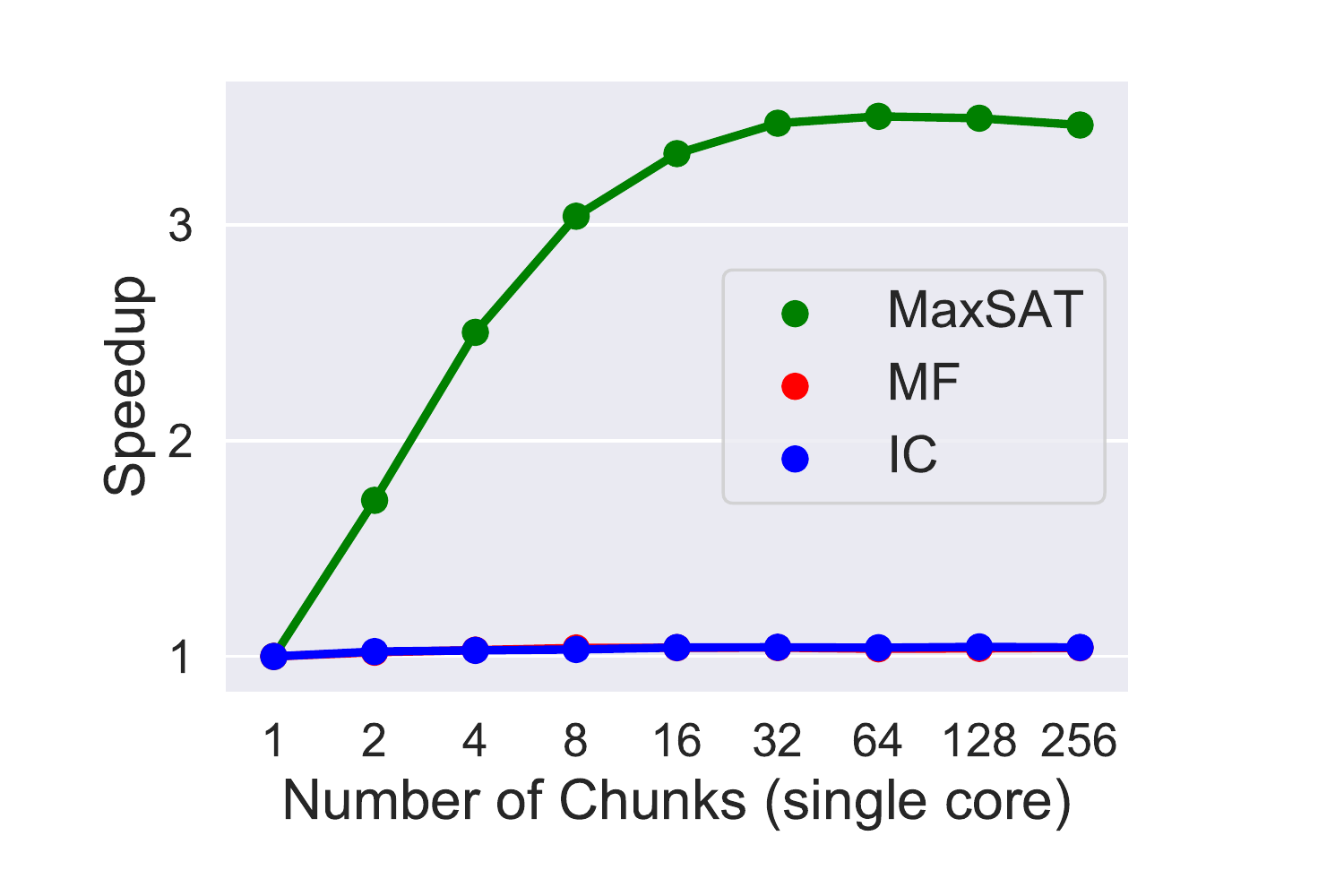}
		\caption*{(b)}
	\end{subfigure}
	\caption{\textmd{Speed up achieved (a) by partitioning  and parallel processing  on 128 cores; (b)  by partitioning  on a single core.}}
	\label{fig:opti}
\end{figure}


To evaluate the effect of partitioning and parallelizing on different methods, we replicated the experiment in sec~\ref{sec:extendtoend} and partitioned Adult data into several chunks of approximately equal sizes; we then repaired the chunks in parallel on a cluster of 128 cores.  Fig~\ref{fig:opti} shows the achieved speed up; all approaches were parallelizable.
Parallel processing was most appealing for MaxSAT since MaxSAT solvers were much more efficient on smaller input sizes. While partitioning had no effect on MF and IC on  a single-core machine, as shown in  Fig~\ref{fig:opti}(b), it sped up MaxSAT approach on even a single core. \revd{Note that partitioning data into several small chunks does not necessarily speed up the MaxSAT approach, since MaxSAT solver must be called for several small inputs. Hence, performance does not increase linearly by increasing the number of chunks. In general partitioning data into several instance of medium size delivers the best performance.}


\subsection{Comparing \sys\ to Other Methods}
\label{sec:comp}
We compared \sys\  with two reference pre-processing algorithms, Feldman et al.~\cite{feldman2015certifying} and  Calmon et al.~\cite{NIPS2017_6988}. Feldman's  algorithm modifies each attribute so that the marginal distributions based on the subsets of the attribute with a given sensitive value are all equal. \ignore{, but it does not modify training labels.} Calmon's algorithm randomly
transforms all variables except for the sensitive attribute to reduce the dependence between training labels and the sensitive attribute subject to the following constraints: (1) the joint distribution of the transformed data is close to the original distribution, and (2) individual attributes are not substantially distorted. {An example of a distorted attribute in the COMPAS dataset would be changing a felon with no prior conviction to 10 prior convictions.} Individual distortion is controlled for using a distortion constraint, which is domain dependent and has to be specified for each dataset. and every single attribute. Note that both approaches are designed to transform training and test data. 

We used these algorithm only to repair training datasets and compare their bias and utility to \sys. In addition, since the distortion function required in Calmon's algorithm is completely arbitrary,  we replicated the same experiments conducted in \cite{feldman2015certifying} using binned Adult data and binned COMPAS data. We note that the analysis in \cite{feldman2015certifying} was restricted to only a few attributes, and the data was excessively binned to few categories (to facilitate the definition of distortion function). As a result, the bias and utility obtained in this experiment was mismatched with Sec~\ref{sec:extendtoend}. For binned Adult data, the analysis was restricted to age, education and gender. COMPAS data used the same attributes as we used in Sec~\ref{sec:extendtoend}. For both datasets, we assumed all attributes  were admissible; hence, the direct effect of direct effect of the protected attribute to outcome was removed.

\begin{figure}[htbp] \centering
	\begin{subfigure}{0.8\textwidth} 
	\hspace*{-.9cm}\includegraphics[scale=0.5]{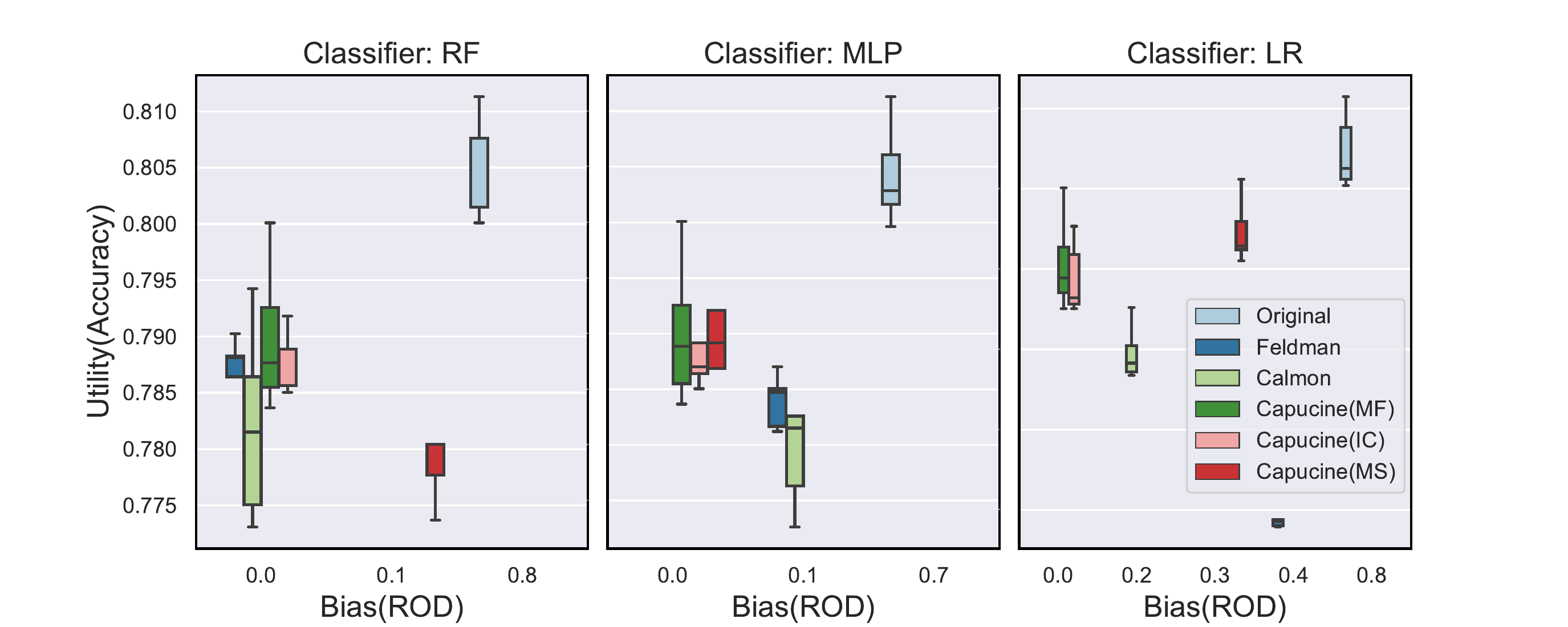}
	\end{subfigure}
	\caption{\textmd{Comparing \sys\ and other methods on Binned Adult data.}}
\label{fig:binnedadult}
\end{figure}

\begin{figure}[htbp] \centering
	\begin{subfigure}{0.8\textwidth} 
	\hspace*{-.9cm}	\includegraphics[scale=0.5]{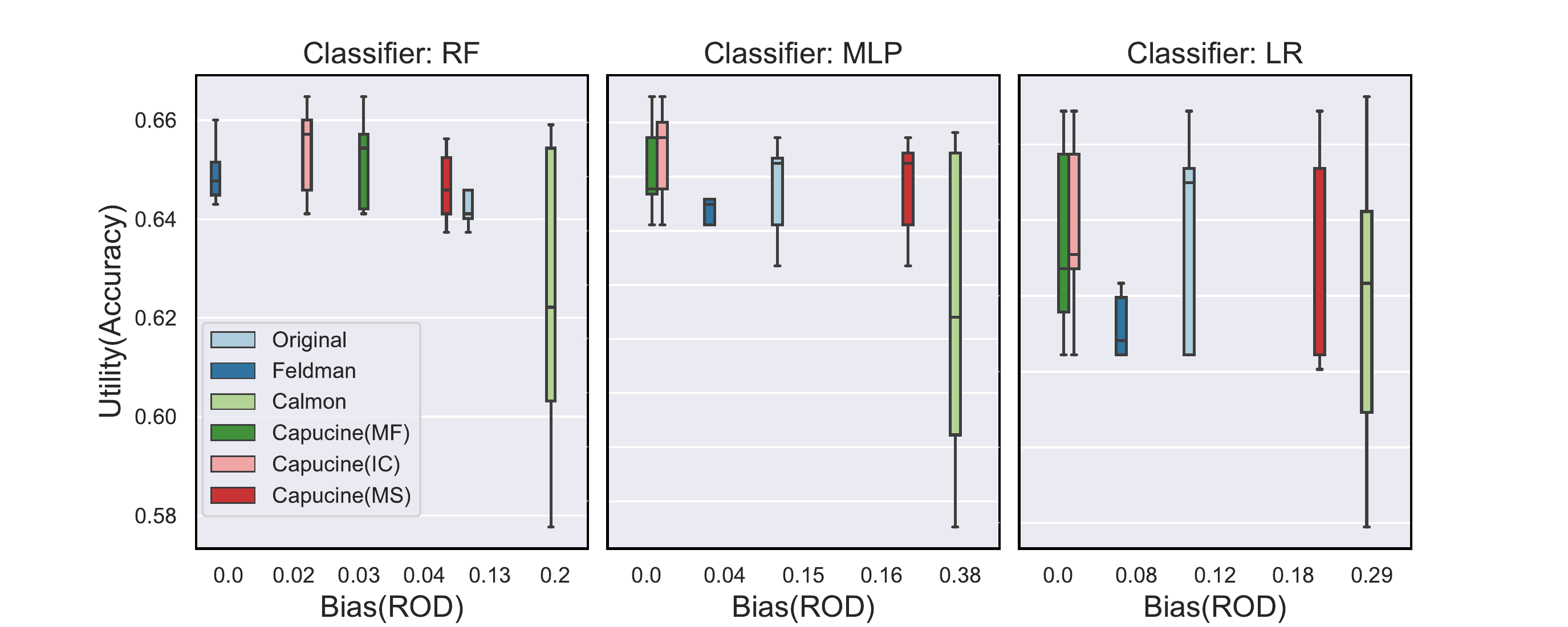}
	\end{subfigure}
	\caption{\textmd{Comparing \sys\ and other methods on Binned COMPAS data.}}
\label{fig:binnedcompas}
\end{figure}

Figs.~\ref{fig:binnedadult} and \ref{fig:binnedcompas}  compares the utility and bias of \sys\ to the reference algorithms. The insights obtained from this experiment follow. For binned Adult data, all methods significantly reduced ROD, even though the goal of Calmon's and Fledman's algorithm is essentially to reduce DP. Similarly, \sys\ reduced DP and other associational metrics as a side effect. However, \sys\ outperformed both methods in terms of utility. Because COMPAS data was excessively binned, the ROD in training labels became insignificant for COMPAS, and accuracy dropped by 2\%. We observe that both reference algorithms enforced DP at the cost of increasing ROD; however, in some cases the introduced bias was not statistically significant. In terms of  utility, all methods of \sys\ (except for MaxSAT) performed better than Feldman's algorithm, and all \sys\ methods outperformed Calmon's algorithm quite significantly. This experiment shows that enforcing DP, while unnecessary, can severely affect the accuracy of a classifier and, even more importantly, introduce bias in sub-populations.

This experiment shows that enforcing DP, while unnecessary, can severely affect the accuracy of a classifier and, even more importantly, introduce bias in sub-populations. In this case, while the overall average of recidivism for protected and privileged groups became more balanced using these approaches, the classifier became unfair toward people with the same number of convictions and charge degrees. We note that the observed bias was toward the majority group. Note that \sys\'s MS approach did not perform well on either of these datasets (as opposed to the original data) because of data density. {Also note that for COMPAS data, \sys\ delivered better overall utility than the original data because, for Calmon's dataset (as opposed to the original data), we observed that dropping race indeed increased the accuracy of both RF and MLP classifiers.}

%% file: conc.tex
\vspace*{-0.1cm}
\section{Conclusions and Future Work}

\ignore{We considered a causal approach for fair ML, reducing it 
to a database repair problem.  We showed that conventional  associational and causal fairness metrics can over- and under-report discrimination.  We defined a new notion of fairness, called  as  {\em justifiable fairness},
that addresses shortcoming of the previous definitions and arguably the strongest notion of fairness that is testable from data. We then proved sufficient properties
for justifiable fairness and use these results to 
translate the properties into saturated conditional independences
that we can be seen as multivalued dependencies  with which to repair the data. We then propose multiple algorithms for implementing these repairs.  Our experimental results show that our algorithms not only outperform state-of-the-art
pre-processing approaches for fairness on our own metrics, but that they are also competitive with existing approaches on conventional metrics. We empirically show that our methods are robust to unseen test data.}

We considered a causal approach for fair ML, reducing it 
to a database repair problem.  We showed that conventional fairness metrics,
including some causal approaches, 
end up as variants of statistical parity
due to the assumptions they make, and that all associational metrics can over-
and under-report discrimination due to statistical anomalies such as Simpson's Paradox.

Instead, we make explicit the assumptions for admissible
variables --- variables through which it is
permissible for the protected attribute to influence the outcome. We use these assumptions to define a new notion of fairness
and to reason about previous definitions.  We then prove sufficient properties
for fairness and use these results to 
translate the properties into saturated conditional independences
 that we can interpret as multivalued dependencies  with which to repair the data. We then propose multiple algorithms for implementing these repairs by casting the problem in 
terms of Matrix Factorization and MaxSAT. 

Our experimental results show that our algorithms not only outperform state-of-the-art
 pre-processing approaches for fairness on our own metrics, but that they are also competitive with existing approaches on conventional metrics. We empirically show that our methods are robust to unseen test data.
   Our results represent an initial attempt to link the language of causality with  database dependencies. 
 
   In future work, we aim to study the effect of training databases that are non-representative of the underling population on our results.
Currently, our proofs assume that the classifier approximates the true distribution, which 
is a common assumption in the machine learning literature.  However, another important source of discrimination  is selection bias or non-representativeness, which we must also correct.  Our
methods do correct for these forms of bias empirically, but we aim to prove bounds on the fairness
metrics based on divergence between training data and test data.

%% file: appx.tex
\vspace*{-0.3cm}
\section{Appendix}

\subsection{Additional Background}
\label{app:addback}

\paragraph{\bf New Proof of Impossibility Result in ~\cite{chouldechova2017fair}}
Chouldechova~\cite{chouldechova2017fair} proves the following {\em
impossibility result}: the Equalized Odds and Predictive Parity are
impossible to achieve simultaneously when prevalence of the two
populations differs, meaning $\pr(Y=1|S=0) \neq \pr(Y=1|S=1)$.  The
proof follows immediately from her observation that, for each
population group $S=i$, the following holds\footnote{EO implies
$FP/(1-FN)$ is the same for both groups,
$\frac{\pr(O=1|S=0,Y=0)}{\pr(O=1|S=0,Y=1)}=\frac{\pr(O=1|S=1,Y=0)}{\pr(O=1|S=1,Y=1)}$,
while PP implies that $(1-PPV)/PPV$ is the same for both groups,
$\frac{\pr(Y=0|O=1,S=0)}{\pr(Y=1|O=1,S=0)}=\frac{\pr(Y=0|O=1,S=1)}{\pr(Y=1|O=1,S=1)}$.
When the prevalence differs, EO and PP cannot hold simultaneously.}:
\begin{align*}
 \frac{\pr(O=1|S=i,Y=0)}{\pr(O=1|S=i,Y=1)} =
 &  \frac{\pr(Y=1|S=i)}{\pr(Y=0|S=i)}  \frac{\pr(Y=0|O=1,S=i)}{\pr(Y=1|O=1,S=i)}
\end{align*}

The following provides a simple alternative proof of the impossibility result using conditional independence.
\begin{prop} \label{prop:imp}  For any probability distribution $\pr$, if $S \indep O| Y$ and $S \indep Y| O$ then $S \indep Y$.
\end{prop}
\begin{myproof}{Proposition}{\ref{prop:imp}}  \em
From  $S \indep O| Y$ and  $S \indep Y| O$  it follows that $\pr(S|Y)=P(S|O)$ (1), which in turns implies $\pr(Y|S)\pr(O)=\pr(O|S)\pr(Y)$ (apply Bayes rule to the both sides of (1)). By summarization over $O$ we get $\pr(Y|S)=\pr(Y)$, which completes the proof.
\end{myproof} 
\paragraph*{\bf Implication Problem for CIs} The implication problem for CI is the problem of deciding whether  a CI $\varphi$ is logically follows from  a set of CIs $\Sigma$, meaning that in every distribution in which $\Sigma$ holds, $\varphi$ also holds. The following set of sound but incomplete axioms, known as {\em Graphoid}, are given in \cite{pearl1985graphoids} for this implication problem.

	Suppose $\att$ consists of a set of protected attributes $\mb P$ such as race and gender;  a set of  attributes $\mb X$ that might be used for decision making, e.g., credit score; a binary outcome attribute $Y$, e.g., good or bad credit score.
	Assume a classifier is trained on $S \subseteq \mb P \cup \mb X$ to predict $Y$.  Suppose $O$ consists of the classifier decisions. Throughout this paper we assume the classifier provides a good appropriation of the conditional distribution of $Y$, i.e.,  $\pr(Y|S) \approx \pr(O|S)$. 

\begin{itemize}

	\item (\bf Symmetry)
	\begin{align}
	(\mb X \indep \mb Y| \mb Z) \rightarrow (\mb Y \indep \mb X| \mb Z)  \label{ax:sy}
	\end{align}
	\item (\bf Decomposition)
	\begin{align}
	(\mb X \indep \mb W \mb Y|  \mb Z)  \rightarrow (\mb X \indep \mb W|  \mb Z)  \label{ax:dec}
	\end{align}
	\item (\bf	Weak Union)
	\begin{align}
	(\mb X \indep \mb W \mb Y|  \mb Z)  \rightarrow 	(\mb X \indep  \mb Y|  \mb Z \mb W)  \label{ax:wu}
	\end{align}

	\item (Contraction)
	\begin{align}
	(	\mb X \indep \mb Y|  \mb W \mb Z) \land  (\mb X \indep \mb W|  \mb Z) \rightarrow (\mb X\indep \mb Y \mb W | \mb Z)	 \label{ax:con}
	\end{align}
\end{itemize}

For strictly positive distribution in addition to the above the following axiom also holds:

{
\begin{itemize}
	\item ({\bf Intersection})
	\begin{align}
	(	\mb X \indep \mb Y|  \mb W \mb Z) \land  (\mb X \indep \mb W| \mb Y \mb Z) \rightarrow (\mb X\indep \mb Y \mb W | \mb Z)	 \label{ax:int}
	\end{align}
\end{itemize}
}

\paragraph*{\bf  Causal Models}  A probabilistic causal model (PCM) is a tuple
$\cm = \langle \mb U, \mb V , \mb F, \pr_{\mb U} \rangle$, where
$\mb U$ is a set of background or exogenous variables that cannot be
observed but which can influence the rest of the model; $\mb V$ is a
set of observable or endogenous variables;
$\mb F = (F_X)_{X \in \mb V}$ is a set of \emph{structural functions}
$F_X : Dom(\mb{Pa}_{\mb U}(X)) \times Dom(\mb{Pa}_{\mb V}(X))
\rightarrow Dom(X)$,
where $\mb{Pa}_{\mb U}(X) \subseteq \mb{U}$ and
$\mb{Pa}_{\mb V}(X) \subseteq \mb V-\set{X}$ are called the exogenous
parents and endogenous parents of $X$ respectively; and $\pr_{\mb U}$
is a joint probability distribution on the exogenous variables
$\mb U$.  Intuitively, the exogenous variables $\mb U$ are not known,
but we know their probability distribution, while the endogenous
variables are completely determined by their parents (exogenous and/or endogenous).

\paragraph*{\bf Causal DAG}  To each PCM $\cm$ we associate a causal graph $\cg$ with nodes
consisting of the endogenous variables $\mb V$, and edges consisting
of all pairs $(Z,X)$ such that $Z \in \mb{Pa}_{\mb V}(X)$; we write
$Z \rightarrow X$ for an edge.  $\cg$ is always assumed to be acyclic,
and called Causal DAG.  One can show that the probability distribution
on the exogenous variables uniquely determined a distribution
$\pr_{\mb V}$ on the endogenous variables and, under the {\em causal
	sufficiency} assumption\footnote{The assumption requires that, for
	any two variables $X, Y \in \mb{V}$, their exogenous parents are
	disjoint and independent
	$\mb{Pa}_{\mb U}(X) \indep \mb{Pa}_{\mb U}(Y)$.  When this
	assumption fails, one adds more endogenous variables to the model to
	expose their dependencies.}, $\pr_{\mb V}$ forms a Bayesian network,
whose graph is exactly $\cg$:

\begin{align}
\pr(\mb V) = & \prod_{X \in \mb V} \pr(X | \mb{Pa}(X)) \label{eq:bayesian}
\end{align}

Thus justifies omitting the exogenous variables from the causal DAG,
and capturing their effect through the probability distribution
Eq.(\ref{eq:bayesian}).  We will only refer to endogenous variables,
and drop the subscript $\mb{V}$ from $\mb{Pa}_{\mb{V}}$ and
$\pr_{\mb{V}}$.  A path in $\cg$ means an undirected path, i.e. we may
traverse edges either forwards or backwards; a directed path is one
where we traverse edges only forwards.

\paragraph*{\bf d-Separation} We review the notion of d-separation,
which is the graph-theoretic characterization of conditional
independence.  A {\em path} $\mb{P}$ from $X$ to $Y$ is a sequence of
nodes $X = V_1, \ldots, V_\ell = Y$ such that
$V_i \rightarrow V_{i+1}$ or $V_i \leftarrow V_{i+1}$ forall $i$.
$\mb P$ is {\em directed} if $V_i \rightarrow V_{i+1}$ forall $i$, and
in that case we write $X \stackrel{*}{\rightarrow} Y$, and say that
$X$ is an {\em ancestor}, or a {\em cause} of $Y$, and $Y$ is a {\em
  descendant} or an {\em effect} of $X$.  If the path contains a
subsequence $V_{k-1} \rightarrow V_k \leftarrow V_{k+1}$ then $V_k$ is
called a {\em collider}.  A path with a collider is {\em closed};
otherwise it is {\em open}; an open path has the form
$X \stackrel{*}{\leftarrow} \stackrel{*}{\rightarrow} Y$, i.e.  $X$
causes $Y$ or $Y$ causes $X$ or they have a common cause.  Given two
sets of nodes $\mb{X},\mb{Y}$ we say that a set $\mb{Z}$ {\em
  d-separates}\footnote{d stands for ``directional''.}  $\mb X$ and
$\mb Y$, denoted by $(\mb{X} \indep \mb{Y} |_d \ \mb{Z})$, if for any
all paths $P$ from $\mb{X}$ to $\mb{Y}$ one of the followings hold:
(1) $\mb P$ is closed at a collider node $V$ such that neither $V$ nor
any of its descendants are in $\mb Z$; (2) $\mb P$ contains a
non-collider node $V'$ such that $V' \in \mb P$. We say that a set
$\mb{Z}$. d-Separation is a sufficient condition for conditional
independence, see Prop.~\ref{prop:d:separation}


\paragraph*{\bf General Identification Criterion} In the presence of unrecorded variables in the causal DAG, the effect of interventions can not be identified using Eq.~\ref{eq:af}. A set of sound and complete axioms known as {\em $do$-calculus} can be used to decide whether the effect of intervention can be identified from the observed distribution \cite{pearl2009causality}. If the effect is identifiable  then by repeatedly applying the rules of do-calculus, one can obtain an statement {\em equivalent} to Eq.~\ref{eq:af}, but free from unobserved variables. Since identification is not the focus of this paper,  we assume all variables are observed hence Theorem \ref{theo:af} is sufficient for identification.

\paragraph*{\bf Markov Blanket}   We briefly review the notion of Markov
blanket, which used in Sec~\ref{sec:jfc}.
\vspace*{-0.1cm}
\begin{defn} \cite{pearl2014probabilistic} Fix a joint probability
	distribution $\pr(\mb v)$ and a variable $X \in \mb V$.  A set of
	variables $\mb  B(X) \subseteq \mb V - \set{X}$ is called a {\em Markov
		Blanket} of $X$ if $(X \indep \mb V - \mb  B(X) - \set{X} | \mb  B(X))$;
	it is called a {\em Markov Boundary} if it is minimal w.r.t. set
	inclusion,  denoted
	$\mmb(X)$.
\end{defn}

In the admission process in Fig~\ref{ex:cmexm}	$\mmb(X)=\{D,H\}$, simply because $O \indep G| H, D$ (since $\set{H,D}$ d-separate $O$ and $G$).
It is known that if $\pr$ is a strictly positive distribution (i.e., forall $\mb v \in Dom(\mb V), P(\mb v)>0$), then $\mmb(V)$ is unique for all $V \in \mb V$  and can be learned from data in polynomial time \cite{margaritis2003learning}. Strictly positive distributions do not allows for logical functional dependencies between their variables. The requirement can be satisfied in data by removing logical dependencies \cite{salimi2018bias}. Note that under the faithfulness assumption, the Markov boundary of a node $X$ in the causal graph consists of the parents of $X$, the children of $X$, and the parents of the children of $X$ \cite{neapolitan2004learning}. 

Note that if $\mb X$ is the set of all inputs of the algorithm included in data, then $\mmb(O)=\mb X$, i.e., parents of $X$ form a boundary for $O$. This is because $(O \indep  \mb V-\mb X|\mb X)$ is implied from the functional dependency (FD) $\mb X \fd O$, which can be discovered from data by a linear search through the attributes. Hence, ROD essentially requires that conditioned on admissible inputs of $\mb X \cap \mb A$, the outcome of algorithm becomes independent of the protected attribute, meaning the algorithm treats similarly individuals that are similar on $\mb X \cap \mb A$ characteristics but  different in $S$. Any imbalance indicated by ROD is worrisome and requires scrutiny.  Also note that in general Markov boundary can be learned from data by the linear number of iterations through the variables \cite{margaritis2003learning}.

\begin{prop}  \label{prop:umb}  Fix a strictly positive probability distribution $\pr_{\mb V}$.   It holds that for any variable $V \in \mb V$, the unique Markov boundary of $V$, $\mmb(V)$ is unique.
\end{prop}

\begin{myproof}{Proposition}{\ref{prop:umb}}{  The uniqueness of the Markov Boundary is implied from the intersection axiom (cf.  Eq~\ref{ax:int}) as follows:
Without loss of generality suppose $\mb V= V\mb X \mb W \mb Z \mb Y$, where $\mb X$ and $\mb Y$ are disjoint. We show that   if two sets $\mb X \mb Z$  and $\mb Y \mb Z$ form a Markov boundary for $V$ then their intersection, i.e, $Z$ is  a Markov blanket for $V$. This contradicts the subset minimality of a Markov boundary.  We show this in the following steps using Graphoid axioms.

			\begin{align}
	  V &\indep \mb W, \mb Y|	\mb X \mb Z  \label{mb1} \\
		  V &\indep \mb W, \mb X|	\mb Y \mb Z   \label{mb2} \\
		  V &\indep \mb Y, \mb X|	\mb Z   \hspace*{0.5cm}  \label{inter}  \text{By  (\ref{mb1}) ,  (\ref{mb2})  Dec. and Inter. axioms} \\
	  V &\indep \mb W|	\mb X \mb Z \mb Y \hspace*{0.3cm} \label{wu}  \text{By  (\ref{mb1}) and Weak Union} \\
	  	  V &\indep \mb W\mb X\mb Y |	\mb Z  \hspace*{0.3cm}    \text{By  (\ref{inter}) ,  (\ref{wu}) and Contraction}  \nonumber
	\end{align}

Therefore, Markov Boundary of $V$ is unique.
}
\end{myproof}

	 \paragraph*{\bf Counterfactual Fairness}   Given a set of features $\mb X$, a protected attribute $S$, an outcome variable $Y$, and a set of unobserved  background  variables $\mb U$, Kusner et
	al.~\cite{kusner2017counterfactual} defined a predictor $\tilde Y$
	to be {\em counterfactually fair} if for any $\mb x \in Dom(\mb X)$:
	{ 
		\begin{align}
		P( \tilde{Y}_{S\leftarrow 0}(\mb U) = 1| \mb X = \mb x,S = 1) = P( \tilde{Y}_{S\leftarrow 1}(\mb U)= 1| \mb X = \mb x;S = 1) \label{eq:cfair}
		\end{align}
	}
	\noindent  where, $\tilde{Y}_{S\leftarrow s}(\mb U)$ means intervening on
	the protected attribute in an unspecified configuration of the exogenous factors.
	The definition meant to capture the requirement that the protected attribute $S$ should not be a cause of $\tilde{Y}$ at individual level.  However, it fails on the simple example in Ex~\ref{ex:intsprox}. This is because, 
	$P({O}_{G\leftarrow g}(U_O) = 1)=  P(U_O=1) P({Y}_{G\leftarrow g}(U_O) = 1 | U_O=1)=\frac{1}{2}$ for $g=\set{0,1}$. We note that the stricter version of counterfactual fairness in \cite{DBLP:journals/corr/KusnerLRS17} also fails to capture the individual-level unfairness in this example. We report that this observation has been confirmed by the authors of \cite{DBLP:journals/corr/KusnerLRS17}.  We defer the full comparison for future work.

\begin{myproof}{Lemma}{\ref{prop:gen}} 
	Because the classifier is a deterministic function trained on $\pr_{B'}$, it follows that $\pr_{T}(o,s,i,a)=\pr_{B'}(o,s,i,a) = \pr_{B'}(o|a) \pr_{B'}(i,s,a)$.
  Hence it is sufficient to show that  $\kld{\pr_T(o|s,i,a)}{\pr_T(o|a)} = 0$ or $\pr_T(o|s,i,a)=\pr_T(o|a)$.
  We show this in the following steps:
%
{ 
	\begin{align}  
	\pr_T(o|s,i,a)=& \frac{\pr_T(o,s,i,a)}{\pr_T(s,i,a)} \\
	=& \frac{\pr_{B'}(o,s,i,a)}{\pr_T(s,i,a)} \\
	=& \frac{\pr_{B'}(o|a) \pr_{B'}(s,i,a) }{\pr_T(s,i,a)} \label{rhs}
	\end{align}
}
{  
	\begin{align}
	\pr_T(o|a)=& \frac{\pr_T(o,a)}{\pr_T(a)} \\
	=& \frac{ \sum_{s,i} \pr_{B'}(o,i,s,a)}{\pr_T(a)} \\
	=& \frac{ \sum_{s,i} \pr_{B'}(o|a) \pr_{B'}(s,i, a) }{\pr_T(a)}   \\
	=& \frac{\pr_{B'}(o|a) \pr_{B'}(a) }{\pr_T(a)} \label{leh}
	\end{align}
}
	Hence,
{  
	\begin{align}   
	\kld{pr_T(o|s,i,a)}{pr_T(o|a)}& \\
	& \hspace*{-3cm} =  - \sum   \pr_T(o|i,s,a) \log  \frac{\pr_T(o|a)}{\pr_T(o|s,i,a)} \\
	& \hspace*{-3cm} =  - \sum  \pr_T(o|i,s,a)  \log \frac{ \frac{\pr_{B'}(o|a) \pr_{B'}(a) }{pr_T(a)}} {\frac{\pr_{B'}(o|a) \pr_{B'}(i,s,a) }{\pr_T(s,i,a)}} \\
	& \hspace*{-3cm} =  - \sum  \pr_T(o|i,s,a) \log \frac{\pr_{B'}(s,i|a)}{\pr_T(s,i|a)}  \label{ratio}
	\end{align}
}
	Thus,  $\kld{\pr_T(o|s,i,a)}{\pr_T(o|a)} = 0$ if $\pr_T(s,i|a)  \pr_{B'}(s,i|a)$,
  which implies $\pr_T(o|s,i,a)=\pr_T(o|a)$
	or equivalently that $(Y \indep S,\mb I|_{\Pr_T} \mb A)$. This completes the proof.

\end{myproof}

\subsection{\bf Proofs and Supplementary Propositions and graphs}
\label{app:proof}

\begin{myproof}{Theorem}{\ref{theo:af}}  Recall that a causal $\cg$ admits the following factorization of the observed distribution:
{ 	\begin{eqnarray}
	\Pr(\mb v)= \prod_{V \in \mb V} \Pr(v| \mb{pa}(V)) \label{eq:fac}
	\end{eqnarray}}
	Now, each atomic intervention $do(X=x)$ modifies the causal DAG  $\cg$ by removing parents of $X$ from $\cg$.  Therefore, the probability distribution $P(\mb v| do(\mb X= \mb x))$ can be obtained from the observed  distribution $P(\mb v)$  by removing all factors
	$\Pr(x|\mb{pa}(X))$, for $X \in \mb X$, from $P(\mb v)$, i.e.,
{ 		\begin{align}
	\Pr(\mb v| do(\mb X= \mb x)) & = &\frac{\Pr(\mb v)}{ \prod_{i=0}^{m} \pr(x_i|\mb{pa}(X_i))}
	\label{ps:inter}
	\end{align}}
	The following holds according to the chain rule of probability:
	{  
	\begin{align}
	\Pr(\mb v) & = & \prod_{i=0}^{m}  \bigg(\Pr\big(\mb{pa}(X_i)|  \bigcup_{j=0}^{i-1} \mb{pa}(X_j ), \bigcup_{j=0}^{i-1}x_j  \big) \bigg) \bigg(  \Pr(x_i|  \bigcup_{j=0}^{i} \mb{pa}(X_j), \bigcup_{j=0}^{i-1}x_j) \bigg) \nonumber \\
	& & \pr(\mb w| \mb x, \mb z)
	\label{ps:chain}
	\end{align}
}
	\noindent where, $\mb Z= \bigcup_{X \in \mb X} \mb{Pa}(X)$, $\mb W=\mb V-(\mb X \cup \mb Z)$ and $j \geq0$. \ignore{For example for $\mb X=\{X_0,X_1\}$ we have:
	\begin{align}
	\Pr(\mb v) & = & \Pr(\mb{pa}(X_0))   \Pr(x_0|\mb{pa}(X_0)) \nonumber  \\ & & \Pr(\mb{pa}(X_1)|\mb{pa}(X_0),x_0)   \Pr(x_1| \mb{pa}(X_0),\mb{pa}(X_1),x_0)   \nonumber  \\
	& &\pr(\mb w| \mb x, \mb z)\nonumber
	\end{align}}
	It holds that in a causal DAG $\cg$, any node $ X \in \mb V $ is independent of its non-descendant condition on it parents $\mb{Pa}(X)$ (known as Markov property \cite{pearl2014probabilistic}). This is simply because  $\mb{Pa}(X)$ d-separates $X$ from its non-descendants.  Therefore, the following is implied from
	the assumption that $X_i$ is an non-descendant of $X_{i+1}$:
		{ 
	\begin{align}
	\Pr(x_i|  \bigcup_{j=0}^{i} \mb{pa}(X_j), \bigcup_{j=0}^{i-1}x_j) = \Pr(x_i|   \mb{pa}(X_i) )
	\ \text{for} \ i=0,m
	\label{ps:indep}
	\end{align}}
	Hence,
	{ 
	\begin{align}
	\Pr(\mb v) & = &   \bigg( \prod_{i=0}^{m} \Pr\big(\mb{pa}(X_i)|  \bigcup_{j=0}^{i-1} \mb{pa}(X_j ), \bigcup_{j=0}^{i-1}x_j  \big) \bigg) \bigg( \prod_{i=0}^{m} \Pr(x_i|  \mb Pa(X_i)) \bigg) \nonumber \\
	& & \pr(\mb w| \mb x, \mb z)
	\label{ps:schain}
	\end{align}}

	The following implied from Eq.~\ref{ps:schain} and \ref{ps:inter}.
		{ 
	\begin{align}
	\Pr(\mb v| do(\mb X= \mb x)) =& \frac{\Pr(\mb v)}{ \prod_{i=0}^{m} \pr(x_i|\mb{pa}(X_i))}  \nonumber \\
	 &= \bigg( \prod_{i=0}^{m} \Pr\big(\mb{pa}(X_i)|  \bigcup_{j=0}^{i-1} \mb{pa}(X_j ), \bigcup_{j=0}^{i-1}x_j  \big) \bigg) \pr(\mb w| \mb x, \mb z)
	\label{ps:simp}
	\end{align}}
	Now,  by summation over all variables except for $Y$ and $\mb X$  in Eq.~\ref{ps:simp} we obtain the following, which proves the theorem.
	{ 	\begin{align}
	P(y|do(\mb X= \mb x ))= \sum_{ \mb z \in \mb Z } \Pr( y | \mb x, \mb z ) \bigg( \prod_{i=0}^{m} \Pr\big(\mb{pa}(X_i)|  \bigcup_{j=0}^{i-1} \mb{pa}(X_j ), \bigcup_{j=0}^{i-1}x_j  \big) \bigg)
	\end{align}}
\end{myproof}

%

	\begin{myproof}{Proposition} {\ref{th:jfcd}} \em In one direction, we note that, for any choice of
		$\mb K$, the causal graph corresponding to an intervention
		$do(\mb K = \mb k)$ disconnects $S$ and $O$, and therefore
		intervening on $S$ does not affect $O$.  In the other direction, let
		$\mb P$ be a path from $S$ to $O$ s.t.
		$\mb P \cap \mb A = \emptyset$, and let $\mb K$ be the set of all
		variables not in $\mb P$; in particular, $\mb A \subseteq \mb K$.
		The causal graph corresponding to an intervention on $\mb K$
		consists of a single path $S \rightarrow^* O$ because all other
		edges are removed by the intervention.  Since $S$ has no parents,
		intervening on $S$ is the same as conditioning on $S$, and, since
		$\pr$ is faithful, we have $\pr(O=o|S=0) \neq \pr(O=0|S=1)$ for some
		outcome $O=o$, contradicting the assumption of
		$\mb K$-fairness.
	\end{myproof}
\begin{myproof}{Theorem}{\ref{theo:suff_jf}}  We show that an algorithm $\mc A$ is $\mb A$-fair if  $\mmb(O) \subseteq \mb A$. From Theorem~\ref{theo:af}, we obtain:
	{ 	\begin{eqnarray}
	\pr(O=o| do(S=i),do(\mb A = \mb a))= \sum_{ \mb z \in Dom(\mb Z)}  \pr( y | S=i, \mb A=\mb a, \mb z )  \nonumber \\  \bigg( \prod_{i=0}^{m} \pr\big(\mb{pa}(X_i)\bigg|  \bigcup_{j=0}^{i-1} \mb{pa}(X_j ), \bigcup_{j=0}^{i-1}x_j  \big) \bigg)
	\end{eqnarray}}
\noindent	where,  $\mb Z= \bigcup_{A \in \mb A} \mb{Pa}(A)$. Without loss of generality assume
	$\mb Z \cap \mb A' = \emptyset$. Let $\mb A=\mmb(O) \cup  \mb A'$ and $ \mb V'=\mb V-\{ \mb A' \cup \mb Z \cup \set{S} \}$. From the definition of Markov boundary we have $(O \indep \mb V', \mb A', S, \mb Z |  \mmb(O))$. It follows from Decomposition and Weak  Union axioms in Graphoid that $(O \indep  S, \mb Z |  \mmb(O), \mb A')$, hence
	$(O \indep  S, \mb Z |  \mb A)$. We obtain the following for $i=\{0,1\}$:
	{ 	\begin{eqnarray}
	\pr(O=o| do(S=i),do(\mb A = \mb a)) &=&  \pr( y | A=\mb a ) \nonumber \\ &&  \hspace*{-2cm}\sum_{ \mb z \in Dom(\mb Z)}     \bigg( \prod_{i=0}^{m} \pr\big(\mb{pa}(X_i)\bigg|   \bigcup_{j=0}^{i-1} \mb{pa}(X_j ), \bigcup_{j=0}^{i-1}x_j  \big) \bigg)  \nonumber \\
	&=&  \pr( y | A=\mb a)  \label{eq:sum}  \noindent
	\end{eqnarray}}
	Note that (\ref{eq:sum}) obtained by the fact that each product inside the summation becomes 1 (simply because $\sum_X \pr(X|Y)=1$) This proves the $\mb A$-fairness of $\mc A$. $\mb K$-fairness for each $\mb K \supseteq \mb A$ can be proved in a similar way.
\end{myproof}

	\begin{myproof}  {Corollary}{\ref{col:fair_learn}}{Without loss of generality, suppose $\mb V=Y\mb D\mb Z \mb W\mb U $ with $\mb X= \mb A \cup \mb Z$ and $\mb A= \mb W \cup \mb Z$. Since the classifier is trained on $\mb X$, there is a functional dependency $\mb X \fd O$, which implies $(O \indep \mb  Y, \mb W, \mb U |  \mb A,\mb Z)$(1), i.e.,  $\mb X$ forms a Markov blanket for $O$. It is also implied from the assumptions $\pr(Y=1| \mb X=\mb x) \approx \pr(O=1| \mb X=\mb x)$ and  $ (Y \indep \mb X-\mb A  | \mb{A}\cap X )$ that $ (O \indep \mb A | \mb Z)$ approximately holds (2). By applying the Contraction axiom in Graphoid to (1) and (2), we obtain  $ (O \indep Y \mb A, \mb  W, \mb U  | \mb Z)$ i.e., $\mmb(O) \subseteq \mb A$. Therefore, $\mc A$ is justifiably fair according to Theorem~\ref{theo:suff_jf}. This completes the proof of part (a). Part (b) is implied from  part(1), definition of Markov boundary and Decomposition axiom in Graphoid.}
\end{myproof}

\begin{prop}  \em \label{prop:deg}Given a \fairapp $(\mc A, S, \mb A, \mb I)$, suppose the probability distribution of $\mc A$ is faithful to the causal DAG.  Then, the application is justifiably fair iff $\delta(S;O| \linebreak \mb  \mmb(O) \cap \mb A)=1$.
\end{prop}

\begin{myproof}{Proposition}{\ref{prop:deg}} 
It is easy to see $\delta(S;O|  \mb \mmb(O)$ $\cap \mb A)=1$ iff  $S \indep O|  \mb   \mmb(O) \cap \mb A$. Under the faithfulness assumption, we obtain  $\mmb(O) \cap \mb A$ and d-separates $S$ and $O$. Hence, all directed paths from $S$ to $O$ go thorough $\mb  \mmb(O) \cap \mb A$. Therefore, the algorithm is justifiably fair according to Theorem \ref{th:jfcd}. The converse is immediate from the  natural assumption that $O$ does not have any descendants in the causal DAG; hence, its Markov boundary consists of the algorithm's inputs.
\end{myproof}

\begin{myproof}{Proposition}{\ref{prop:min_rep_lin}} 
	The proposition follows from three facts, all easily verified. (1)
	$D \subseteq \hb$, (2) $\hb$ satisfies the MVD $\mb Z \mvd \mb X$,
	and (3) If two databases $D_1, D_2$ satisfy the MVD then so does
	$D_1 \cap D_2$.  Indeed, the three facts imply that, for any repair
	$D'$, the database $\hb \cap D'$ is also a repair and
	$|\Delta(D, \hb \cap D')| \leq \Delta(D, D')|$, hence, if $D'$ is a
	minimal repair, then $D' \subseteq \hb$.
\end{myproof}

%% file: main.bbl
\begin{thebibliography}{10}

\bibitem{AbiteboulHVBook}
Serge Abiteboul, Richard Hull, and Victor Vianu.
\newblock {\em Foundations of Databases}.
\newblock Addison-Wesley, 1995.

\bibitem{avin2005identifiability}
Chen Avin, Ilya Shpitser, and Judea Pearl.
\newblock Identifiability of path-specific effects.
\newblock 2005.

\bibitem{DBLP:series/synthesis/2011Bertossi}
Leopoldo~E. Bertossi.
\newblock {\em Database Repairing and Consistent Query Answering}.
\newblock Synthesis Lectures on Data Management. Morgan {\&} Claypool
  Publishers, 2011.

\bibitem{bodie2017law}
Matthew~T Bodie, Miriam~A Cherry, Marcia~L McCormick, and Jintong Tang.
\newblock The law and policy of people analytics.
\newblock {\em U. Colo. L. Rev.}, 88:961, 2017.

\bibitem{calders2009building}
Toon Calders, Faisal Kamiran, and Mykola Pechenizkiy.
\newblock Building classifiers with independency constraints.
\newblock In {\em Data mining workshops, 2009. ICDMW'09. IEEE international
  conference on}, pages 13--18. IEEE, 2009.

\bibitem{calders2010three}
Toon Calders and Sicco Verwer.
\newblock Three naive bayes approaches for discrimination-free classification.
\newblock {\em Data Mining and Knowledge Discovery}, 21(2):277--292, 2010.

\bibitem{NIPS2017_6988}
Flavio Calmon, Dennis Wei, Bhanukiran Vinzamuri, Karthikeyan
  Natesan~Ramamurthy, and Kush~R Varshney.
\newblock Optimized pre-processing for discrimination prevention.
\newblock In I.~Guyon, U.~V. Luxburg, S.~Bengio, H.~Wallach, R.~Fergus,
  S.~Vishwanathan, and R.~Garnett, editors, {\em Advances in Neural Information
  Processing Systems 30}, pages 3992--4001. Curran Associates, Inc., 2017.

\bibitem{chang2017meta}
Bei-Hung Chang and David~C Hoaglin.
\newblock Meta-analysis of odds ratios: Current good practices.
\newblock {\em Medical care}, 55(4):328, 2017.

\bibitem{chouldechova2017fair}
Alexandra Chouldechova.
\newblock Fair prediction with disparate impact: A study of bias in recidivism
  prediction instruments.
\newblock {\em Big data}, 5(2):153--163, 2017.

\bibitem{corbett2017algorithmic}
Sam Corbett-Davies, Emma Pierson, Avi Feller, Sharad Goel, and Aziz Huq.
\newblock Algorithmic decision making and the cost of fairness.
\newblock In {\em Proceedings of the 23rd ACM SIGKDD International Conference
  on Knowledge Discovery and Data Mining}, pages 797--806. ACM, 2017.

\bibitem{courtland2018bias}
Rachel Courtland.
\newblock Bias detectives: the researchers striving to make algorithms fair.
\newblock {\em Nature}, 558, 2018.

\bibitem{amazonhire2018}
Jeffrey Dastin.
\newblock Rpt-insight-amazon scraps secret ai recruiting tool that showed bias
  against women.
\newblock {\em Reuters}, 2018.
\newblock
  {\tiny{\url{https://www.reuters.com/article/amazoncom-jobs-automation/rpt-insight-amazon-scraps-secret-ai-recruiting-tool-that-showed-bias-against-women-idUSL2N1WP1RO}}}.

\bibitem{dwork2012fairness}
Cynthia Dwork, Moritz Hardt, Toniann Pitassi, Omer Reingold, and Richard Zemel.
\newblock Fairness through awareness.
\newblock In {\em Proceedings of the 3rd innovations in theoretical computer
  science conference}, pages 214--226. ACM, 2012.

\bibitem{feldman2015certifying}
Michael Feldman, Sorelle~A Friedler, John Moeller, Carlos Scheidegger, and
  Suresh Venkatasubramanian.
\newblock Certifying and removing disparate impact.
\newblock In {\em Proceedings of the 21th ACM SIGKDD International Conference
  on Knowledge Discovery and Data Mining}, pages 259--268. ACM, 2015.

\bibitem{fevotte2011algorithms}
C{\'e}dric F{\'e}votte and J{\'e}r{\^o}me Idier.
\newblock Algorithms for nonnegative matrix factorization with the
  $\beta$-divergence.
\newblock {\em Neural computation}, 23(9):2421--2456, 2011.

\bibitem{galhotra2017fairness}
Sainyam Galhotra, Yuriy Brun, and Alexandra Meliou.
\newblock Fairness testing: testing software for discrimination.
\newblock In {\em Proceedings of the 2017 11th Joint Meeting on Foundations of
  Software Engineering}, pages 498--510. ACM, 2017.

\bibitem{hardt2016equality}
Moritz Hardt, Eric Price, Nati Srebro, et~al.
\newblock Equality of opportunity in supervised learning.
\newblock In {\em Advances in neural information processing systems}, pages
  3315--3323, 2016.

\bibitem{hartung1999note}
Joachim Hartung.
\newblock A note on combining dependent tests of significance.
\newblock {\em Biometrical Journal: Journal of Mathematical Methods in
  Biosciences}, 41(7):849--855, 1999.

\bibitem{amazonrace2016}
David Ingold and Spencer Soper.
\newblock Amazon doesn't consider the race of its customers. should it?
\newblock {\em Bloomberg}, 2016.
\newblock \url{www.bloomberg.com/graphics/2016-amazon-same-day/}.

\bibitem{kamiran2009classifying}
Faisal Kamiran and Toon Calders.
\newblock Classifying without discriminating.
\newblock In {\em Computer, Control and Communication, 2009. IC4 2009. 2nd
  International Conference on}, pages 1--6. IEEE, 2009.

\bibitem{kamishima2012fairness}
Toshihiro Kamishima, Shotaro Akaho, Hideki Asoh, and Jun Sakuma.
\newblock Fairness-aware classifier with prejudice remover regularizer.
\newblock In {\em Joint European Conference on Machine Learning and Knowledge
  Discovery in Databases}, pages 35--50. Springer, 2012.

\bibitem{kilbertus2017avoiding}
Niki Kilbertus, Mateo~Rojas Carulla, Giambattista Parascandolo, Moritz Hardt,
  Dominik Janzing, and Bernhard Sch{\"o}lkopf.
\newblock Avoiding discrimination through causal reasoning.
\newblock In {\em Advances in Neural Information Processing Systems}, pages
  656--666, 2017.

\bibitem{kusner2017counterfactual}
Matt~J Kusner, Joshua Loftus, Chris Russell, and Ricardo Silva.
\newblock Counterfactual fairness.
\newblock In {\em Advances in Neural Information Processing Systems}, pages
  4069--4079, 2017.

\bibitem{DBLP:journals/corr/KusnerLRS17}
Matt~J. Kusner, Joshua~R. Loftus, Chris Russell, and Ricardo Silva.
\newblock Counterfactual fairness.
\newblock {\em CoRR}, abs/1703.06856, 2017.

\bibitem{larson2016we}
Jeff Larson, Surya Mattu, Lauren Kirchner, and Julia Angwin.
\newblock How we analyzed the compas recidivism algorithm.
\newblock {\em ProPublica (5 2016)}, 9, 2016.

\bibitem{adult}
M.~Lichman.
\newblock Uci machine learning repository, 2013.

\bibitem{DBLP:conf/pods/LivshitsKR18}
Ester Livshits, Benny Kimelfeld, and Sudeepa Roy.
\newblock Computing optimal repairs for functional dependencies.
\newblock In {\em Proceedings of the 37th {ACM} {SIGMOD-SIGACT-SIGAI} Symposium
  on Principles of Database Systems, Houston, TX, USA, June 10-15, 2018}, pages
  225--237, 2018.

\bibitem{loftus2018causal}
Joshua~R Loftus, Chris Russell, Matt~J Kusner, and Ricardo Silva.
\newblock Causal reasoning for algorithmic fairness.
\newblock {\em arXiv preprint arXiv:1805.05859}, 2018.

\bibitem{loux2017comparison}
Travis~M Loux, Christiana Drake, and Julie Smith-Gagen.
\newblock A comparison of marginal odds ratio estimators.
\newblock {\em Statistical methods in medical research}, 26(1):155--175, 2017.

\bibitem{luong2011k}
Binh~Thanh Luong, Salvatore Ruggieri, and Franco Turini.
\newblock k-nn as an implementation of situation testing for discrimination
  discovery and prevention.
\newblock In {\em Proceedings of the 17th ACM SIGKDD international conference
  on Knowledge discovery and data mining}, pages 502--510. ACM, 2011.

\bibitem{margaritis2003learning}
Dimitris Margaritis.
\newblock Learning bayesian network model structure from data.
\newblock Technical report, Carnegie-Mellon Univ Pittsburgh Pa School of
  Computer Science, 2003.

\bibitem{martins2014open}
Ruben Martins, Vasco Manquinho, and In{\^e}s Lynce.
\newblock Open-wbo: A modular maxsat solver.
\newblock In {\em International Conference on Theory and Applications of
  Satisfiability Testing}, pages 438--445. Springer, 2014.

\bibitem{nabi2018fair}
Razieh Nabi and Ilya Shpitser.
\newblock Fair inference on outcomes.
\newblock In {\em Proceedings of the... AAAI Conference on Artificial
  Intelligence. AAAI Conference on Artificial Intelligence}, volume 2018, page
  1931. NIH Public Access, 2018.

\bibitem{neapolitan2004learning}
Richard~E Neapolitan et~al.
\newblock {\em Learning bayesian networks}, volume~38.
\newblock Pearson Prentice Hall Upper Saddle River, NJ, 2004.

\bibitem{pearl2009causality}
Judea Pearl.
\newblock {\em Causality}.
\newblock Cambridge university press, 2009.

\bibitem{pearl2014probabilistic}
Judea Pearl.
\newblock {\em Probabilistic reasoning in intelligent systems: networks of
  plausible inference}.
\newblock Morgan Kaufmann, 2014.

\bibitem{pearl2009causal}
Judea Pearl et~al.
\newblock Causal inference in statistics: An overview.
\newblock {\em Statistics Surveys}, 3:96--146, 2009.

\bibitem{pearl1985graphoids}
Judea Pearl and Azaria Paz.
\newblock {\em Graphoids: A graph-based logic for reasoning about relevance
  relations}.
\newblock University of California (Los Angeles). Computer Science Department,
  1985.

\bibitem{rubin1970thesis}
Donald~B Rubin.
\newblock {\em The Use of Matched Sampling and Regression Adjustment in
  Observational Studies}.
\newblock Ph.D. Thesis, Department of Statistics, Harvard University,
  Cambridge, MA, 1970.

\bibitem{rubin1986statistics}
Donald~B Rubin.
\newblock Statistics and causal inference: Comment: Which ifs have causal
  answers.
\newblock {\em Journal of the American Statistical Association},
  81(396):961--962, 1986.

\bibitem{rubin2008comment}
Donald~B Rubin.
\newblock Comment: The design and analysis of gold standard randomized
  experiments.
\newblock {\em Journal of the American Statistical Association},
  103(484):1350--1353, 2008.

\bibitem{russell2017worlds}
Chris Russell, Matt~J Kusner, Joshua Loftus, and Ricardo Silva.
\newblock When worlds collide: integrating different counterfactual assumptions
  in fairness.
\newblock In {\em Advances in Neural Information Processing Systems}, pages
  6414--6423, 2017.

\bibitem{salimi2018bias}
Babak Salimi, Johannes Gehrke, and Dan Suciu.
\newblock Bias in olap queries: Detection, explanation, and removal.
\newblock In {\em Proceedings of the 2018 International Conference on
  Management of Data}, pages 1021--1035. ACM, 2018.

\bibitem{salimi2019interventional}
Babak Salimi, Luke Rodriguez, Bill Howe, and Dan Suciu.
\newblock Interventional fairness: Causal database repair for algorithmic
  fairness.
\newblock In {\em Proceedings of the 2019 International Conference on
  Management of Data}, pages 793--810. ACM, 2019.

\bibitem{selbst2017disparate}
Andrew~D Selbst.
\newblock Disparate impact in big data policing.
\newblock {\em Ga. L. Rev.}, 52:109, 2017.

\bibitem{simoiu2017problem}
Camelia Simoiu, Sam Corbett-Davies, Sharad Goel, et~al.
\newblock The problem of infra-marginality in outcome tests for discrimination.
\newblock {\em The Annals of Applied Statistics}, 11(3):1193--1216, 2017.

\bibitem{tramer2017fairtest}
Florian Tramer, Vaggelis Atlidakis, Roxana Geambasu, Daniel Hsu, Jean-Pierre
  Hubaux, Mathias Humbert, Ari Juels, and Huang Lin.
\newblock Fairtest: Discovering unwarranted associations in data-driven
  applications.
\newblock In {\em Security and Privacy (EuroS\&P), 2017 IEEE European Symposium
  on}, pages 401--416. IEEE, 2017.

\bibitem{valentino2012websites}
Jennifer Valentino-Devries, Jeremy Singer-Vine, and Ashkan Soltani.
\newblock Websites vary prices, deals based on users’ information.
\newblock {\em Wall Street Journal}, 10:60--68, 2012.

\bibitem{vavasis2009complexity}
Stephen~A Vavasis.
\newblock On the complexity of nonnegative matrix factorization.
\newblock {\em SIAM Journal on Optimization}, 20(3):1364--1377, 2009.

\bibitem{Veale:2018:FAD:3173574.3174014}
Michael Veale, Max Van~Kleek, and Reuben Binns.
\newblock Fairness and accountability design needs for algorithmic support in
  high-stakes public sector decision-making.
\newblock In {\em Proceedings of the 2018 CHI Conference on Human Factors in
  Computing Systems}, CHI '18, pages 440:1--440:14, New York, NY, USA, 2018.
  ACM.

\bibitem{Verma:2018:FDE:3194770.3194776}
Sahil Verma and Julia Rubin.
\newblock Fairness definitions explained.
\newblock In {\em Proceedings of the International Workshop on Software
  Fairness}, FairWare '18, pages 1--7, New York, NY, USA, 2018. ACM.

\bibitem{personalitywsj2017}
Lauren Weber and Elizabeth Dwoskin.
\newblock Are workplace personality tests fair?
\newblock {\em Wall Strreet Journal}, 2014.

\bibitem{wong2000implication}
SK~Michael Wong, Cory~J. Butz, and Dan Wu.
\newblock On the implication problem for probabilistic conditional
  independency.
\newblock {\em IEEE Transactions on Systems, Man, and Cybernetics-Part A:
  Systems and Humans}, 30(6):785--805, 2000.

\bibitem{pmlr-v65-woodworth17a}
Blake Woodworth, Suriya Gunasekar, Mesrob~I. Ohannessian, and Nathan Srebro.
\newblock Learning non-discriminatory predictors.
\newblock In Satyen Kale and Ohad Shamir, editors, {\em Proceedings of the 2017
  Conference on Learning Theory}, volume~65 of {\em Proceedings of Machine
  Learning Research}, pages 1920--1953, Amsterdam, Netherlands, 07--10 Jul
  2017. PMLR.

\bibitem{xu2018provenance}
Jane Xu, Waley Zhang, Abdussalam Alawini, and Val Tannen.
\newblock Provenance analysis for missing answers and integrity repairs.
\newblock {\em Data Engineering}, page~39, 2018.

\bibitem{zafar2017fairness}
Muhammad~Bilal Zafar, Isabel Valera, Manuel Gomez~Rodriguez, and Krishna~P
  Gummadi.
\newblock Fairness beyond disparate treatment \& disparate impact: Learning
  classification without disparate mistreatment.
\newblock In {\em Proceedings of the 26th International Conference on World
  Wide Web}, pages 1171--1180. International World Wide Web Conferences
  Steering Committee, 2017.

\bibitem{pmlr-v54-zafar17a}
Muhammad~Bilal Zafar, Isabel Valera, Manuel~Gomez Rogriguez, and Krishna~P.
  Gummadi.
\newblock {Fairness Constraints: Mechanisms for Fair Classification}.
\newblock In Aarti Singh and Jerry Zhu, editors, {\em Proceedings of the 20th
  International Conference on Artificial Intelligence and Statistics},
  volume~54 of {\em Proceedings of Machine Learning Research}, pages 962--970,
  Fort Lauderdale, FL, USA, 20--22 Apr 2017. PMLR.

\bibitem{zemel2013learning}
Rich Zemel, Yu~Wu, Kevin Swersky, Toni Pitassi, and Cynthia Dwork.
\newblock Learning fair representations.
\newblock In {\em International Conference on Machine Learning}, pages
  325--333, 2013.

\bibitem{vzliobaite2011handling}
Indre {\v{Z}}liobaite, Faisal Kamiran, and Toon Calders.
\newblock Handling conditional discrimination.
\newblock In {\em Data Mining (ICDM), 2011 IEEE 11th International Conference
  on}, pages 992--1001. IEEE, 2011.

\end{thebibliography}
